\documentclass[10pt,a4]{article}
\usepackage[cp866]{inputenc}
\usepackage[T2A]{fontenc}
\usepackage[dvips]{graphicx}
\usepackage{amsmath,euscript,amssymb}
\newcommand{\be}{\begin{equation}}
\newcommand{\ee}{\end{equation}}
\newcommand{\bea}{\begin{eqnarray}}
\newcommand{\eea}{\end{eqnarray}}
\def\vh{\varphi}

\begin{document}

\begin{titlepage}

\title{\LARGE{Hamiltonian Unification of General Relativity
and Standard Model}}

\author{L.A. Glinka\footnote{Email to: glinka@theor.jinr.ru, laglinka@gmail.com}
~~and V.N. Pervushin\footnote{E-mail to: pervush@theor.jinr.ru}
\vspace*{10pt}\\
\textit{N.N. Bogoliubov Laboratory of Theoretical Physics}\\
\emph{Joint Institute for Nuclear Research}\\
\emph{6 Joliot-Curie Street, 141980 Dubna, Russia}}

\date{\today}
\maketitle

\begin{abstract}
The Hamiltonian approach to the General Relativity and the
Standard Model  is studied in the context of its consistency with
the Newton law,
 the Higgs effect,   the Hubble
cosmological evolution and the Cosmic Microwave Background
radiation physics.

The version of the Higgs potential is proposed, where its constant
parameter is replaced by the dynamic zeroth Fourier harmonic of
the very Higgs field. In this model, the extremum of the
Coleman--Weinberg effective potential obtained from the unit
vacuum--vacuum transition amplitude immediately predicts mass of
Higgs field and removes tremendous vacuum cosmological density.

We show that the relativity principles unambiguously treat
 the Planck epoch, in
the General Relativity, as the present-day one. It was shown that
there are initial data of the Electro-Weak epoch compatible with
supposition that all particles in the Universe
 are  final  products of decays of primordial Higgs particles and W-, Z-vector bosons
  created from vacuum  at the
 instant    treated as
the "Big-Bang".
\end{abstract}

\end{titlepage}

\tableofcontents
\newpage
\section*{Introduction}

The unification problem of the General Relativity (GR) and the
Standard Model (SM) is one of main questions of modern physics.
The main difficulty of this unification lies in the different
theoretical levels of their presentation: quantum for SM and
classical for GR. However, both these theories have common roots
of their origin (mechanics and electrodynamics) and
\emph{principles of relativity}, and both they are in agreement
with observational and experimental data.

It is worthwhile to recall here these common \emph{principles of
relativity} and their relation to observational and experimental
data. Actually, physics arises as a science about measurements and
observations. It supposes two distinguished reference frames - the
observer \emph{rest frame} and the observable \emph{comoving frame}.
In particular, in modern cosmology, comoving frame of the Universe
is identified with the Cosmic Microwave Background radiation frame
that differs from the rest frame by the nonzero dipole component of
the temperature fluctuations \cite{WMAP01}.

\emph{Differences} between these two frames underlie all
principles of relativity including the Copernicus -- Galilei
relativity as a \emph{difference} of initial positions and
velocities, and the Poincar\'e -- Einstein special relativity (SR)
\cite{poi,ein} as a \emph{difference} of measurable times in
different frames. Principles of relativity mean that there are
\emph{degrees of freedom} together with their motion equations and
initial data that are free from these equations. The Copernicus --
Galilei relativity means that these \emph{degrees of freedom} are
spatial coordinates of a particle. Equations of motion as
invariants of the Galilei group and the manifold of initial data
are the main concepts of the \emph{first physical theory} created
by Isaac Newton.

The Poincar\'e -- Einstein Special Relativity (SR) means that the
time coordinate is the \emph{degree of freedom} of a particle too,
so that the complete set of \emph{degrees of freedom} forms the
Minkowski \emph{space of events}. A geometric interval on the line
of a particle in this space of events is formed by its metric with
a single component (the \emph{lapse function}), and there are
\emph{reparametrizations of a coordinate evolution parameter}. The
Hilbert action-interval variational principle provides the
\emph{lapse function} equation as the \emph{energy constraint}. A
solution of this \emph{constraint} with respect to the time-like
variable momentum  gives \emph{energy in space of events} and the
relation of this time-like variable with \emph{geometric
interval}. The primary and secondary quantization of the
\emph{energy constraint} give Quantum Field Theory (QFT) where the
vacuum as the state with minimal \emph{energy in space of events}
is postulated with definite traffic rules of the motion of a
particle in its \emph{space of events}. The complete set of these
results can be described by the geometro-dynamic action principle
formulated by David Hilbert in his \emph{Foundations of Physics}
\cite{H} in the Einstein \emph{General Relativity} (GR).

Recall that the Hilbert geometro-dynamics includes a geometric
interval as an additional reference quantity, and \emph{general
coordinate transformations} are considered as diffeomorphisms of
coordinates and variables \cite{H,einsh} similar to
\emph{reparametrizations of a coordinate evolution parameter} in SR
\cite{pp,bpp,zpz}.

Actually, Hilbert's "Foundations  of Physics" for General
Relativity and Quantum Field Theory give hopes for an opportunity
to construct a realistic quantum theory for GR. These hopes are
based, on the one hand, on the existence of the Hilbert-type
geometric formulation \cite{H} of SR with the energy constraint
considered as the simplest model of GR and, on the other hand, on
the contemporary QFT based on the primary and secondary
quantization of this energy constraint \cite{zpz,origin,Bog}.

One can reconstruct a direct pathway from geometry of a relativistic
particle in SR to the causal operator quantization of fields of
these particles and their quantum creation from a vacuum in order to
formulate a similar direct way from geometry of GR \cite{H} to the
causal operator quantization of universes and to their quantum
creation from a vacuum treated as a state with the minimal
\emph{energy of events}. This formulation includes
\begin{enumerate}
 \item
 the Wheeler--DeWitt definition of the \emph{field
space of events} \cite{WDW}, where diffeomorphisms are split from
transformations of the frames of references using the Fock
\emph{simplex of reference} \cite{fock29};

\item the choice of the Dirac
specific \emph{reference frame} \cite{dir};

\item resolving the energy constraint in the class of
functions of the gauge transformations established by Zel'manov
\cite{vlad};

\item a treatment of the cosmological scale factor
as a zeroth mode field variable \cite{pp,bpp};

\item the constraint-shell values of the
 action and geometric interval \cite{bpzz}
  in terms of diffeo-invariant variables;

\item and the notions of \emph{energy, time, particle and universe, number
of particles and number of universes} defined by the low-energy
expansion of this \emph{reduced} action following the correspondence
principle with nonrelativistic theory in SR \cite{poi,ein}.
\end{enumerate}

Thus, further theoretical developments of GR and QFT  are
convergent, in spite of the accepted opinion that quantum
formulation of GR can not exist. At the present time, there is a
set of theoretical and observational arguments in  favor of the
opposite opinion: GR \cite{H,einsh} has a consistent
interpretation only in the form of quantum theory of the type of
the microscopic theory of superfluidity \cite{Lon,Lan,B} with the
Bogoliubov transformations used for construction of integrals of
motion and stable physical states including a vacuum
\cite{origin,Bog}.

In any case, the Hamiltonian pathway of SR towards QFT can be
repeated for GR, because the dimension of the diffeomorphism group
of Dirac--Arnowitt--Deser--Misner Hamiltonian approach to GR
coincides with the dimension of constraints removing a part of
canonical momenta in accordance with the second N\"other theorem
\cite{origin,Bog}.

Thus, \emph{the Hamiltonian approach to both the GR and SM can be
the basis of their unification}.

This Hamiltonian unification of GR and SM is just the topic of the
present paper. We show that the final GR\&SM theory is a
\emph{conformal relativistic brane} in $D=4$ space-time (internal
coordinates) moving in field space of events forming by dilaton
and scalar, spinor, and vector fields of SM. The zeroth mode
sector of the GR\&SM theory forms a cosmological model with free
initial data at the beginning of the Universe. There are initial
data that describe the vacuum creation of SM particles in
agreement with Supernovae (SN) data, CMB physics, and the present
day energy budget of the Universe.

We show that  the Hamiltonian presentation of the GR\&SM brane
differs from the acceptable approach to a relativistic brane
\cite{bn,2,3}. The first difference is  that the zeroth modes
(zeroth Fourier harmonics) are completely separated from the
nonzero ones and their interference term disappears in the energy
constraint, so that the constraint algebra differs from the
Virasoro one (in the string theory the Hamiltonian method
corresponds to the
 R\"ohrlich gauge \cite{4,fr}). In the
opposite case of the Virasoro algebra, we have the double counting
of the zeroth mode destructing the Hamiltonian presentation of the
GR\&SM brane. The second difference are  the diffeo-invariant
observables including the conformal time, in contrast to the naive
diffeo-variant formulation of GR \cite{linde,mukh}, where the
coordinate time as the object of reparametrizations is confused
with the reparametrization-invariant conformal time treated as an
observable quantity in cosmology. The third difference is the Weyl
principle of relativity of units \cite{we}. In accord with this
principle we can measure only a ratio of a measurable interval and
units of measurement of the interval.

Thus, the Hamiltonian formulation of the GR\&SM theory keeps all
concepts that were worked out in modern relativistic and quantum
physics, including the first and second N\"other theorems
\cite{Noter}, \emph{space of events, energy, time of events,
time-interval, vacuum postulate, Wigner representation of the
Poincar\'e group, Hamiltonian reduction}.

All these relativity principles mean that the cosmological scale
factor can be a "degree of freedom" with free initial data fitted
by observations \cite{zpz,Bog,bpzz,242,242a}. Recall that, in the
Inflationary Model \cite{linde},  the initial data of the
cosmological scale factor is identified  with the Planck scale.

The topic of the present paper is  the Hamiltonian GR\&SM
unification, where the Universe is  identified  with classical and
quantum solutions of equations of motion with ``free
diffeo-invariant initial data'' in the CMB reference frame.

Section 1 is devoted to the Hamiltonian approach to SM. Section 2
is devoted to the Hamiltonian approach to GR. In Section 3, we
consider the problem of  unification of GR and SM compatible with
the Newton law in GR and spontaneous symmetry breaking in SM. The
identification of GR and SM with a brane is considered in Section
4. Section 5 is devoted to observational tests of unified theory.
 \newpage
\renewcommand{\theequation}{1.\arabic{equation}}
\section{Hamiltonian  approach to gauge theories} \setcounter{equation}{0}

The Hamiltonian  approach to gauge theories was considered as the
mainstream of the development of gauge theories beginning with the
pioneer papers by Paul Dirac \cite{d}, Werner Heisenberg, Wolfgang
Pauli \cite{hp}, and finishing by Julian Schwinger's quantization of
the non-Abelian theory \cite{sch2} (see in detail
\cite{pol,f1,hpp,6}). They postulated the higher priority of the
quantum principles, in particular, in accordance with the
uncertainty principle, one counted that we cannot quantize ''field
variables'' whose velocities are absent in the Lagrangian.
Therefore, vector field time components with negative contributions
to energy are eliminated, as it was accepted in the Dirac approach
to QED \cite{d}. This illumination leads to  static interactions and
instantaneous bound states.

Remember that the Dirac Hamiltonian approach generalized to the
non-Abelian theory \cite{sch2,6} and the massive vector fields
\cite{hpp} provide the fundamental operator quantization   and
correct relativistic transformations of states of quantized fields.
This Hamiltonian approach is considered \cite{f1} as the foundation
of all heuristic methods of quantization of gauge theories,
including the Faddeev-Popov (FP) method \cite{fp1}  used now for
description of the Standard Model of elementary particles \cite{db}.
Moreover, Schwinger {\it ... rejected all Lorentz gauge formulations
as unsuited to the role of providing the fundamental operator
quantization} (see \cite{sch2} {p.324}). However, a contemporary
reader could not find the Hamiltonian presentation of the Standard
Model (SM) because there is an opinion \cite{f1} that this
presentation is completely equivalent to the accepted version of SM
based on the FP  method \cite{db}.

\subsection{Quantum Electrodynamics}

\subsubsection{Action and reference frame}

 Let us recall the Dirac approach \cite{d} to QED. The theory is given by
 the well known action
 \be\label{1e} S=\int d^4x\bigg\{-\frac{1}{4}F_{\mu\nu}F^{\mu\nu}+
 \bar \psi [i\rlap/\partial -m]\psi+A_\mu j^{\,\mu}\bigg\},
 \ee
where $F_{\mu\nu}=\partial_\mu A_\nu -\partial_\nu A_\mu$ is
tension, $A_{\mu}$ is a vector potential, $\psi$ is the
 Dirac electron-positron bispinor field, and
 $j_{\mu}=e  \bar {\psi}  \gamma_{\mu} \psi$ is the charge
 current,  $\rlap/\partial =
\partial^{\mu} \gamma_{\mu}$. This action is invariant with respect to the collection of gauge
transformations
 \bea
 \label{3e1}
 A^{\lambda}_\mu=A_\mu+\partial_\mu\lambda,~~~
\psi^{\lambda}=e^{+\imath e\lambda}\psi.
 \eea
 The action principle used for the action (\ref{1e}) gives the Euler-Lagrange equations of motion - known as the Maxwell equations
 \be\label{vp}\partial_\nu F^{\mu\nu}+j^\mu=0,
 \ee
Physical solutions of the Maxwell equations are obtained in a fixed
{\it inertial reference frame} distinguished by a unit timelike
vector \mbox{$n_{\mu}$}. This vector splits the
 gauge field $A_\mu$ into the timelike $A_0=A_\mu n_{\mu}$
 and spacelike \mbox{$A^{\bot}_\nu=A_\nu - n_{\nu}(A_\mu n_{\mu})$}
 components. Now we rewrite the Maxwell equations in terms of
  components\bea\label{c1}
 \Delta A_0-\partial_0\partial_{k}A_k &=&j_{0},\\\label{jc1}
\Box A_k-\partial_k[\partial_{0}A_0-\partial_iA_i ]&=&-j_{k}.
  \eea
The field component $A_0$
  cannot be a {\it degree of freedom}
   because its canonical conjugate momentum vanishes.
The Gauss constraints (\ref{c1}) have the solution \bea\label{c2}
  A_0+\partial_0\Lambda=-\frac{1}{4\pi}\int d^3y\frac{j_0(x_0,y_k)}{|\mathbf{x}-\mathbf{y}|},
  \eea
 where
  \bea\label{lc2}
  \Lambda=-\frac{1}{\Delta}\partial_{k}A_{k}=\frac{1}{4\pi}\int{d^3y}\frac{\partial_{k}A_{k}}{|\mathbf{x}-\mathbf{y}|}
  \eea
  is a longitudinal component.
  The result (\ref{c2}) is treated as the {\it Coulomb potential field} leading to
 the {\it static} interaction.

\subsubsection{Elimination of time component}

 Dirac \cite{d} proposed to eliminate the time component by
 substituting  the manifest resolution of the Gauss constraints given by
(\ref{c2}) into the
 initial action (\ref{1e}). This substitution - known as the reduction
 procedure -
  allows us to eliminate nonphysical pure gauge degrees of
  freedom \cite{gpk}. After this step the action (\ref{1e}) takes
the form \bea\label{2.3-1}S\!=\!\int d^4x
\left\{\frac{1}{2}(\partial_{\mu}A^{\rm T}_k)^2
 \!+\!
 \bar \psi
 [i\rlap/\partial\!-\!m]\psi\!-\!j_0\partial_0 \Lambda\!-\!
 A^{\rm T}_kj_{k}\!+\!\frac{1}{2}
 j_0\frac{1}{\triangle}j_0\right\},
 \eea
 where
 \be\label{lc1}
 A^{\rm T}_k=\left(\delta_{ij}-
\frac{\partial_i\partial_j}{\triangle}\right)A_j.
 \ee
This substitution leaves the longitudinal component
  $\Lambda$
 given by Eq. (\ref{lc2}) without any kinetic term.

 There are two possibilities. The first one is to treat
 $\Lambda$ as the Lagrange factor that leads to
 the conservation law (\ref{vp}).  In this approach, the longitudinal component
is treated
 as an independent variable. This treatment violates gauge invariance because
 this component is gauge-variant and it  cannot  be measurable.
 Moreover, the time derivative of the longitudinal  component in Eq. (\ref{c2}) looks like
 a physical source of the Coulomb potential. By these reasons we will not consider this
approach in this paper.

 In the second possibility, a measurable
 potential stress is identified with the gauge-invariant quantity  (\ref{c2})
 \be\label{2-3-4}
 A_0^{\rm R}=A_0-\frac{\partial_0\partial_k }{\triangle}A_k~,
 \ee
 This approach  is consistent with    the
 principle of gauge invariance
  that identifies
 observables  with gauge-invariant quantities.
 Therefore, according to the gauge-invariance,
   the longitudinal
 component should be eliminated from the set of degrees of freedom of QED too.

\subsubsection{Elimination of longitudinal  component}

 This elimination is fulfilled by the choice of the "radiation variables"
 as gauge invariant functionals
 of the initial fields, \textit{i.e.} "dressed fields" \cite{d}
 \bea
 A^{\rm R}_\mu=A_\mu+\partial_\mu\Lambda,~~\psi^{\rm R}=e^{\imath e\Lambda}\psi,\label{d2}
 \eea
 In this case, the linear term $\partial_k A_k$ disappears
 in the Gauss law (\ref{c1})
 \be\label{1c1}
 \Delta A^{\rm R}_0=j^{\rm R}_{0}\equiv e
 \bar \psi^{\rm R}\gamma_0\psi^{\rm R}.
  \ee
 The source of the gauge-invariant {\it potential
field} $A^{\rm R}_0$
 can be only an
 electric  current $j^{\rm R}_0$, whereas
 the spatial components  of the vector field  $A^{\rm R}_k$
 coincide with the  transversal one
\be\label{kc1}
 \partial_k A^{\rm R}_k=\partial_k A^{\rm T}_k\equiv {0}.
  \ee
In this manner the frame-fixing $A_\mu=(A_0,A_k)$ is compatible with
understanding of $A_0$ as a classical field and the use of
  the Dirac  dressed fields (\ref{d2}) of the Gauss constraints (\ref{c1})
leads to understanding of the  variables (\ref{d2}) as
gauge-invariant functionals of the initial fields.

\subsubsection{Static interaction}

 Substitution of the manifest resolution
 of the Gauss constraints (\ref{c1})
  into the initial action (\ref{1e})
  calculated on constraints leads to that the initial action
 can be expressed
 in terms of the gauge-invariant radiation variables
  (\ref{d2})
 \cite{d,pol}
 \bea\label{14-2}
S\!=\!
 \int d^4x \left\{\frac{1}{2}(\partial_{\mu}A^{\rm R}_k)^2
 +
 \bar \psi^{\rm R}
 [i\rlap/\partial -m]\psi^{\rm R}-
 A^{\rm R}_kj^{\rm R}_{k}+\frac{1}{2}
 j_0^{\rm R}\frac{1}{\triangle}j_0^{\rm R}\right\}
 .
 \eea
The Hamiltonian, which corresponds to this action, has the form
 \bea\label{2-5-2}
 {\cal
 H}\!\!=\!\!\frac{(\Pi_k^{\rm R})^2\!+\!(\partial_jA_k^{\rm R})^2}{2}+p^{R}_{\psi}\gamma_{0}
 [i\gamma_k\partial_k\!+\!m]\psi^{\rm R}+
 A^{\rm R}_kj^{\rm R}_{k}-\frac{1}{2}
 j_0^{\rm R}\frac{1}{\triangle}j_0^{\rm R},
 \eea
  where $\Pi_{k}^{R}$, $p_\psi^{\rm R}$
  are the canonical conjugate momentum fields of the theory calculated
  in a standard way. Hence the
  vacuum can be defined as a state with minimal energy obtained
  as the value of the Hamiltonian for the equations of motion.
 Relativistic covariant transformations of the
 gauge-invariant fields are proved on the level of the fundamental
 operator quantization in the form of
 the Poincar\'e
 algebra generators \cite{sch2}.
 The status of the theorem of equivalence between the
 Dirac radiation variables and the Lorentz gauge formulation
 is considered in \cite{6,hpp}.

\subsubsection{Comparison of radiation variables with the Lorentz gauge ones}

 The static interaction and the corresponding bound states
  are lost in any frame free formulation including the
  Lorentz gauge one. The action (\ref{2.3-1}) transforms into
\bea\label{2-6-1} S=\int d^4x
\left\{-\frac{1}{2}(\partial_{\mu}A^{\rm L}_\nu)^2
 +
 \bar \psi^{\rm L}
 [i\rlap/\partial -m]\psi^{\rm L}+
 A^{\rm L}_\mu j^{\rm L\mu}\right\} ,
 \eea
 where
\bea\label{2-6-2} A^{\rm L}_\mu=A_\mu+\partial_\mu\Lambda^{\rm
L},~~\psi^{\rm L}=e^{ie\Lambda^{\rm L}}\psi,~~\Lambda^{\rm
L}=-\frac{1}{\Box}\partial^\mu A^{\rm L}_\mu
 \eea
 are the manifest gauge-invariant functionals satisfying
 the equations of motion
 \be\label{2-6-3}
 \Box A^{\rm L}_\mu=-j^{\rm L}_\mu,
 \ee
 with the current $j^{\rm L}_\mu=-e\bar{\psi}^{\rm L}\gamma_{\mu}\psi^{\rm L}$ and
 the gauge constraints
 \be\label{2-6-4}
\partial_\mu A^{\rm L\mu}\equiv 0.
 \ee
 Really, instead of the potential (satisfying the Gauss constraints
  $\triangle A^{\rm R}_0= j^{\rm R}_0$)
 and two transverse variables
 in QED in terms of the
 radiation variables (\ref{d2}) we have here  three independent  dynamic
 variables,
 one of which $A^{\rm L}_0$ satisfies the equation
 \be\label{2-6-5}
\Box A_0^{\rm L}= -j_0,
 \ee
   and
 gives a negative contribution
 to the energy.

 We can see that there are two distinctions of the
 ``Lorentz gauge formulation'' from the
 radiation variables.
 The first is the loss of Coulomb poles (\textit{i.e.} static
 interactions). The second is the treatment of the time component
 $A_0$ as an independent variable with the negative contribution
 to the energy; therefore, in this case,
 the vacuum as the state with the
 minimal energy is absent.  In other words,
 one can say that the static interaction
 is the consequence of the vacuum postulate
 too. The inequivalence between the radiation variables
  and the Lorentz ones
 does not mean
violation
 of the gauge invariance,
 because both the variables  can be defined as
 the gauge-invariant functionals of the initial gauge fields
   (\ref{d2}) and (\ref{2-6-2}).

In order to demonstrate the inequivalence between the radiation
variables
  and the Lorentz ones, let us consider
 the electron-positron scattering amplitude
$T^R=\langle e^+,e^-|\hat S|e^+,e^-\rangle$. One can see that the
Feynman rules in the radiation gauge give the amplitude in terms of
the current $j_\nu=\!\bar e \gamma_\nu e$
 \bea\label{1wr}
 T^R&=&\frac{j^2_0}{\mathbf{q}^2}\!+\!
 \left(\!\delta_{ik}-\frac{q_iq_k}{\mathbf{q}^2}\!\right)
 \frac{j_ij_k}{q^2+i\varepsilon}\!\\\nonumber
 &\equiv&\frac{-j^2}{q^2+i\varepsilon}+
 \fbox{$\dfrac{(q_0j_0)^2-
 (\mathbf{q}\cdot\mathbf{j})^2}{\mathbf{q}^2[q^2+i\varepsilon]}$}~.
 \eea
 This amplitude coincides with the Lorentz gauge one
 \be
 \label{2wr}
 T^L =
 -\frac{1}{q^2+i\varepsilon}
 \left[j^2-\fbox{$\dfrac{(q_0j_0-
 \mathbf{q}\cdot\mathbf{j})^2}{q^2+i\varepsilon}$}\,\,\right]~
 \ee
 when the box terms  in Eq. (\ref{1wr}) can be
 eliminated. Thus, the Faddeev equivalence theorem \cite{f1} is valid
  if the currents
 are conserved
 \be \label{3wr}
 q_0{j}_0-\mathbf{q}\cdot\mathbf{j}=qj=0,
 \ee
 However, for the action with the external sources   the currents
 are not conserved. Instead of the classical conservation laws
 we have the Ward--Takahashi identities for
 Green functions, where the currents are not conserved
\be \label{3wr1}
 q_0{j}_0-\mathbf{q}\cdot\mathbf{j}\not =0.
 \ee
 In particular, the
Lorentz gauge perturbation theory (where the propagator has only the
light cone singularity $q_{\mu}q^{\mu}=0$) can not describe
instantaneous Coulomb atoms; this perturbation theory contains only
the Wick--Cutkosky bound states whose spectrum is not observed in
the Nature.

Thus, we can give a response to the question: What are new physical
results that follow from the
 Hamiltonian approach to QED in comparison with the
 frame-free Lorentz gauge formulation? In the framework of the perturbation
 theory, the Hamiltonian presentation of  QED contains the static
 Coulomb interaction  (\ref{1wr})  forming
 instantaneous bound states observed in the Nature, whereas
 all  frame free formulations lose this static interaction
 together with instantaneous bound states in the lowest order of
 perturbation theory on retarded interactions
   called the radiation correction. Nobody has proved that
   the sum of these retarded radiation corrections with the
   light-cone singularity    propagators (\ref{2wr}) can restore
   the Coulomb interaction that was removed from
   propagators (\ref{1wr})
   by  hand
   on the level of the action.

\subsection{Vector bosons theory}

\subsubsection{Lagrangian and reference frame}

The classical Lagrangian of massive QED is
\begin{equation}
\label{3-1-1} {\cal L}=
-\frac{1}{4}F_{\mu\nu}F^{\mu\nu}+\frac{1}{2}M^2V_\mu^2+
\bar\psi(i\rlap/\partial-m)\psi +V_{\mu}j^{\mu}~,
\end{equation}
In a fixed reference frame this Lagrangian takes the form
\begin{eqnarray}
 \label{3-1-2}\mathcal{L}\!&=&\!\frac{(\dot V_k\!-\!\partial_kV_0)^2\!-\!
 (\partial_jV_k^{\rm T})^2\!+\!M^2(V_0^2\!-
 \!V_k^2)}{2}\!+\\\nonumber
 &+&\!\bar\psi(i\rlap/\partial\!-\!m)\psi\!+\!V_{0}j_{0}\!-\!V_{k}j_{k},
\end{eqnarray}
where $\dot V=\partial_0V$ and $V_k^{\rm T}$ is the transverse
component defined by the action of the projection operator given in
Eq. (\ref{lc1}). In contrast to QED this action is not invariant
with respect to gauge transformations. Nevertheless,
 from the Hamiltonian viewpoint the massive theory
has the same problem as QED. The time component of the massive boson
has a vanishing canonical momentum.

\subsubsection{Elimination of time component}

In \cite{hpp} one supposed to eliminate the time component from the
set of degrees of freedom like the Dirac approach to QED,
\textit{i.e.}, using the action principle. In the massive case it
produces the equation of motion
\begin{equation}\label{3-1-3}
(\triangle-M^2)V_0=\partial_i\dot{V}_i+j_0.
\end{equation}
which is understood as constraints and has the solution \bea
\label{3-1-4} V_0
=\left(\frac{1}{\triangle-M^2}\,\partial_i{V}_i\right)^{\cdot}
+\frac{1}{\triangle-M^2}\,j_0.\eea In order to eliminate the time
component, let us insert (\ref{3-1-4}) into the Lagrangian
(\ref{3-1-2}) \cite{d,hpp}
\begin{eqnarray} \nonumber
 {\cal L}\!&=&\!\frac{1}{2}\left[(\dot V_k^{\rm
 T})^2\!+\!V_k^{\rm T}(\triangle\!-\!M^2)V^{\rm T}_k
 \!+\!j_0\frac{1}{\triangle\!-\!M^2}j_0\right]\!+\\\nonumber
 &+&\!\bar\psi(i\not{\!\partial}\!-\!m)\psi\!-\!V^{\rm T}_kj_k\\\label{3-2-7}
                \!&+&\! \frac{1}{2}\left[\dot V_k^{\rm ||}
 M^2\frac{1}{\triangle\!-\!M^2}\dot V_k^{\rm ||}\!-\!M^2
 (V_k^{\rm ||})^2
 \right]\!-\!V^{\rm
 ||}_kj_k\!+\!\\\nonumber
&+&j_0\frac{1}{\triangle\!-\!M^2}\partial_k \dot V_k^{\rm ||},
\end{eqnarray}
where we decomposed the vector field
 $V_k=V_k^{\rm T}+V_k^{\rm ||}$ by means of the  projection
 operator by analogy with (\ref{lc1}). The last two terms are the contributions of the longitudinal
component only. This Lagrangian contains the longitudinal component
  which is  the dynamical variable described by the bilinear term.
  Now we propose the following transformation:
 \bea
\bar\psi(i\not{\!\partial}\!-\!m)\psi\!&-&\!V^{\rm ||}_kj_k\!+\!j_0
\frac{1}{\triangle\!-\!M^2}\partial_k \dot V_k^{\rm ||}=\\\nonumber
&=&\bar\psi^{\rm R}(i\not{\!\partial}\!-\!m)\psi^{\rm R}\!-\!V^{\rm
R ||}_kj_k\label{3-2-8b},
 \eea
where
  \bea\label{3-2-9}
  V^{\rm R||}_k&=&V^{\rm||}_k-\partial_k\frac{1}{\triangle-M^2}\,\partial_i{V}_i=
-M^2\frac{1}{\triangle-M^2}V^{\rm ||}_k~,\\
 \psi^{\rm R}&=&\exp\left\{-ie\frac{1}{\triangle-M^2}\,\partial_i{V}_i\right\}\psi
 \eea
are the radiation-type variables. It removes the linear term
$\partial_i\dot{V}_i$ in the Gauss law
 (\ref{3-1-3}). If the mass $M\not = 0$, one can
 pass from the initial variables $V^{\rm ||}_k$
 to the radiation ones $V^{\rm R ||}_k$ by the change
 \be\label{3-2-9a} V^{\rm ||}_k= \hat{Z}V^{\rm R ||}_k,~~\hat{Z}=\frac{M^{2}-\triangle}{M^{2}}
  \ee
Now the Lagrangian (\ref{3-2-7}) goes into
 \begin{eqnarray}\label{3-2-10}
 {\cal L}\!&=&\!\frac{1}{2}\left[(\dot V_k^{\rm
 T})^2\!+\!V_k^{\rm T}(\triangle\!-\!M^2)V^{\rm T}_k
 \!+\!j_0\frac{1}{\triangle\!-\!M^2}j_0\right]\!+\!\bar\psi^{\rm R}(i\not{\!\partial}\!-\!m)\psi^{\rm R}\nonumber\\
 \!&+&\!\frac{1}{2}\left[\dot V_k^{\rm R\rm ||}
 \hat{Z}\dot V_k^{\rm R\rm ||}\!+\!
 V_k^{\rm R\rm ||}(\triangle\!-\!M^{2})\hat{Z}V_k^{\rm R\rm ||}
 \right]\!-\!V^{\rm T}_kj_k\!-\!V^{\rm R ||}_kj_k.
\end{eqnarray}
The Hamiltonian corresponding to this Lagrangian can be constructed
in the standard canonical way. Using the rules of the Legendre
transformation and canonical conjugate momenta $\Pi_{V^{\rm T}_k}$,
$\Pi_{V^{\rm R ||}_k}$, $\Pi_{\psi^{R}}$  we obtain
\begin{eqnarray}\label{3-2-14}
\cal{H}\!&=&\!\frac{1}{2}\left[\Pi_{V^{\rm T}_k}^{2}\!+\!V_k^{\rm
T}(M^2\!-\!\triangle)V^{\rm T}_k
 \!+\!j_0\frac{1}{M^2\!-\!\triangle}j_0\right]\!-\!\\\nonumber
 &-&\Pi_{\psi^{R}}\gamma_0(i\gamma_{k}\partial_{k}\!+\!m)\psi^{\rm R}\nonumber\\
 \!&+&\!\frac{1}{2}\left[\Pi_{V^{\rm R ||}_k}\hat{Z}^{-1}\Pi_{V^{\rm R
 ||}_k}\!+\!
 V_k^{\rm R\rm ||}(M^{2}\!-\!\triangle)\hat{Z}V_k^{\rm R\rm
 ||}\right]\!+\\\nonumber
 &+&\!V^{\rm T}_kj_k\!+\!V^{\rm R
 ||}_kj_k.
\end{eqnarray}
One can be convinced \cite{hpp} that the corresponding
 quantum system has a vacuum as a state with
 minimal energy and correct relativistic transformation
 properties.

\subsubsection{Quantization}

We start the quantization procedure from the canonical quantization
by using the following equal time canonical commutation relations
(ETCCRs):
\begin{eqnarray}\label{3-4-1}
 \left[\hat{\Pi}_{V^{\rm T}_k},\hat{V}^{\rm
 T}_k\right]=i\delta_{ij}^{\rm T}\delta^3(\mathbf{x}-\mathbf{y}),\\
\left[\hat{{\Pi}}_{V^{\rm R ||}_k}, \hat{V}^{\rm R ||}_k\right]=
 i\delta_{ij}^{||}\delta^3(\mathbf{x}-\mathbf{y}).
\end{eqnarray}
The Fock space of the theory is built by the ETCCRs
\begin{eqnarray}
\left[{a^{-}_{(\lambda)}\left({\pm
k}\right),a_{(\lambda')}^{+}\left({\pm k'}\right)}\right]&=&\delta
^{3}\left({{\bf k}-{\bf k'}}\right)\delta_{(\lambda)(\lambda')};\\
\left\{b^{-}_\alpha\left({\pm k}\right),b_{\alpha'}^{+}\left({\pm
k'}\right)\right\}&=&\delta^{3}\left({{\bf k}-{\bf
k'}}\right)\delta_{\alpha\alpha'};\\
\left\{{c^{-}_\alpha\left({\pm k}\right),c_{\alpha'}^{+}\left({\pm
k'}\right)}\right\}&=&\delta^{3}\left({{\bf k}-{\bf
k'}}\right)\delta_{\alpha\alpha'}.
\end{eqnarray}
with the vacuum state $|0\rangle$ defined by the relations
 \be \label{3-4-7}
 a_{(\lambda)}^-|0\rangle=b_\alpha^-|0\rangle=c_\alpha^-|0\rangle=0.
 \ee
The field operators have the Fourier decompositions in the plane
wave basis
\begin{eqnarray}\nonumber
 V_j\left(x\right)\!=\!
 \!\int[d\mathbf{k}]_{v}
 \epsilon_{j}^{(\lambda)}{\left[{a_{(\lambda)}^{+}\left({\omega,{\bf k}}
 \right)e^{-i\omega t + i{\bf kx}}\!+\!a_{(\lambda)}^{-}
 \left({\omega,-\!{\bf k}}\right)e^{i\omega t -i{\bf kx}}}\!\right]}&\\\nonumber
 \psi\left(x\right)=\sqrt{2m_{s}}\!\int[d\mathbf{k}]_{s}
 {\left[{b^+_\alpha\left(k\right)u_\alpha
 e^{-i\omega t + i{\bf kx}}+c^-_\alpha\left({-k}\right)\nu _\alpha
 e^{i\omega t -i{\bf kx}}}\!\right]}&\\\nonumber
 \psi^{+}\left(x\right)=\sqrt{2m_{s}}\!\int[d\mathbf{k}]_{s}
 {\left[{b^-_\alpha\left(k\right)u_\alpha^{+}e^{i\omega t -i{\bf kx}}+c^+_\alpha
 \left({-k}\right)\nu _\alpha^{+}e^{-i\omega t +i{\bf
 kx}}}\!\right]}&
\end{eqnarray}
with the integral measure
$[d\mathbf{k}]_{v,s}=\dfrac{1}{\left({2\pi}\right)^{3/2}}\dfrac{{d^3\bf
k}}{{\sqrt{2\omega_{v,s}(\bf k)}}}$ and the frequency of
oscillations
$\omega_{v,s}(\mathbf{k})=\sqrt{\,\mathbf{k}^2+m^2_{v,s}}$. One can
define the vacuum expectation values of the
 instantaneous products of the field operators
\bea \label{3-4-8}
 V_i(t,\vec x)V_j(t,\vec y)&=:V_i(t,\vec x)V_j(t,\vec y):
 +\langle V_i(t,\vec x)V_j(t,\vec y) \rangle,\\
 \overline{\psi}_\alpha(t,\vec x) \psi_\beta(t,\vec y)&=:
 \overline{\psi}_\alpha(t,\vec x) \psi_\beta(t,\vec y):
 +\langle \overline{\psi}_\alpha(t,\vec x) \psi_\beta(t,\vec y),
 \eea
where
 \bea \label{3-4-9}
 \langle V_i(t,\vec x)V_j(t,\vec y) \rangle=\frac{1}
 {(2\pi)^3}\int\frac{d^3\bf{k}}{2\omega_v(\bf{k})}
 \sum\limits_{(\lambda)}^{}
 \epsilon_{i}^{(\lambda)}\epsilon_{j}^{(\lambda)}e^{-i\bf{k}(\bf{x}-\bf{y})},
\\\label{3-4-9a}
 \langle \overline{\psi}_\alpha(t,\vec x) \psi_\beta(t,\vec y) \rangle=
 \frac{1}{(2\pi)^3}\int\frac{d^3\bf{k}}{2\omega_s(\bf{k})}
 (\mathbf{k}\vec{\gamma}+m)_{\alpha\beta}\,e^{-i\bf{k}(\bf{x}-\bf{y})}
\eea
 are the  Pauli -- Jordan functions.

\subsubsection{Propagators and condensates}

The vector field in the Lagrangian (\ref{3-2-10}) is given by the
formula
\begin{equation}\label{3-5-1}
V^{\rm R}_i= \left[\delta_{ij}^{\rm T}+\hat Z^{-1}\delta_{ij}^{\rm
||}\right]V_j=V_i^{\rm T}+\hat Z^{-1}V_i^{\rm ||}.
\end{equation}
Hence, the propagator of the massive vector field in radiative
variables is
\begin{eqnarray}\label{3-5-2}
&&D^R_{ij}(x-y)=\langle
0|TV^R_{i}(x)V^R_{j}(y)|0\rangle=\nonumber\\&=&-i\int
\frac{d^4q}{(2\pi)^4}\frac{e^{-iq\cdot
(x-y)}}{q^2-M^2+i\epsilon}\left(\delta_{ij}-\frac{q_{i}q_{j}}{\mathbf{q}^2+M^2}\right)~.
\end{eqnarray}
 Together with the instantaneous  interaction described by
  the current--current term in
 the Lagrangian (\ref{3-2-10})  this propagator
 leads to the amplitude
\bea\label{3-2-11} T^{\rm R}&=& D^{\rm
R}_{\mu\nu}(q)\widetilde{j}^\mu \widetilde{j}^\nu =\\\nonumber
\frac{\widetilde{j}_0^2}{\mathbf{q}^2+M^2}&+&\left(\delta_{ij}-\frac{q_i
q_j}{\mathbf{q}^2+M^2}\right) \frac{\widetilde{j}_i
\widetilde{j}_j}{{q}^2-M^2+i\epsilon}~ \eea
 of the current-current interaction which differs from the acceptable one
\begin{equation}\label{3-1-8}
 T^{\rm L}=\widetilde{j}^{\mu}D^{\rm L}_{\mu\nu}(q)\widetilde{j}^{\nu}=
-\widetilde{j}^{\mu}\frac{g_{\mu\nu}-\dfrac{q_\mu q_\nu}{M^2}
}{q^2-M^2+i\epsilon}\widetilde{j}^{\nu}.
\end{equation}
The amplitude given by Eq. (\ref{3-2-11})  is the generalization of
the  radiation amplitude  in QED. As it was shown in  \cite{hpp},
the Lorentz transformations of classical radiation variables
coincide with the  quantum ones  and they both (quantum and
classical) correspond to the transition to another Lorentz frame of
reference distinguished by another time-axis, where the relativistic
covariant propagator takes the form
\begin{eqnarray}\label{3-5-4}
D^{R}_{\mu\nu}(q|n)\!=\!\frac{-g_{\mu\nu}}{q^2\!-\!M^2\!+\!i\epsilon}\!+\!
\frac{n_{\mu}n_{\nu}(qn)^2\!-\![q_{\mu}\!-\!n_{\mu}(qn)][q_{\nu}\!-\!n_{\nu}(qn)]}
 {(q^2\!-\!M^2\!+\!i\epsilon)(M^2\!+\!|q_{\mu}\!-\!n_{\mu}(qn)|^2)}\label{3-5-4a},
\end{eqnarray}
where $n_{\mu}$ is determined by the external states. Remember that
 the conventional local field massive vector propagator
 takes the form  (\ref{3-1-8})
 \begin{equation}\label{3-5-5}
 D^L_{\mu\nu}(q)=-
\dfrac{g_{\mu\nu}-\dfrac{q_\mu q_\nu}{M^2}}{q^2-M^2+i\epsilon}~.
\end{equation}
In contrast to this conventional  massive vector propagator the
 radiation-type propagator (\ref{3-5-4})  is regular in the
limit $M\rightarrow 0$ and is well behaved for large momenta,
whereas the propagator (\ref{3-5-5}) is singular. The radiation
amplitude (\ref{3-2-11}) can be rewritten   in the alternative form
\begin{equation}
T^{\rm R}=-\frac{1}{q^2-M^2+i\epsilon}\left[\widetilde{j}_{\nu}^2
+\frac{(\widetilde{j}_iq_i)^2-(\widetilde{j}_0q_0)^2}{\vec{q}^2+M^2}\right]~,
\label{mvecprop2}
\end{equation}
for comparison with the conventional amplitude defined by the
propagator (\ref{3-5-5}). One can find that for a massive vector
field coupled to a conserved current
$(q_{\mu}\widetilde{j}^{\mu}=0)$ the collective current-current
interactions mediated by the radiation propagator  (\ref{3-5-4}) and
by the conventional propagator (\ref{3-5-5}) coincide
\begin{equation}\label{3-5-52}
\widetilde{j}^{\mu}D^{\rm R}_{\mu\nu}\widetilde{j}^{\nu}=
\widetilde{j}^{\mu}D^{\rm L}_{\mu\nu}\widetilde{j}^{\nu}=T^{\rm L}~.
\end{equation}
 If the  current is not conserved $\widetilde{j}_0q_0\not =\widetilde{j}_kq_k$,
 the collective
 radiation field variables
 with the propagator (\ref{3-5-4})
 are inequivalent to   the initial local  variables
 with the propagator  (\ref{3-5-5}),  and the amplitude
 (\ref{3-2-11}). The amplitude (\ref{3-5-52}) in the Feynman gauge is
 \be\label{3-5-52a}
 T^{\rm L} =
 -\frac{j^2}{q^2-M^2+i\varepsilon},
 \ee
and corresponds to the Lagrangian
 \be\label{brst}
\mathcal{L}_{F}=\frac{1}{2}(\partial_{\mu}V_{\mu})^{2}-j_{\mu}V_{\mu}+
\frac{1}{2}M^2 V_\mu^2 \ee
 In this theory the time component has a negative contribution
 to the energy. According to  this a correctly defined
 vacuum state could not exist. Nevertheless, the vacuum expectation
 value $\langle V_\mu(x)V_\mu(x) \rangle$ coincides with the values
 for two  propagators (\ref{3-5-4a}) and
 (\ref{3-5-5})
 because in both these propagators the longitudinal part
 does not give a contribution if one treats them as derivatives of constant
 like
 $\langle \partial V_\mu(x)V_\mu(x) \rangle
 =\partial \langle  V_\mu(x)V_\mu(x) \rangle=0$.
 In this case, we have
\bea \label{v3-4-8}
 \langle V_\mu(x)V_\mu(x) \rangle &=&-\frac{2}
 {(2\pi)^3}\int\frac{d^3\bf{k}}{\omega_v(\bf{k})}= 2 L_v^2(M_v),
 \\\label{s3-4-8}
 \langle \overline{\psi}_\alpha(x)  \psi_\alpha(x)\rangle&=&-
 \frac{m_s}{(2\pi)^3}\int\frac{d^3\bf{k}}{\omega_s(\bf{k})}=m_s
 L_s^2(m_s),
 \eea
where $m_s$, $M_{v}$ are masses of the spinor and vector fields, and
$L^2_{s,v}$ are values of the integrals.

\subsection{Electroweak Standard Model}
\subsubsection{The SM action}

 The Standard Model constructed on  the Yang--Mills theory \cite{ym}
 with the symmetry group ${SU(2)}\times{U(1)}$ is known as the
Glashow-Weinberg-Salam theory of electroweak interactions
\cite{weak}.  The action of the Standard Model in the electroweak
sector with presence of the
 Higgs field can be written in the form
 \be\label{1-sm}
 S_{\rm SM}=\int d^4x
 {\cal L}_{\rm SM}=\int d^4x\left[{\cal L}_{\rm Ind}
 +{\cal L}_{\rm Higgs}
 \right],
 \ee
where
 \bea\label{M}
&&{\cal L}_{\rm Ind}=-\frac{1}{4}G^a_{\mu\nu}
G^{\mu\nu}_{a}-\frac{1}{4}F_{\mu\nu} F^{\mu\nu}\\\nonumber
&+&\sum_s\bar{s}_1^{R}\imath \gamma^{\mu}\left(D_{\mu}^{(-)}
 +\imath
g^{\prime}B_{\mu}\right)s_1^{R}+\sum_s\bar{L}_s\imath
\gamma^{\mu}D^{(+)}_{\mu}L_s, \eea is the Higgs field independent
part of the Lagrangian and \bea\label{Ma}
{\mathcal{L}_{\rm{Higgs}}}= {\partial_\mu\phi\partial^\mu\phi}
-\phi\sum_sf_s\bar s s+\frac{{\phi^2}}{4}\sum_{\rm v} g^2_{\rm
v}V^2-
\underbrace{{\lambda}\left[\phi^2-\textrm{C}^2\right]^2}_{V_{\rm
Higgs}}
 \eea
is the Higgs field dependent
 part. Here
\begin{eqnarray}\label{1-6a}
 \sum_s f_s\bar s s&\equiv&\sum_{s=s_1,s_2} f_{s}\left[\bar s_{sR}s_{sL}
 +\bar s_{sL}s_{sR}\right],\\
\label{w1-6a} \frac{1}{4}\sum_{\rm v=W_1,W_2,Z} g^2_{\rm
v}V^2&\equiv&\frac{g^{2}}{4}W_\mu^{+}W^{-\mu}+\frac{g^{2}+g'^{2}}{4}Z_{\mu}Z^{\mu}
\end{eqnarray}
  are
  the mass-like terms  of  fermions  and W-,Z-bosons coupled with the Higgs field,
 \mbox{$G^a_{\mu\nu}=\partial_{\mu}A^{a}_{\nu}-\partial_{\nu}A^{a}_{\mu}+g\varepsilon_{abc}A^{b}_{\mu}A^{c}_{\nu}$}
is the field strength of non-Abelian $SU(2)$ fields  and
\mbox{$F_{\mu\nu}=\partial_{\mu}B_{\nu}-\partial_{\nu} B_{\mu}$} is
the field strength of Abelian $U(1)$ (electromagnetic interaction)
ones,
$D_{\mu}^{(\pm)}=\partial_{\mu}-i{g}\frac{\tau_{a}}{2}A^a_{\mu}\pm\frac{i}{2}g^{\prime}B_{\mu}$
are the covariant derivatives, $\bar L_s=(\bar s_1^{L}\bar s_2^{L})$
are the fermion doublets,  $g$ and $g'$ are the Weinberg coupling
constants, and
 measurable gauge bosons $W^+_{\mu},~W^-_{\mu},~Z_{\mu}$ are defined by the relations:
\bea
W_{\mu}^{\pm}&\equiv&{A}_{\mu}^1\pm{A}_{\mu}^2={W}_{\mu}^1\pm{W}_{\mu}^2,\\
Z_{\mu}&\equiv&-B_{\mu}\sin\theta_{W}+A_{\mu}^3\cos\theta_{W},\\
\tan\theta_{W}&=&\frac{g'}{g},\eea where $\theta_{W}$ is the
Weinberg angle.

The crucial meaning has a distribution of the Higgs field $\phi$ on
the zeroth Fourier harmonic
 \be\label{h0-1}
 \langle\phi\rangle=\frac{1}{V_0}\int d^3x \phi
 \ee and the nonzeroth ones $h$, which we will call
 the Higgs boson
  \be\label{h-1}
 \phi=\langle\phi\rangle+\frac{h}{\sqrt{2}},~~~~\int d^3x h=0.
 \ee
In the acceptable  way, $\langle\phi\rangle$ satisfies the
 particle vacuum classical equation ($h=0$)
 \be\label{hc-1}
 \frac{\delta V_{\rm Higgs}(\langle\phi\rangle)}
 {\delta\langle\phi\rangle}=4{\langle\phi\rangle}
 [{\langle\phi\rangle}^2-\textrm{C}^2]=0
 \ee
 that has two solutions
\be\label{hc-2}
 {\langle\phi\rangle}_1=0,~~~~~~~~~~~~~~~
 {\langle\phi\rangle}_2=\textrm{C}\not =0.
 \ee
 The second solution corresponds to the spontaneous vacuum symmetry
 breaking
 that determines the masses of all elementary particles
 \bea
\label{0W-1}M_{W}&=&\frac{\langle\phi\rangle}{\sqrt{2}}g\\\label{0Z-1}
M_{Z}&=&\frac{\langle\phi\rangle}{\sqrt{2}}\sqrt{g^2+g'^2}
\\\label{0s-1}m_{s}&=& \langle\phi\rangle y_s,
\eea according to the definitions of the masses of vector (v) and
fermion (s) particles
 \be\label{vs-1}
{\cal L}_{\rm mass~ terms}= \frac{M_v^2}{2}V_\mu V^\mu-m_s\bar s s.
 \ee

\subsubsection{Hamiltonian approach to SM}

 The  accepted SM (\ref{1-sm})
 is bilinear with
 respect to the time components of the vector fields $V^K_0=(A_0,Z_0,W_0^{+},W_0^{-})$
 in the ``comoving frame'' $n^{\rm cf}_\mu=(1,0,0,0)$
 \be\label{sm4}
 S_{V}=\int d^4x \left[\frac{1}{2}V^K_0\hat L^{KI}_{00}V^I_0+V^K_0J^K+...
 \right]~,
 \ee
  where $\hat L^{KI}_{00}$ is
 the matrix of differential operators. Therefore, the Dirac approach
to SM can be realized. This means that the
 problems of the reduction and
  diagonalization of the set of the Gauss laws are solvable, and
  the Poincar\'e algebra of gauge-invariant observables can be proved \cite{hpp}.
In any case,  SM in the lowest order of  perturbation theory is
reduced to
 the sum of the Abelian massive vector fields, where
  Dirac's  radiation variables were considered
 in Section 3.
\subsubsection{The conformal vacuum Higgs effect}

 The Hamiltonian approach to the Standard Model  considered in \cite{252} leads to
 fundamental operator quantization that allows
a possibility of  dynamic spontaneous symmetry breaking
 based on
the Higgs potential  (\ref{Ma}), where instead of a dimensional
parameter $\textrm{C}$ we substitute the zeroth Fourier harmonic
(\ref{h0-1}) \bea\label{0-Ma}{\mathcal{L}_{\rm{Higgs}}}=
{\partial_\mu\phi\partial^\mu\phi} -\phi\sum_sf_s\bar s
s+\frac{{\phi^2}}{4}\sum_{\rm v} g^2_{\rm v}V^2
-\underbrace{{\lambda}\left[\phi^2-\langle\phi
\rangle^2\right]^2}_{V_{\rm Higgs}}.
 \eea
 After the separation of the zeroth mode (\ref{h-1}) the
 bilinear part of the Higgs Lagrangian takes the form
 \bea\label{cMa}
 {\mathcal{L}_{\rm{Higgs}}^{\textrm{bilinear}}}\!=\!
 \frac{1}{2}{\partial_\mu h\partial^\mu h}
 \!-\!\langle\phi\rangle\sum_sf_s\bar s s
 \!+\!\frac{{\langle\phi\rangle^2}}{4}\sum_{\rm v} g^2_{\rm
 v}V^2\!-\!2{\lambda}{\langle\phi\rangle^2}h^2. \eea
 In the lowest order in the coupling constant,
 the bilinear Lagrangian of the sum of all
 fields
\mbox{$S^{(2)}=\sum_F S_F^{(2)}[\langle\phi\rangle]$} arises with
the masses of vector (\ref{0W-1}), (\ref{0Z-1}), fermion (s)
(\ref{0s-1}) and
 Higgs (h) particles:
 \bea
\label{0h-1}
 m_h&=&2\sqrt{\lambda}\langle\,\phi\rangle.
\eea

 The sum of all
 vacuum-vacuum transition amplitude diagrams of the theory is known
 as the effective Coleman -- Weinberg
potential \cite{coleman79}
 \bea\label{eff-1}\textsf{V}^{\rm conf}_{eff}=-i
 \mathrm{Tr}\log <0|0>_{(\langle\,\phi\rangle)}=-i \mathrm{Tr}\log
  \prod_{F} G^{-A_F}_F[\langle\,\phi\rangle]G^{A_F}_F[\phi_{\rm I}],\eea
 where $G^{-A_F}_F$ are the Green-function operators with $A_F=1/2$ for bosons and
 $A_F=-1$ for fermions.
 In this case, the unit vacuum-vacuum transition amplitude $<0|0>
~\Big|_{\langle\,\phi\rangle =\phi_{\rm I}}=1$ means that
 \bea
 \label{va-2}
 \textsf{V}^{\rm conf}_{eff} (\phi_{\rm I})=0,
 \eea
where
   $\phi_{\rm
 I}$ is a  solution of the variation equation
\bea\label{va-1}
 &&\partial^2_0\langle\phi\rangle+\frac{d\textsf{V}^{\rm
 conf}_{eff}(\langle\,\phi\rangle)}{d\langle\,\phi\rangle}~
 \Bigg|_{\langle\,\phi\rangle=\phi_{\rm I}}
 =\partial^2_0\langle\phi\rangle+\\\nonumber
 &+&\sum_sf_s<\!\!\bar s
 s\!\!>-\frac{{\langle\,\phi\rangle}}{2}\sum_{\rm v} g^2_{\rm
 v}<\!\!V^2\!\!>\,+\,4\lambda{\langle\,\phi\rangle}<\!\!h^2\!\!>=0,
 \eea
 here $<V^2>,<\bar ss>,<h^2>$ are the condensates
 determined by   the Green
 functions in \cite{252}  \bea \label{v3-4-8x}
 <V^2>=\langle V_\mu(x)V_\mu(x) \rangle &=&
 -2 L_{\rm v}^2(M_{R\,\rm v}),
 \\\label{s3-4-8x}
 <\bar ss>=\langle \overline{\psi}_\alpha(x)  \psi_\alpha(x)\rangle&=&
 -m_{R\,s}L_s^2(m_{R\,s})
 ;\\
\label{h3-4-8}< h^2> =\langle h(x)h(x) \rangle &=&
 \frac{1}{2} L_h^2(m_{R\,h});
 \eea
here $L^2_{p}(m_p^2)$ are values of the integral \bea\label{v-5}
 L^2_{p}(m_p^2)=\frac{1}{(2\pi)^3}\int\frac{d^3\bf{k}}{\sqrt{m^2_p+\bf{k}^2}}.
 \eea

 Finally, using the definitions of the condensates and
 masses (\ref{0W-1}), (\ref{0Z-1}),(\ref{0s-1}),(\ref{0h-1}) we
 obtain the equation of motion
\bea\label{0-Ma-3}
\langle\phi\rangle\partial^2_0\langle\phi\rangle=\sum_sm_s^2
L_s^2-2\sum_{\rm v} M_{\rm v}^2L^2_{\rm v}-\frac{1}{2}m_h^2L^2_{\rm
h} . \eea
 In the class of  constant solutions $\partial^2_0\langle\phi\rangle\equiv 0$
 this equation has two solutions
\be\label{hc-2a}
 {\langle\phi\rangle}_1=0,~~~~~~~~~~~~~~~
 {\langle\phi\rangle}_2=\textrm{C}\not =0.
 \ee
 The nonzero solution means that there is
 the Gell-Mann--Oakes--Renner type relation
\bea\label{31-t}
 L_h^2 m^2_h=2\sum_{s=s_1,s_2}{L_s^2}m^2_s
 -4[2M_W^2L_W^2+M_Z^2L_Z^2].
 \eea

 If we suppose that the condensates $L^2_p(m^2_{\rm R\,p})$
 are defined by the subtraction
 procedure associated with  the renormalization of masses and wave functions
 leading to the finite value
 \bea\nonumber
 L^2_{\rm R p}(m^2_{\rm R\,p})\!=\!L^2_{\rm p}(m^2_{\rm Rp})-
 L^2_{\rm p}(\Lambda^2)\!-(m^2_{\rm R p}-\Lambda^2)\!
 \frac{d}{d \Lambda^2} L^2_{\rm p}(\Lambda^2)=\\\label{log}
 =\frac{m^2_{\rm R\,p}}{2(2\pi)^2}\log\frac{m^2_{\rm R\,p}}{e\Lambda^2},
 \eea
where $\Lambda$ is a subtraction constant.

In this case,
 the sum rule (\ref{31-t})
 takes the form
\bea\nonumber
 L^2_{R\, h}(m^2_{R\,h}) m^2_h=
 2\sum_{f=f_1,f_2}{L^2_{R\, f}(m^2_{R\,f})}m^2_f-\\\label{31-tr}
 -4[2M_W^2L^2_{R\, W}(M^2_{R\,W})+M_Z^2L^2_{R\, Z}(M^2_{R\,Z})].
 \eea

We substitute  the experimental data by the  values of masses of
bosons
 $M_W=80.403\pm0.029$ GeV, $M_Z=91.1876\pm0.00021$ GeV \cite{pdg},
 and t-quark $m_t=170.9\pm 1.8$ GeV \cite{tquark}. In the minimal SM \cite{db}, the three color t-quark dominates
$\sum_f m_f^2\simeq 3m_t^2$ because
  contributions of other fermions $\sum_{f\not=t} m_f^2/2m_t\sim 0.17$ GeV
 are very small.

In Fig.~\ref{fig1} the solution of the above equation is plotted
for the range $0.3~{\rm GeV} < \Lambda < 100$ GeV. One can see
that the Higgs mass is the order of $215\div 255$ GeV and is not
very sensitive to the choice of the parameter, because the
dependence is logarithmic. The measurement of the mass at an
experiment would provide us the proper value of $\Lambda$
according to Fig.~\ref{fig1}.

\begin{figure}[ht]
\begin{center}
\includegraphics*[width=10cm,height=6.6cm,
angle=0]{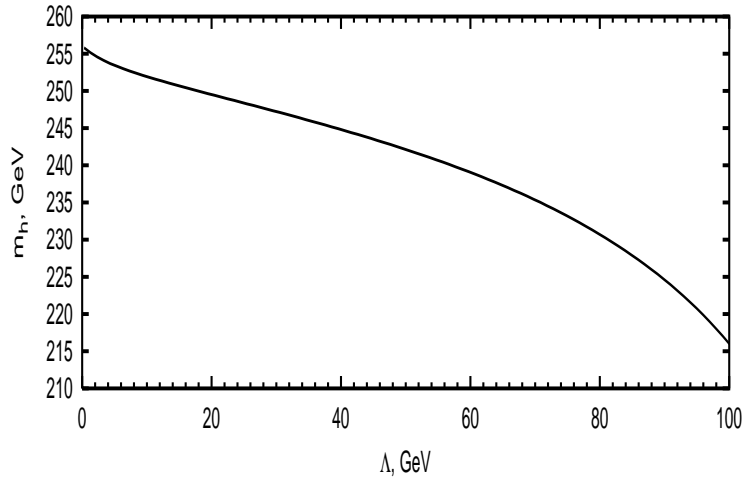}
\end{center}
\caption{Value of the Higgs mass from Eq.~(\ref{31-tr}) with the
condensates 
defined by Eq.~(\ref{log}) as a function of $\Lambda$. }
\label{fig1}
\end{figure}

Radiative corrections  to this quantity in the Standard Model are
not small, first of all due to a large coupling constant of Higgs
with top quark.
 Note that relation~(\ref{31-tr}) in our model
should be valid in all orders of the perturbation theory.


The choice of the parameters in the inertial Higgs potential in
our model can be motivated by the cosmological reasons.
 Even so that the resulting
Lagrangian of the model is practically the same as the on of SM,
we get a prediction for value of the the Higgs boson mass to be in
the range $215\div255$ GeV. In this range of $m_h$ the width of
the Higgs particle is between 5 and 10~GeV. Here the main decay
modes are $W\to ZZ$ and $H\to WW$ (since $M_Z<m_h<2m_t$), which
are quite convenient for experimental studies~\cite{hhg}. The
so-called ``gold--plated'' channel $H\to 4\mu$ should allow a
rather accurate measurement of $m_h$ with at least $0.1\%$
relative error~\cite{Delmeire:2007mq}. So it is important to
provide adequately precise theoretical predictions for this
quantity. As concerns the production mechanism, the sub-process
with gluon-gluon fusion dominates~\cite{Hahn:2006my} for the given
range of $m_h$ and the corresponding cross section of about
$10^4$~fb provides a good possiblity to dicover the Higgs boson at
the high-luminosity LHC machine.

In this way  the potential free Higgs mechanism gives the
possibility to solve the question about a consistence of the
nonzero vacuum value of the scalar field with the zero vacuum
 cosmological energy as a consequence of the unit
vacuum-vacuum transition amplitude. The inertial motion of a
scalar field corresponds to the dominance of the most singular
rigid state at the epoch of the intensive vacuum creation of the
primordial  bosons \cite{bpzz}. 

\subsubsection{The static interaction mechanism of
  the enhancement of the $\triangle T=1/2$ transitions}

\begin{figure}
  \begin{center}
  \includegraphics[scale=0.5]{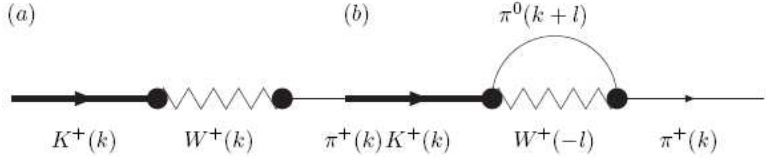}
  \caption{Axial (a) and vector (b) current contribution into $K^+\to
\pi^+$ transition} \label{1ac}
  \end{center}
\end{figure}

 Let us consider the $K^+\to \pi^+$ transition amplitude
\bea \nonumber\left\langle\pi^+\left|-i\int d^4xd^4y
 J^\mu(x)D_{\mu\nu}^WJ^\nu(y)\right|K^+\right\rangle=i(2\pi)^{4}\delta^{4}(k-p)G_{\rm EW}\Sigma(k^{2})
 \eea
 where $D_{\mu\nu}^W\equiv{D}_{\mu\nu}^W(x-y)$in the first order of the EW perturbation theory in the Fermi
 coupling constant \be\label{gsd}
  G_{\rm EW}= \frac{\sin\theta_{C} \cos\theta_{C}}{8
  M^{2}_{W}}\frac{e^{2}}{\sin^{2}\theta_{W}}\equiv
  \sin\theta_{C} \cos\theta_{C}\frac{G_F}{\sqrt{2}},
 \ee
 comparing two different W-boson field propagators,
 the accepted Lorentz (L)
 propagator (\ref{3-5-5})
 and
 the  radiation (R) propagator (\ref{3-5-4a}).
 These propagators give the expressions
 corresponding to the diagrams in Fig. \ref{1ac}
\bea\nonumber
  \Sigma^R(k^{2})&=&\!2F^2_{\pi}{k^2}
  -2i\int \!\! \frac{d^4q{M_W^2}}{(2\pi)^4}
 \frac{k^2+(k_0+q_0)^2}{(|\vec q|^2+M_W^2)[(k+q)^2
 \!-\!m^2_\pi+i\epsilon]}, \\
 \nonumber
  \Sigma^L(k^{2})&=&2F^2_{\pi}{k^2}
 +2i\int \frac{d^4q{M_W^2}}{(2\pi)^4}
 \frac{(2k_\mu+q_\mu)D^{L}_{\mu\nu}(-q)(2k_\nu+q_\nu)}{(k+q)^2
 -m^2_\pi+i\epsilon}.
 \eea
 The versions R and L coincide in the case of the axial contribution
 corresponding to the first diagram in Fig. \ref{1ac},
 and they both reduce to the static interaction contribution
 because
\be k^\mu k^\nu D^F_{\mu\nu}(k)\equiv k^\mu k^\nu
D^R_{\mu\nu}(k)=\frac{k_0^2}{M^2_W}. \ee
 However, in the case of the vector contribution
 corresponding to the second diagram in Fig. \ref{1ac}
 the radiation version differs from the  Lorentz gauge version
  (\ref{3-5-5}).

  In contrast to the Lorentz gauge version  (\ref{3-5-5}),
  two radiation variable diagrams  in Fig. \ref{1ac} in  the
  rest kaon frame $k_\mu=(k_0,0,0,0)$
  are reduced to
  the static interaction contribution
 \be\label{kp1}
 i(2\pi)^{4}\delta^{4}(k-p)G_{\rm EW}\Sigma^R(k^{2})=
 \left\langle\pi^+\left|-i\int d^4x
 \frac{J_0^2(x)}{\triangle-M_W^2}\right|K^+\right\rangle\nonumber
 \ee
  with the normal ordering of the pion fields which are
  at their mass-shell,
   so that
 \be\nonumber
 \Sigma^R(k^{2}) = 2k^2{F^2_{\pi}}\left[1+\frac{M_{W}^{2}}{F^2_{\pi}(2\pi)^3}\int\frac{d^3l}{2E_\pi(\vec{l})}\frac{1}{M^2_W+\vec{l}^2}\right]
 \equiv 2k^2{F^2_{\pi}}g_8.
 \ee
Here $E_{\pi}(\vec{l})=\sqrt{m^{2}_{\pi}+\vec{l}^{2}}$ is the
 energy of $\pi$-meson and $g_8$ is
 the parameter of the enhancement of the probability
 of the axial $K^+ \to \pi^+$ transition.
 The pion mass-shell justifies
 the application of the low-energy ChPT~\cite{fpp,vp1},
 where the summation of the chiral series can be
 considered here as
  the meson form factors
  \cite{bvp,ecker,belkov01}
  \be\int\frac{d^3l}{2E_\pi(\vec{l})}\to
  \int\frac{d^3l f^V_K(-(\vec l)^2)f^V_\pi(-(\vec l)^2)}{2E_\pi(\vec{l})}\ee
 Using
  the covariant perturbation theory \cite{pv}\ developed as the
 series
 \be\underline{J}_\mu^k(\gamma
 \oplus\xi)=\underline{J}_\mu^k(\xi)+ F^2_\pi
 \partial_\mu \gamma^k
 +\gamma^if_{ijk}\underline{J}_\mu^j(\xi)+O(\gamma^2)\ee
 with respect to quantum fields $\gamma$ added to $\xi$
 as  the product $e^{i\gamma}e^{i\xi}\equiv e^{i(\gamma \oplus\xi)}$, one can see that the normal
  ordering \be<0|\gamma^i(x)\gamma^{i'}(y)|0>=\delta^{ii'}N(\vec
  z),~~~~
  N(\vec z)=\int \frac{d^3l e^{i\vec l\cdot (\vec z)}}{(2\pi)^32E_\pi(\vec
  l)},\ee
  where $\vec z=\vec x-\vec y$,
  in the product of the currents $\underline{J}_\mu^k(\gamma
 \oplus\xi)$,
   leads to an effective Lagrangian with the rule $\triangle T=1/2$
\bea
 M_W^2\int d^3z g_8(z)\frac{e^{-M_W|\vec z|}}{4\pi|\vec z|}
 \Bigg[\underline{J}_\mu^j(x)\underline{J}_\mu^{j'}(z+x)(f_{ij1}+if_{ij2})\nonumber\\
\times(f_{i'j'4}-if_{i'j'5})\delta^{ii'}+h.c\Bigg], \eea
 where
\be{g}_8(|z|)= 1+\sum\limits_{I\geq 1}c^I N^I(\vec z)\ee
 is a series over the multipaticle intermediate states known as
 the Volkov superpropagator \cite{vp1,S}.
 In the  limit $M_W\to \infty$,
  in the lowest order with respect to $M_W$,  the
  dependence of $g_8(|\vec z|)$ and the currents on $\vec z$
 disappears in the integral of the type of
\be M^2_W \int d^3z \frac{g_8(|\vec z|)e^{-M_W|\vec z|}}{4\pi|\vec
z|}=
 \int_0^{\infty} drr e^{-r}g_8({r}/{M_W})\simeq g_8(0).\ee
 In the next order, the amplitudes $K^0(\bar K^0)\to \pi^0$ arise.
  Finally, we get
  the effective  Lagrangians  \cite{kp}
 \bea \label{ef1}\mathcal{L}_{(\Delta T=\frac{1}{2})}&=&
\frac{G_{F}}{\sqrt{2}}g_{8}(0)\cos\theta_{C}\sin\theta_{C}
\Bigg[(\underline{J}^1_{\mu}+i\underline{J}^2_{\mu})
(\underline{J}^4_{\mu}-i\underline{J}^5_{\mu})-\nonumber\\
&&\left(\underline{J}^3_{\mu}+\frac{1}{\sqrt{3}}\underline{J}^8_{\mu}\right)
(\underline{J}^6_{\mu}-i\underline{J}^7_{\mu})+h.c.\Bigg], \eea
\be\nonumber \mathcal{L}_{(\Delta T=\frac{3}{2})}=
\frac{G_{F}}{\sqrt{2}}\cos\theta_{C}\sin\theta_{C}
\left[\left(\underline{J}^3_{\mu}+\frac{1}{\sqrt{3}}\underline{J}^8_{\mu}\right)
(\underline{J}^6_{\mu}-i\underline{J}^7_{\mu})+h.c.\right]. \ee

 This result shows that the enhancement can be explained
  by static vector interaction that
  increases the $K^+\to \pi^+$ transition
 by a factor of $g_8=g_8(0)$, and yields a new term describing the
  $K^0\to \pi^0$ transition proportional to $g_8-1$.

 This Lagrangian with the fit parameter $g_8=5$ describes the nonleptonic decays in
 satisfactory agreement with experimental data
 \cite{vp1,kp,06}.
 Thus, for normal ordering of
 the weak static interaction in the Hamiltonian SM
 can explain the rule $\triangle T=1/2$ and universal factor $g_8$.

 On the other hand, contact character of
 weak static interaction in the Hamiltonian SM
  excludes  all retarded diagram contributions in the effective
   Chiral Perturbation
  Theory considered in \cite{da98} that destruct
  the form factor structure of the kaon radiative decay rates
 with the amplitude \be\nonumber
  T_{(K^+\to\pi^+l^+l^-)}=g_8 t(q^2)
   2F^2_\pi  \sin\theta_{C} \cos\theta_{C}\frac{G_F}{\sqrt{2}}
 \frac{(k_{\mu }+p_{\mu })}{q^2}\bar{l}\gamma_{\mu}l\,\,\,
 \ee
where $q^2=(k-p)^2$, and
 \be\nonumber
 t(q^2)=
 \frac{f^{A}_{K}(q^{2})\!+\!f^{A}_{\pi}(q^{2})}{2}
 - f^{V}_{\pi}(q^{2})+
 \Big[\!f^{V}_{K}(q^{2})\!-\!
 f^{V}_{\pi}(q^{2})\!\Big]\frac{m_\pi^2}{M_K^2-m_\pi^2},
 \ee
 and $f_{K}^{V}\simeq f_{\pi}^{V}(q^2)=1+M^{-2}_\rho q^2+...$,
$f_{K}^{A}(q^2)\simeq f_{\pi}^{A}(q^2)=1+M^{-2}_a
 q^2+\ldots$ are form factors determined by
 the masses of the nearest resonances for meson -- gamma --
 meson vertex.

 Therefore, the static interaction mechanism of
  the enhancement of the $\triangle T=1/2$ transitions
 predicts \cite{06}  that  the meson form factor
 resonance parameters explain the experimental values of rates of
  the radiation kaon decays  $K^+ \to
 \pi^+e^{+}e^{-}(\mu^{+}\mu^{-})$.
 Actually,  substituting  the PDG data on the resonance masses
 $M_\rho=775.8$ {\rm MeV}, $1^+(1^{--})=I^G(J^{PC})$
 and meson -- gamma -- W-boson one $M_a=984.7$ {\rm MeV},
  $1^-(0^{++})=I^G(J^{PC})$ into the decay amplitudes
 one can obtain
 the decay branching fractions \cite{06}
\begin{eqnarray}\nonumber
{\rm Br}(K^+\to\pi^+e^+e^-) &=& 2.93 \times 10^{-7},~~[2.88\pm0.13
\times 10^{-7}]_{\rm PDG }\\\nonumber{\rm Br}(K^+\to\pi^+\mu^+\mu^-)
&=& 0.73 \times 10^{-7},~~[0.81\pm0.14 \times 10^{-7}]_{\rm PDG }
\end{eqnarray}
 in satisfactory agreement with  experimental data \cite{ap99,pdg}.
  Thus, the
  off-mass-shell kaon-pion transition in the radiation weak kaon decays
     can be a good probe of the weak static interaction
     revealed by the
     radiation propagator (\ref{3-5-4a}) of the Hamiltonian presentation of SM.

\subsection{Summary}

\emph{Physical consequences of the Hamiltonian
   approach to the Standard Model are the weak  static interactions,
   like
    the Coulomb static interaction is
a consequence of the Hamiltonian  approach in QED.} The static
interactions can be  omitted if we restrict ourselves to the
scattering processes of  elementary particles
 where static interactions
 are not important. However, the static poles
 play a crucial role in the mass-shell
 phenomena of the    bound state type,  spontaneous symmetry breaking,
  kaon - pion transition in the weak decays, etc.
  \emph{Static interactions follow from the  spectrality principle that means
 existence of a vacuum
 defined as a state with the minimal energy.}
  We discussed physical effects testifying  to the static
 interactions omitted by the  accepted version of SM.

 One of these effects is revealed by the loop
  meson diagrams in the low-energy weak static interaction.
 These  diagrams lead to the enhancement coefficient $g_8$ in weak kaon
 decays and  the rule $\triangle T=\frac{1}{2}$.
    The loop pion diagrams
     in the Chiral Perturbation Theory  \cite{vp1} in
     the framework of the Hamiltonian approach
      with
  the weak static interaction
 lead to a definite relation
  of the vector form factor with the differential radiation kaon decay
   rates in agreement with the present day PDG data \cite{06},
 in  contrast to the acceptable renormalization group
     analysis based on the Lorentz gauge formulation omitting weak static
     interaction \cite{da98}, where loop pion diagrams destroy
     the above-mentioned  relation
  of the vector form factor with the differential radiation kaon decay
   rates.
 \emph{Therefore,
   the radiation kaon decays can be a good probe
   of the  weak static  interaction.}

We considered the consequence of the spontaneous symmetry breaking
in the Standard Model, where the parameter of the Higgs potential
is replaced by the initial data of the zeroth Fourier harmonic. In
this case, the Hamiltonian  approach   and its operator
fundamental
   formulation immediately  lead to the effective quantum potential
     predicting the mass of Higgs particle.
\newpage
\renewcommand{\theequation}{2.\arabic{equation}}
\section{Hamiltonian General Relativity
} \setcounter{equation}{0}

The statement of the problem given at the beginning is to unify SM
and GR on equal footing using as a basis the Hamiltonian approach to
both these theories, in order to describe the Universe in its
comoving frame. The Hamiltonian approach to GR is well known. It is
the Dirac -- ADM constrained method \cite{dir} formulated for
infinite space-time in a definite frame of reference, where the
observable time is distinguished and the observable space is
foliated. This Hamiltonian comoving frame of the Universe can be
identified with the  frame of the Cosmic Microwave Background (CMB)
as the evidence of the Early Universe creation.

The present-day measurement of the dipole component of CMB radiation
temperature $T_0(\theta)=T_0[1+(\beta/c)\cos\theta]$, where
$\beta=390\pm 30$ km/s, \cite{WMAP01} testifies to a motion of an
Earth observer to the Leo with the velocity $|\vec v|$= 390$\pm$ 30
km/s with respect to CMB, where 30 km/s rejects the copernican
annual motion of the Earth around the Sun, and 390 km/s to the Leo
is treated as the parameter of the Lorentz transformation from the
the Earth frame to the CMB frame.

This relativistic treatment of the observational data in the context
of the Hamiltonian approach produces the definite questions to the
GR and the modern cosmological models destined for description of
the processes  of origin of the Universe and its evolution:

\begin{enumerate}
\item How  the CMB {\it inertial frame} can be separated  from the
general coordinate transformations?
\item How  the cosmic evolution can be separated
from the dynamics of the local scalar component in the CMB reference
frame?
\item What is the version of the canonical approach to
the General Relativity and the Standard Model in the finite
space-time, because the observable Universe in the finite space and
has a finite life-time?
\end{enumerate}

In this Section, we discuss  possible responses to these issues that
follow from the principles of General Relativity and Quantum Field
Theory.

\subsection{Canonical General Relativity}

\subsubsection{The Fock separation of the frame transformations
 from diffeomorphisms}

   Recall that the Einstein--Hilbert theory is given by
   two fundamental quantities; they are a
 {\it geometric interval}
\be \label{1-2}
 ds^2=g_{\mu\nu}dx^\mu dx^\nu
 \ee
   and  the  {\it dynamic} Hilbert action
 \be\label{1-1}
 S_{\rm GR}=\int d^4x\sqrt{-g}\left[-\frac{\vh_0^2}{6}R(g)
 \right]
 \ee
 where $\varphi^2_0=\dfrac{3}{8\pi}M^2_{\rm Planck}=
 \dfrac{3}{8\pi G}$, $G$ is the Newton
 constant in units ($\hbar=c=1$).

 Quantities (\ref{1-2}) and (\ref{1-1}) are invariant with respect to action diffeomorphisms
 \be \label{1-5}
 x^{\mu} \rightarrow  \tilde x^{\mu}=\tilde
 x^{\mu}(x^0,x^{1},x^{2},x^{3}),
 \ee
 Separation of  the diffeomorphisms from
 the Lorentz transformations in GR is fulfilled
 by linear invariant forms $
 \omega_{(\alpha)}(x^{\mu})~\to ~\omega_{(\alpha)}(\tilde x^{\mu})=
 \omega_{(\alpha)}(x^{\mu})
 $ \cite{fock29}
\be \label{1-3}
 ds^2\equiv\omega_{(\alpha)}\omega_{(\alpha)}=
 \omega_{(0)}\omega_{(0)}-
 \omega_{(1)}\omega_{(1)}-\omega_{(2)}\omega_{(2)}-\omega_{(3)}\omega_{(3)},
 \ee
 where $\omega_{(\alpha)}$ are diffeo-invariants. These forms are
treated as components of an orthogonal reference simplex with the
following Lorentz transformations: \bea \label{1-4}
{\omega}_{(\alpha)}~\to ~\overline{\omega}_{(\alpha)}=
\overline{\omega}_{(\alpha)}=L_{(\alpha)(\beta)}{\omega}_{(\beta)}.
\eea

 There is an essential difference between diffeomorphisms (\ref{1-5}) and
 the Lorentz
 transformations  (\ref{1-4}).
  Namely,
 the parameters of the Lorentz
 transformations  (\ref{1-4}) are measurable quantities, while the parameters
 of diffeomorphisms (\ref{1-5}) are unmeasurable one. Especially, the simplex
 components ${\omega}_{(\alpha)}$ in the Earth frame
   moving with respect
  to Cosmic Microwave Background (CMB) radiation with
 the velocity $|\vec v|$= 390 km/s to the Leo
 are connected with the simplex
 components in the CMB frame $\overline{\omega}$ by the following
 formulae:
 \bea \label{1-6} \overline{\omega}_{(0)}&=&\frac{1}{\sqrt{1-\vec
 v^2}}\left[\omega_{(0)}- v_{(c)}\omega_{(c)}\right],\\\nonumber
 \overline{\omega}_{(b)}&=&\frac{1}{\sqrt{1-\vec
 v^2}}\left[\omega_{(b)}- v_{(b)}\omega_{(0)}\right], \eea
 where the velocities $\vec v$ are measured \cite{WMAP01} by the the modulus of
 the dipole component
  of CMB temperature $T_0(\theta)=T_0[1+(\beta/c)\cos\theta]$
   and its direction in
   space\footnote{Frame transformations invariance of action means that there are
 integrals of motion (1st Noether theorem \cite{Noter}),
 while diffeoinvariance of action leads to the Gauss type
 constraints between the integrals of motion
 (2nd Noether theorem \cite{Noter}). These constraints are derived in a
 {\it specific reference frame to the initial data}. The constraints mean that only
 a part of metric components becomes {\it degrees of freedom} with the initial data.
 Another part corresponds to the diffeo-invariant {\it static  potentials}
 that does not have initial data because their equations
 contain the Beltrami-Laplace operator. The third part of
 metric components after the resolution of constraints becomes
 diffeo-invariant non-dynamical variables that can be excluded
 by the gauge-constraints \cite{dir} like the longitudinal fields in QED \cite{d}.}.

\subsubsection{The Dirac -- ADM approach to GR}

 The problem of specific frame destined for description of
 evolution of the
 Universe in GR  was formulated by Dirac  and Arnovitt,
 Deser and Misner \cite{dir} as 3+1 foliated space-time (see also \cite{vlad}).
 This foliation can be
 rewritten in terms of the Fock simplex components as follows:
 \be
\label{1-7}
 \omega_{(0)}=\psi^6N_{\rm d}dx^0,
 ~~~~~~~~~~~
 \omega_{(b)}=\psi^2 {\bf e}_{(b)i}(dx^i+N^i dx^0),
 \ee
 where triads ${\bf e}_{(a)i}$ form the spatial metrics with $\det |{\bf
 e}|=1$, $N_{\rm d}$ is the Dirac lapse function, $N^k$ is the shift
 vector, and $\psi$ is a determinant of the spatial metric.

 The Hilbert action (\ref{1-1})
  in terms of the Dirac -- ADM variables
  (\ref{1-7})
 is as follows: \bea \label{Asv11}
 S_{\rm GR}&=& -\int d^4x\sqrt{-g}\frac{\vh_0^2}{6}~{}^{(4)}R(g)=\\\nonumber
 &=&\int d^4x
 ({\mathcal{K}}[\vh_0|
 {g}]-{\mathcal{P}}[\vh_0|{g}]+{\mathcal{S}}[\vh_0|{g}]),
 \eea
 where
\bea
 {\mathcal{K}}[\vh_0|e]\!\!\!&=&\!\!\!{{N}_d}\vh_0^2\left(-{\vphantom{\int}}4
 {  {v_\psi}}^2+\frac{v^2_{(ab)}}{6}\right),
 \label{k1}\\
 {\mathcal{P}}[\vh_0|e]\!\!\!&=&\!\!\!\frac{{N_d}\varphi_0^2{\psi}^{7}}{6}\left(
 {}^{(3)}R({\bf e}){\psi}+
 {8}\triangle{\psi}\right),
 \label{p1}\\
 {\cal S}[\vh_0|e]\!\!\!&=&\!\!\!2\varphi_0^2\left[\partial_0{v_{\psi}}-
 \partial_l(N^l{v_{\psi}})\right]-\frac{\varphi^2_0}3 \partial_j[\psi^2\partial^j (\psi^6
 N_d)]\label{0-s1}
 \eea
 are the kinetic and  potential terms,
 respectively,
  \bea\label{proi1}
 {v_\psi}\!\!\!&=&\!\!\!\frac{1}{{N_d}}\left[
 (\partial_0-N^l\partial_l)\log{
 {\psi}}-\frac16\partial_lN^l\right],\\
 v_{(ab)}\!\!\!&=&\!\!\!\frac{1}{2}\left({\bf e}_{(a)i}v^i_{(b)}+{\bf
 e}_{(b)i}v^i_{(a)}\right),\\\label{proizvod}
 v_{(a)i}\!\!\!&=&\!\!\!
 \frac{1}{{N_d}}\left[(\partial_0-N^l\partial_l){\bf e}_{(a)i}
+ \frac13 {\bf
 e}_{(a)i}\partial_lN^l-{\bf e}_{(a)l}\partial_iN^l\right]
 \eea
 are velocities of the metric components,
   ${\triangle}\psi=\partial_i({\bf e}^i_{(a)}{\bf
 e}^j_{(a)}\partial_j\psi)$ is the covariant Beltrami--Laplace operator,
 ${}^{(3)}R({\bf{e}})$ is a three-dimensional curvature
 expressed in terms of triads
   ${\bf e}_{(a)i}$:
\be \nonumber\label{1-17}
 {}^{(3)}R({\bf e})=-2\partial^{\phantom{f}}_{i}
 [{\bf e}_{(b)}^{i}\sigma_{{(c)|(b)(c)}}]-
 \sigma_{(c)|(b)(c)}\sigma_{(a)|(b)(a)}+
 \sigma_{(c)|(d)(f)}^{\phantom{(f)}}\sigma^{\phantom{(f)}}_{(f)|(d)(c)}.
 \ee
 Here
 \be\label{1-18} \sigma_{(a)|(b)(c)}=
 {\bf e}_{(c)}^{j}
 \nabla_{i}{\bf e}_{(a) k}{\bf e}_{(b)}^{\phantom{r}k}=
 \frac{1}{2}{\bf e}_{(a)j}\left[\partial_{(b)}{\bf e}^j_{(c)}
 -\partial_{(c)}{\bf e}^j_{(b)}\right]
  \ee
  are the coefficients of the spin-connection (see \cite{242a}),
  \be\nabla_{i}{\bf e}_{(a) j}=\partial_{i}{\bf e}_{(a)j}
  -\Gamma^k_{ij}{\bf e}_{(a) k},~~~~\Gamma^k_{ij}=\frac{1}{2}{\bf e}^k_{(b)}(\partial_i{\bf e}_{(b)j}
  +\partial_j{\bf e}_{(b)i})\ee are covariant derivatives. The  canonical conjugated momenta are \bea
\label{1-32as}{p_{\psi}}&=&\frac{\partial {\cal
K}[\vh_0|e]}{\partial
 (\partial_0\ln{{{\psi}}})}~=-8\vh_0^2{{v}},
 \\\label{1-33as}
 p^i_{(b)}&=&\frac{\partial {\cal K}[\vh_0|e]
 }{\partial(\partial_0{\bf e}_{(a)i})}
 =\frac{\vh^2}{3}{\bf e}^i_{(a)} v_{(a b)}.
 \eea
The Hamiltonian action takes the form \cite{242,242a}
 \be\label{1-16} S_{\rm GR}=\int d^4x
 \left[\sum\limits_{{F=e,\log\psi,Q}
 } P_{F}\partial_0F
 -{\cal H}_{\rm d}\right] \ee
 where
 \be\label{1-17c}
{\cal H}_{\rm d}={N_d} T_{\rm d}+N_{(b)}
  {T}^0_{(b)} +\lambda_0{p_\psi}+
  \lambda_{(a)}\partial_k{\bf e}^k_{(a)}
 \ee
is the sum of constraints with  the Lagrangian multipliers ${N_d}$,
$N_{(b)}={\bf e}_{k(b)}N^k$, $\lambda_0$ $,\lambda_{(a)}$, including
 the additional (second class) Dirac gauge conditions -- the local transverse
  $\partial_k{\bf e}^k_{(a)} = 0$ and
  the minimal 3-dimensional hyper-surface  too
\be\label{c1-42}
 p_{{\widetilde{\psi}}}=0 \to
 (\partial_0-N^l\partial_l)\log{
 {\widetilde{\psi}}}=\frac16\partial_lN^l,
 \ee and three  first class constraints
\bea\label{t1-37a}
 T^0_{\rm (a)}=-e^l_{\rm (a)}\frac{\delta S}{\delta
N^{l}} &=&
 -{p_{\psi}}\partial_{(a)}
 {\psi}+\frac{1}{6}\partial_{(a)}
 ({p_{\psi}}{\psi})+\\\nonumber
 &+&2p_{(b)(c)}\gamma_{(b)|(a)(c)}-\partial_{(b)}p_{(b)(a)}+T^0_{\rm
 (a)m}=0
 \eea
 are the components of the total
energy-momentum tensor ${T}^0_{(a)}=-\frac{\delta S}{\delta
N_{k}}{\bf e}_{k(a)}$ (we included  here the matter field
contribution $T^0_{\rm (a)m}$  considered in Appendix C using as an
example a massive electrodynamics),
 and the first class energy constraint
\bea\label{1-37a} T_{\rm d}[\vh_0|\psi]&=&-\frac{\delta S}{\delta
N_{\rm d}} = \frac{4\vh_0^2}{3}{\psi}^{7} \triangle {\psi}+
  \sum\limits_{I} {\psi}^I{\cal T}_I=0,
\eea
 here $ \triangle
 {\psi}\equiv
 \partial_{(b)}\partial_{(b)}{\psi}$ is
 the Beltrami--Laplace operator,
 $\partial_{(a)}={\bf e}^k_{(a)}\partial_k$,
 and $\cal{T}_I$ is partial energy density
 \bea
 \label{h32}
 {\cal T}_{I=0}&=&\frac{6{p}_{(ab)}{p}_{(ab)}}{\vh_0^2}
 -\frac{16}{\vh_0^2}{p_{\psi}}^2
\\\label{h35aa}
 {\cal T}_{I=8}&=&\frac{\varphi_0^2}
  {6}R^{(3)}({\bf e})
, \eea
  here ${p}_{(ab)}=\frac{1}{2}({\bf e}^i_{(a)}\widetilde{p}_{(b)i}+
  {\bf e}^i_{(b)}{p}_{(a)i})$
  marked by the index $I$.

 The Newton law is determined by the energy constraints $T_{\rm d}=0$ (\ref{1-37a})
 and the equation of motion of the spatial determinant takes the form
 \bea \label{1-37ab}
   T_{\psi}[\vh_0|\psi]&=&-\psi\frac{\delta S}{\delta \psi}\equiv
  (\partial_0-N^l\partial_l)p_\psi+\\\nonumber &+&
 \frac{4\varphi_0^2}{3}\left[7N_d{\psi}^{7}\!
  \triangle
{\psi}+{\psi}\! \triangle(N_d{\psi}^{7})\right]+\!
N_d\sum\limits_{I}I {\psi}^I{\cal T}_I=0.
 \eea
It is not embarrassing to check that in the region of  empty space,
where
 two dynamic variables are absent ${\bf e}_{(a)k}=\delta_{(a)k}$
 (i.e. ${\cal T}_{I}=0$), one can get the Schwarzschild-type
 solution of these equations
  in the form
 \be\nonumber \label{h-c4}
 \triangle {\psi}=0,~~~\triangle [N_d{\psi}^{7}]=0
 ~~~\to~~~{\psi}=1+\frac{r_g}{r},~~
 [N_d{\psi}^{7}]=1-\frac{r_g}{r},~~N^k=0,
 \ee
 where $r_g$ is the constant of the
 integration given by the boundary conditions that take into account
 massive fields and sources.

\subsubsection{The Lichnerowicz variables and cosmological models
}

In the general case of  massive electrodynamics considered in detail
in Appendix C,
 the dependence of the energy momentum tensor
(\ref{1-37a})
    on the
   spatial determinant {\it potential} $\psi$ is completely determined by
   the Lichnerowicz (L) transformation to the
   conformal-invariant variables
  \bea \label{1-12}
 \omega_{(\mu)}&=&\psi^{2}\omega^{(L)}_{(\mu)},\\\label{1-13}
 g_{\mu\nu}&=&\psi^{4}\,g_{\mu\nu}^{(L)} ,\\\label{1-14}
 F^{(n)}&=&\psi^{2n}\,F_{(L)}^{(n)},
 \eea
 where $F^{(n)}$ is any field with the one of conformal weights
 $(n)$: $n_{\rm scalar}=-1$, $n_{\rm spinor}=-3/2$, \mbox{$n_{\rm
 vector}=0$}, and $n_{\rm tensor}=2$. In the case,
the index $I$ in the energy momentum tensor (\ref{1-37a})
$\sum\limits_{I} {\psi}^I{\cal T}_I$ runs a set of values
   I=0 (stiff), 4 (radiation), 6 (mass), 8 (curvature)
   $I=12$ ($\Lambda$-term)
 in correspondence with a type of matter field contributions.

 This $\psi$-independence of L-variables is compatible with the
 cosmological dependence of the energy density on the scale factor $a$
 in the homogeneous
 approximation
 \bea\label{ca-1}
 \psi^2&\simeq& a(\eta),\\\label{ca-2}
 ds^2&=&a^2(\eta)[(d\eta)^2-(dx^k)^2]
  \eea
 where the energy constraint (\ref{1-37a}) takes the form
\bea\label{ca-3}
 \vh^2_0a'^2&=&\sum\limits_{I} {a}^{-2+I/2}\langle{\cal T}_I\rangle;
 \eea
here
 \bea\label{ca-4}
 \langle {\cal T}\rangle&=&\frac{1}{V_0}\int d^3x{\cal T}
 \eea
 is averaging over the finite volume of the coordinate
 space $V_0=\int d^3x$.

 The Newton law (\ref{h-c4}) is compatible with the cosmological
 approximation (\ref{ca-1}) if the spacial
 determinant variable takes the form of a product of two factors
\bea\label{ca-5}
 \psi^2=a(\eta)\widetilde{\psi}^2.
 \eea
 This means that
 the logarithm of the spacial
 determinant can be given as the sum of
 the zeroth Fourier harmonic and nonzero ones
 \be\label{d-t1}
  \log\psi^2(x^0,x^k)=\log a(x^0)+\log \widetilde{\psi}^2(x^0,x^k),
  \ee
   with the additional constraints
 \be\label{3-20}
 \int d^3x \log\widetilde{\psi}=\int d^3x \left[\log{\psi}
 -\left\langle{ \log{\psi}}\right\rangle\right]\equiv 0,
 \ee
 where $V_0=\int d^3x  < \infty$ is the finite
 Lichnerowicz volume.

 This presentation  of   the spacial
 determinant variable (\ref{ca-5}) is well known as the Lifshits cosmological
 perturbation theory \cite{lif,MFB}.

The question arises about the consistence of this
 cosmological
 perturbation theory \cite{lif,MFB} defined in the
  finite space-time of observable coordinate space and conformal
  time with the Dirac -- ADM Hamiltonian \cite{dir}
  approach proposed for infinite space-time.

How  the Hubble evolution  can be included into the canonical GR?
and How
  the Dirac -- ADM Hamiltonian formalism   can be generalized
 for finite space-time in order to give the
 Hamiltonian version of  cosmological
 perturbation theory? The responses to these questions were given in
   \cite{242,bpzz} using the exact solution of the energy constraint
  in accord with the group of the diffeomorphisms of
  the Dirac -- ADM foliation
  and second N\"other theorem.

\subsubsection{Global energy constraint and
 dimension of diffeomorphisms ($3L+1G\neq4L$)}

The Dirac -- ADM approach to the Einstein--Hilbert theory \cite{d}
states that five components $\psi,N_{\rm d},N^k$ are
 treated  as {\it
 potentials} satisfying the Laplace type equations in curved space
  without the initial data, three components
  are excluded by the gauge constraints $\partial_k{\bf e}^k_{(b)}=0$,
  and only  two rest  transverse gravitons
  are considered as
  independent {\it degrees of freedom}
  satisfying the d'Alambert type equations with the
  initial data.
  This Dirac -- ADM classification is not compatible with
the group of general coordinate transformations
 that conserves
  a family  of constant coordinate time hypersurfaces $x^0=\rm{const}$.
 The group of these transformations, known as
  {\it
kinemetric} subgroup \cite{vlad}, contains only homogeneous
 reparameterizations of the coordinate evolution parameter $(x^0)$ and three local transformations of the spatial coordinates:
  \bea\label{1-8}
  \left[\begin{array}{c}x^0\\x^i\end{array}\right]\to \left[\begin{array}{c}\widetilde{x}^0(x^0)\\\widetilde{x}^i(x^0,{x}^i)\end{array}\right]
 \eea
 This means that dimension of the kinemetric subgroup of
 diffeomorphisms (three local functions and one global one)
 does not coincide with the dimension of the
  constraints in the canonical approach to the classical theory of
  gravitation
  wich remove four local variables (the law $3L+1G\neq4L$).
  In accord with the second N\"other theorem, the dimension of
  the diffeomorphism group $3L+1G$ determines the dimension
  of manifold of canonical momenta that can be removed by
  the first class constraints. This means that the energy
  constraint can remove the zeroth mode
  (i.e. cosmological scale factor) and the zeroth mode momentum
  from the phase space.
  However, it does not mean that both the scale factor and its
  momentum are removed from the set  of physical quantities.
  They become the evolution parameter in the field space of events
  and the event-energy.

 Recall that according to the definition of
 all measurable quantities  as diffeo-invariants \cite{d}, in
 finite space-time the non diffeo-invariant quantity (\ref{1-8})
 $(x^0)$ is not measurable.
  Wheeler and DeWitt \cite{WDW} drew attention to that in this case
   evolution of a universe in GR
  is in full analogy with a relativistic particle
  given by the action
 \bea\nonumber
 \widetilde{S}_{\rm SR}[X^0|X^k]\!&=&\!-\frac{m}{2}
 \int d\tau~ \frac{1}{e_p}\left[\left(\frac{dX^0}{d\tau}\right)^2
 -\left(\frac{dX^k}{d\tau}\right)^2+e^2_p\right]=
 \\\label{3-25}&=&\!\int\! d\tau \left[
 -P_\mu \frac{dX^\mu}{d\tau}+\frac{e_p}{2m}(P^2_\mu-m^2)\right]
 \eea
   in the Minkowski space
  of events $[X^0|X^k]$ and the interval $ds=e_pd\tau$,
  because both the actions (\ref{3-25}) in SR  and (\ref{h-1}) in GR
 are  invariant with respect to reparametrizations of the
 coordinate evolution parameters
 $\tau\to \widetilde{\tau}=\widetilde{\tau}(\tau)$ and
 $x^0\to \widetilde{x}^0=\widetilde{x}^0(x^0)$, respectively, see Table
 \ref{tab1}.

 In any  relativistic theory given by an action and
 a geometrical interval \cite{H}
 there are two diffeo-invariant
 time-like parameters:
 the diffeo-invariant geometrical proper time interval (g-time)
 $e_pd\tau=ds$ and the one of dynamical variables $X^0$ in the
 {\it space of events} $[X^0|X^k]$ (d-time).
 Thus, in accord with the cosmological perturbation theory
 \cite{lif},
 there is a possibility to
  identify the dynamic evolution parameter $a$ in the field space
 of events with the zeroth Fourier harmonic of the metric scalar component
 logarithm if the Hamiltonian formalism in finite volume is consistent with
 the cosmological perturbation theory. Let us consider
 the Hamiltonian formalism in finite volume and find a condition
 of this consistence.

\subsubsection{The separation of the zeroth mode
 in finite space}

 Reparametrizations of the coordinate evolution parameter $(x^0)$
  mean that in  finite space-time the quantity $(x^0)$ is
 not observable, and one should distinguish
 a diffeo-invariant homogeneous {\it``time-like variable''}.
 Modern observational data in astrophysics and cosmology
  \cite{linde, MFB}
 are the irrefutable arguments in favor of identification of
  such  a  diffeo-invariant homogeneous
{\it``evolution parameter''}  with the  cosmological
 scale factor  $a(x_0)$ introduced
 by the scale transformation of the metrics
 $g_{\mu\nu}=a^2(x^0){\widetilde{g}}_{\mu\nu}$
and any field $F^{(n)}$ with the conformal weight $(n)$:
 \be\label{F}
 F^{(n)}=a^n(x_0) {\widetilde{F}}^{(n)}.
 \ee
 In particular, the
   curvature
 \be \label{cur}
 \sqrt{-g}R(g)=a^2\sqrt{-{\widetilde{g}}}R({\widetilde{g}})-6a
 \partial_0\left[{\partial_0a}\sqrt{-{\widetilde{g}}}~{\widetilde{g}}^{00}\right]\ee
 can be expressed in terms of
   the new lapse
 function ${\widetilde{N}_d}$ and spatial determinant ${\widetilde{\psi}}$ in
 the Fock simplex (\ref{1-7})
 \be \label{lfsd}
 {\widetilde{N}}_d=[\sqrt{-{\widetilde{g}}}~{\widetilde{g}}^{00}]^{-1}=a^{2}{N}_d,~~~~~~~~
 {\widetilde{\psi}}=(\sqrt{a})^{-1}\psi.
 \ee
 In order to keep the number of variables, we identify $\log \sqrt{a}$ with
 the  spatial volume ``averaging'' of $\log{\psi}$
 \be\label{1non1}
 \log \sqrt{a}=\langle \log{\psi}\rangle\equiv\frac{1}{V_0}\int
 d^3x\log{\psi}.
 \ee
After the separation of  the zeroth mode the action
 (\ref{Asv11}) takes the form
  (\ref{1-7})
  as follows: \bea \label{Asv110}
 &&S_{\rm GR}[\vh_0|\psi]=
 S_{\rm GR}[\vh|\widetilde{\psi}]+S_{\rm
 int}+S_0,
 \eea
 where
 \bea\label{Asv-110}
 S_{\rm GR}[\vh|\widetilde{\psi}]=\int d^4x({\mathcal{K}}[\vh|
 {g}]-{\mathcal{P}}[\vh|{g}]+{\mathcal{S}}[\vh|{g}])
 \eea
 is the action $S_{\rm GR}[\vh_0|{\psi}]$ with the change
 $[\vh_0|{\psi}]\to[\vh|\widetilde{\psi}]$
 \bea\vh&=&\vh_0a\label{vv-01}
 \\\label{proi10}
 {v_{\widetilde{\psi}}}&=&\frac{1}{{\widetilde{N}_d}}\left[
 (\partial_0-N^l\partial_l)\log{
 {\widetilde{\psi}}}-\frac16\partial_lN^l\right],\\
 {\mathcal{K}}[\vh|e]&=&{{\widetilde{N}}_d}
 \vh^2\left(-{\vphantom{\int}}4
 {{v_{\widetilde{\psi}}}}^2+\frac{v^2_{(ab)}}{6}\right),
 \label{k10}\\
 {\mathcal{P}}[\vh|e]&=&\frac{{N_d}\varphi^2{\widetilde{\psi}}^{7}}{6}\left(
 {}^{(3)}R({\bf e}){\widetilde{\psi}}+
 {8}\triangle{\widetilde{\psi}}\right),
 \label{p10}\\
 {\cal S}[\vh|e]&=&2\varphi^2\left[\partial_0{v_{\widetilde{\psi}}}-
 \partial_l(N^l{v_{\psi}})\right]-\frac{\varphi^2_0}3 \partial_j[\psi^2\partial^j (\psi^6
 N_d)]\label{0-s10}
 \eea
 are the kinetic and  potential terms,
 respectively,
\bea \label{INT-0}
 S_{\rm
 int}=-2 \int dx^0 \partial_0\vh(x^0)\int d^3x v_{\widetilde{\psi}}
 \eea
  is the interference between the zeroth mode and nonzero ones,
 \bea \label{1-0}
 S_0=-\int dx^0\int d^3x\frac{(\partial_0\vh)^2 }{\widetilde{N}_{\rm
 d}}\equiv-V_0\int dx^0\frac{(\partial_0\vh)^2 }{N_0},
 \eea
 is the zeroth mode action, and \bea \label{2-0}
 \frac{1 }{N_0}\equiv \frac{1}{V_0}\int d^3x\frac{1 }{\widetilde{N}_{\rm
 d}},
 \eea
is the global lapse function.

\subsubsection{The superfluidity condition}

 Thus, after the separation of  the zeroth mode in the action
 (\ref{Asv11}) its  part
   describing the spatial metric determinant takes the form
\be\label{sd-1}
 S_{\rm D}=-\int d^4x{N}_d
 \left[{\vphantom{\int}}4\vh^2~
 {  (v_{\psi})}^2
 +4\vh~ v_\vh~  v_{\psi}+
 ({v_\vh)^2}{\vphantom{\int}}\right],
 \ee
 where
 \bea \label{sd-2}\vh=\vh_0a(x^0)\\
 \label{sd-3} v_\vh=\partial_0\vh/N_d,
 \eea
 the first term in the Lagrangian arises from the kinetic part
 Eq. (\ref{k10}), the second goes from the ``quasi-surface'' one (\ref{0-s10}),  and
the third term goes from the zeroth mode action (\ref{1-0}). The
canonical momentum of the scale factor can be obtained by variation
of Lagrangian  (\ref{sd-1}) with respect to  the time derivative of
scale factor $\partial_0\varphi$
$$
 P_\vh\equiv
  \frac{\partial {L}_{SD}}{\partial(\partial_0 \vh)}=
 -\int d^3x\left[ 4\vh~v_\psi~+2v_\vh\right]\equiv -[4\vh
 V_\psi+2V_\vh],
 $$
 while the zeroth Fourier harmonics of  canonical momentum of the spatial
 metric determinant  is
\bea\label{pi}
 P_\psi&\equiv&-\int d^3x
 \frac{\partial{\cal L}_{SD}}{\partial(\partial_0 \log
 \overline{\psi})}= - \int d^3x{ \bar p}_{\psi}=\\\nonumber&=&
-\int d^3x
 \left[8\vh^2 v_\psi+4\vh~v_\vh\right]\equiv -2\vh[4\vh
 V_\psi+2V_\vh],
 \eea
 where $V_\vh=\int d^3x ~v_\vh,V_\psi=\int d^3x ~v_\psi$. These two equations
 have no solutions as
  the matrix  of the transition
from ``velocities'' to momenta
   has the zeroth  determinant.
 This means that the ``velocities''
  $[V_\vh,V_\psi]$ could not
  be expressed in terms of
  the canonical momenta
  $[P_\vh,P_{\psi}]$  and
 the Dirac Hamiltonian approach becomes a failure.
 To be consistent with identity
 (\ref{3-20}) and to keep the number of variables of GR, we should
 impose the strong constraint
 \be
 V_\psi\equiv\int d^3 x {v_\psi}\equiv 0,
 \ee
 otherwise we shall have the double counting of the zeroth-Fourier
 harmonics of spatial metric determinant.

  A {\it``double counting''} is replacement of
  $L_1=(\dot x)^2/2$ by $L_2=(\dot x+\dot y)^2/2$.
 The second theory is not mathematically equivalent to the first.
 The test of this nonequivalence is the failure
 of the Hamiltonian approach to $L_2=(\dot x+\dot y)^2/2$.
Therefore,
  the replacement $L_1\to L_2 $ is nonsense in the context
  of the Hamiltonian approach.

 The interference term plays role of friction. If we accept the Landau condition for
superfluidity
 \be\label{land0}\int d^3x {v}_{\overline{\psi}}=0,
 \ee
then the interference term vanishes. The Landau condition
(\ref{land0}) is consistent with the Hamiltonian system, because in
the opposite case we have the double counting of the zeroth mode
component which destroys the  Hamiltonian structure of the theory
and leads to problems in expressing  the velocity by canonical
conjugate momentum.

 The next example   is Lifshitz's perturbation theory
 given by  Eq. (3.21) p. 217 in \cite{MFB}
$$
ds^2= a^2(\eta)[(1+2\Phi)d\eta^2-(1-2\Psi)\gamma_{ij}dx^idx^j].
$$
 This
formula contains the double counting of the zeroth Fourier harmonics
of the spatial metrics determinant presented by two variables: the
scale factor $a$ and $<\Psi>=\int d^3x\Psi(\eta,x_i)$ instead of
one. Thus, the accepted cosmological perturbation theory is not
consistent with the Hamiltonian approach to GR considered above.
\begin{table}[h!]\label{tab1}
\centering
\begin{tabular}{|c|c|c|}
\hline $\mathrm{N}^{\underline{\mathrm{o}}}$&UNIVERSE&PARTICLE\\[.1cm]
\hline 1.&$S[\vh_0|F]$~~(\ref{3-21})&$\widetilde{S}_{\rm SR}[X^0|X^k]$~~(\ref{3-25})\\[.1cm]
\hline 2.&$d\zeta=dx^0e_u$&$ds=d\tau e_p$\\[.1cm]
\hline 3.&$x^0 \to \widetilde{x}^0=\widetilde{x}^0(x^0)$&$\tau\to\widetilde{\tau }=\widetilde{\tau }(\tau )$\\[.1cm]
\hline 4.&$\vh(x^0)=\vh_0a(x^0)$&$X_0(\tau)$\\[.1cm]
\hline 5.&$\vh~|~\widetilde{F}$&$X_0~|~X_k$\\[.1cm]
\hline 6.&$P^2_\vh-E^2_\vh=0$&$P^2_0-E^2_0=0$\\[.1cm]
\hline 7.&$\zeta_{(\pm)}=\pm\int^{\vh_0}_{\vh_I}{d\vh}~{\langle{(\widetilde{T}_{\mathrm{d}}})^{-1/2}\rangle}\geq0$&$s_{\pm}=\pm\frac{m}{E}[X_0^0-X^0_I]\geq 0$\\[.2cm]
\hline 8.&$P_\vh=\pm 2 \int d^3x(\widetilde{T}_{\rm d})^{1/2}$&$P_0=\pm\sqrt{m^2c^4+|\vec p|^2}$\\[.1cm]
\hline 9.&$[\hat P^2_\vh-E^2_\vh]\Psi_{\rm WDW}=0$&$[\hat P^2_0-E^2_0]\Psi_{\rm KG}=0$\\[.1cm]
\hline 10.&$\Psi_{\rm WDW}=\dfrac{A^++A^-}{\sqrt{2E_\vh}}$&$\Psi_{\rm KG}=\dfrac{a^++a^-}{\sqrt{2E_0}}$\\[.1cm]
\hline 11.&$A^+=\alpha B^+\!+\!\beta^*B^-$&$a^+=\alpha b^+\!+\!\beta^*b^-$\\[.1cm]
\hline 12.&$B^-|0>=0$&$b^-|0>=0$\\[.1cm]
\hline 13.&$<0|A^+A^-|0>\not=0$&$<0|a^+a^-|0>\not=0$\\[.1cm]
\hline
\end{tabular}
\caption{\textbf{The $3L+1G$ diffeomorphisms \& universe-particle
correspondence} \cite{bpzz,242}. \small{This universe-particle
correspondence rejects Hilbert's {\it Foundations of relativistic
physics} of 1915 \cite{H} that based on the action principle
($\mathrm{N}^{\underline{\mathrm{o}}}$1) with a geometric interval
($\mathrm{N}^{\underline{\mathrm{o}}}$2) and the group of
diffeomorphisms ($\mathrm{N}^{\underline{\mathrm{o}}}$3), in
contrast to the classical physics based  only on an action and the
group of the data transformations. The group of diffeomorphisms
($\mathrm{N}^{\underline{\mathrm{o}}}$3) leads to the energy
constraint ($\mathrm{N}^{\underline{\mathrm{o}}}$6). Resolution of
the energy constraint gives the Hubble type relation
($\mathrm{N}^{\underline{\mathrm{o}}}$7) between the time-variable
($\mathrm{N}^{\underline{\mathrm{o}}}$4) in space of events
($\mathrm{N}^{\underline{\mathrm{o}}}$5) and the time-interval
($\mathrm{N}^{\underline{\mathrm{o}}}$2) and determines the energy
of events ($\mathrm{N}^{\underline{\mathrm{o}}}$8) that can take
positive and negative values. With the aim to remove the negative
value, one can use the  experience of QFT, i.e., the primary
quantization ($\mathrm{N}^{\underline{\mathrm{o}}}$9) and the
secondary one ($\mathrm{N}^{\underline{\mathrm{o}}}$10). This
quantization procedure leads immediately to creation from stable
Bogoliubov vacuum state ($\mathrm{N}^{\underline{\mathrm{o}}}$12)
of both quasiuniverses and quasiparticles
($\mathrm{N}^{\underline{\mathrm{o}}}$13) obtained by the
Bogoliubov transformation
($\mathrm{N}^{\underline{\mathrm{o}}}$11)
\cite{Schweber,Logunov}.}}
\end{table}
\subsubsection{The Hamiltonian formalism in finite space-time}

 The  cosmological perturbation theory is
consistent with the Dirac -- ADM Hamiltonian approach to GR
considered above, if the zeroth time-like variable in the field
space of events  (\ref{d-t1}) is extracted on the level of action
(\ref{Asv110}), (\ref{1-0}), (\ref{2-0}) \cite{bpzz,242}
 \be\label{3-21}
 S[\vh_0|F]=\widetilde{S}[\vh|\widetilde{F}]-
 V_0\int dx^0 \frac{1}{N_0}\left(\frac{d\vh}{dx^0}\right)^2=\int dx^0 L;
\ee
 where  $\widetilde{S}[\varphi|\widetilde{F}]$
  is the action (\ref{1-1})  in
 terms of metrics ${\widetilde{g}}$, where  $\vh_0$ is replaced by
 the running scale
 $\vh(x^0)=\vh_0a(x^0)$
  of all masses  of the
 matter fields.  The action (\ref{3-21}) leads to
 the energy constraints
 \be\label{3-22}
 \frac{\delta S[\vh_0]}{\delta
 \widetilde{N}_{\rm d}}=-{T}_{\rm d}=
 \frac{(\partial_0\varphi)^2}{\widetilde{N}_{\rm d}^2}-
 \widetilde{T}_{\rm d}=0,~~ \widetilde{T}_{\rm d}\equiv-
 \dfrac{\delta \widetilde{S}[\vh]}{\delta  \widetilde{N}_{\rm
 d}}\geq 0
 \ee

The kinemetric subgroup (\ref{1-8}) essentially
  simplifies the solution of the energy constraint (\ref{3-22})
  if the homogeneous variable is extracted from
  the spacial determinant.

 Therefore, one should point out in the finite volume
 the homogeneous variable $\vh(x^0)$
 as the evolution parameter (time-variable) in the field space of events
 $[\vh|\widetilde{F}]$
 and diffeo-invariant time-interval $N_0 dx^0=d\zeta$
 (g-time), where $N_0[\widetilde{N}_{\rm d}]$ as functional of
 $\widetilde{N}_{\rm d}$ can be defined as the spacial averaging (\ref{2-0})
 \be\label{3-21-2}
 \frac{1}{N_0[\widetilde{N}_{\rm d}]}=\frac{1}{V_0}\int \frac{d^3x}{\widetilde{N}_{\rm d}}
 \equiv\langle\widetilde{N}_{\rm d}^{-1} \rangle.
 \ee

  According to the Wheeler -- DeWitt \cite{WDW} there is the
 universe -- particle correspondence  given in the Table \ref{tab1} \cite{bpzz,242}.

 This QFT experience illustrates  the possibility to  solve
 the problems of the quantum origin of all matter fields
  in the Early Universe,  its evolution,
  and the present-day energy budget \cite{bpzz,origin,114:a}.
  In order to use this possibility, one should impose a set of
 requirements on the cosmic motion in the field space of events
 that follow from the general principles of QFT.

    The  QFT experience  supposes that the action
 (\ref{3-21})  can be represented in the {\it canonical}
 Hamiltonian form like (\ref{h-1})
 \bea\nonumber
 S[\vh_0|F]&=&\int dx^0\left\{-P_{\vh}\partial_0\vh+
 N_0[\widetilde{N}_{\rm d}]\frac{P^2_\vh}{4V_0}\right\}+\\\label{1-36a}
 &+&\int d^4x
 \left[\sum\limits_{\widetilde{F}
 } P_{\widetilde{F}}\partial_0\widetilde{F}
 +{\cal C}-\widetilde{N}_d \widetilde{T}_{\rm d}\right].
 \eea
In this case, the energy constraint (\ref{3-22}) takes the form of
\emph{the Friedmann equation}
 \be\label{6-1-113ec}
 \left[\frac{d\varphi}{d\zeta}\right]^2\equiv\vh'^2=
 {\left\langle(\widetilde{T}_{\rm d})^{1/2}\right\rangle}^2 ,
 \ee
 and the algebraic equation for
the diffeo-invariant lapse function
 \be\label{3-29}
  {\cal N}=\langle(\widetilde{N}_{\rm d})^{-1} \rangle
 \widetilde{N}_d=
 \left\langle (\widetilde{T}_{\rm d})^{1/2}\right\rangle
 ({{\widetilde{T}_{\rm d}}})^{-1/2}.
 \ee
We see that the energy constraint (\ref{3-22}) removes
  only one  global momentum $P_\vh$
 in accord to the dimension
  of the kinemetric diffeomorphisms (\ref{1-8})
  that is consistent with the second N\"other theorem.

 One can find the
 evolution of all field variables $F(\vh,x^i)$  with respect to
 $\vh$ by variation of the ``reduced'' action
 \be\label{2ha2} S[\vh_0]{}_{{}_{{P_\vh=\pm E_\vh}}} =
 \int\limits_{\vh_I}^{\vh_0}d\widetilde{\vh} \left\{\int d^3x
 \left[\sum\limits_{  F}P_{  F}\partial_\vh F\mp2\sqrt{\widetilde{T}_{\rm d}(\widetilde{\vh})}\right]\right\} \ee
 obtained as the constraint-shell (\ref{6-1-113ec})
   values of the Hamiltonian form of the initial action
   (\ref{1-36a}) \cite{pp}.

 The energy constraints (\ref{3-22}) and the Hamiltonian reduction  (\ref{2ha2})
 lead to the  definite
 {\it  canonical   rules} of the Universe evolution in the field space of
 events $[\vh|\widetilde{F}]$.

\begin{description}
 \item[Rule 1: Causality Principle in the WDW space] $\dfrac{d\vh_I}{d\vh_0} =0${\textit{
  follows from the Hamiltonian reduction
 (\ref{2ha2}) that gives us the solution of the Cauchy problem
 and means that initial data $\vh_I,\vh'_I$ do not
 depend on the Planck value $\vh_0,\vh'_0$.
 }}
 \item[Rule 2: Positive
 Energy Postulate] \emph{follows from
 the energy constraint} (\ref{3-22})
  $\dfrac{(\partial_0\varphi)^2}{\widetilde{N}_{\rm d}^2}=
 \widetilde{T}_{\rm d}=-\dfrac{16}{\vh^2}
p^2_{\widetilde{\psi}}+...\geq 0${
 \begin{equation}\label{rule-I}
 \widetilde{T}_{\rm d}\geq 0 ~ \to~
 p_\psi=-\frac{4\vh^2}{3{\widetilde{\psi}^6N_{\rm
 inv}}}[\partial_j(\widetilde{\psi}^6{\cal
 N}^j)-(\widetilde{\psi}^6)']=0,
 \end{equation}
 \emph{where
  $ ({\cal N}^j=N^j\langle
 N^{-1}_d\rangle\not =0)$.}
 }
 \item[Rule 3: Vacuum Postulate]{\textit{ $B^-|0>=0$
 restricts  the  Universe motion in the field space of events
 \begin{eqnarray}
 P_\vh\geq 0~~\mathrm{for}~~\vh_I\leq\vh_0\\\nonumber P_\vh\leq
 0~~\mathrm{for}~~\vh_I\geq\vh_0.
 \end{eqnarray}
 }}
 \item[Rule 4: Lapse Function ]{\textit{${\cal N}>0$
 follows from
 the nonzero energy density $\widetilde{T}_{\rm d}\not =0$.
 }}
\end{description}

The Rule 1  is not compatible with the Planck epoch \cite{linde}
 \be\label{pl-e}
 \vh_0 \cdot a_I=\vh_0 \cdot \frac{a'_0}{\vh_0} \to \vh_I=\frac{\vh'_0}{\vh_0}={\cal H}_0
 \ee
 in the beginning of the Universe as
$\dfrac{d\vh_I}{d\vh_0}\not =0$.

The Rule 2 (\ref{rule-I}) means that the local scalar component
$\widetilde{\psi}^2=\psi^2/a$ has zero momentum and satisfies the
 equation with Laplacian (instead of the D'alambertian) in accord with
 the Dirac classification of the radiation-like variables in GR \cite{dir}.
  In other words, the local scalar component cannot be the
local dynamic variable as it is proposed for the description of
the CMB power spectrum in the acceptable $\Lambda$CDM model
\cite{mukh}.

 The Rule 3 leads to the arrow of the geometric
time-interval.

The Rule 4 forbids any zero values of the local lapse function
(\ref{3-29}), so that penetration into a internal region of black
hole is not possible because this penetration is accompanied the
change of a sign of the local lapse function (\ref{3-29}) that
proposes zero values of the local lapse function.

  Let us check
 the correspondence of the canonical GR
  with both the QFT in the flat space-time and
  the classical Newton theory.

\subsection{Correspondence principle and QFT limits}

 The correspondence principle \cite{pp}
 as the low-energy
 expansion of the {\it``reduced action''} (\ref{2ha2}) over the
 field density ${T}_{{\rm s}}$
 \be
 2d\vh \sqrt{\widetilde{T}_{\rm d}}= 2d\vh
 \sqrt{\rho_{0}(\vh)+{T}_{{\rm s}}}
 =
 d\vh
 \left[2\sqrt{\rho_0(\vh)}+
 \frac{{T}_{{\rm s}}}{\sqrt{\rho_0(\vh)}}\right]+...
 \ee
 gives the following sum:
 \be
 S^{(+)}|_{\rm
 constraint}= S^{(+)}_{\rm cosmic}+S^{(+)}_{\rm
 field}+\ldots,
 \ee where
\be
 S^{(+)}_{\rm cosmic}[\varphi_I|\varphi_0]= -
 2V_0\int\limits_{\vh_I}^{\vh_0}\!
 d\vh\!\sqrt{\rho_0(\vh)}
 \ee is the reduced  cosmological action (\ref{2ha2}),
 and
 \be\label{12h5} S^{(+)}_{\rm field}=
 \int\limits_{\eta_I}^{\eta_0} d\eta\int\limits_{V_0}^{} d^3x
 \left[\sum\limits_{ F}P_{ F}\partial_\eta F
 -{T}_{{\rm s}}\right]
 \ee
 is the standard field action
 in terms of the conformal time:
 $d\eta=\dfrac{d\vh}{\sqrt{\rho_0(\vh)}}$,
 in the conformal flat space--time with running masses
 $m(\eta)=a(\eta)m_0$.

 This expansion shows that the Hamiltonian approach to the General
 Theory of Relativity
 in terms of the Lichnerowicz scale-invariant variables
 (\ref{adm-2}) identifies the ``conformal quantities''
  with the observable ones including the conformal time $d\eta$,
  instead of $dt=a(\eta)d\eta$, the coordinate
 distance $r$, instead of the Friedmann one $R=a(\eta)r$, and \emph{the conformal
 temperature $T_c=Ta(\eta)$, instead of the standard one $T$}.
 Therefore,
 the scale-invariant variables  distinguish the conformal
 cosmology (CC) \cite{039},
  instead of the standard cosmology (SC) \cite{linde}.

\subsection{Canonical Cosmological Perturbation Theory}

 In diffeo-invariant formulation of GR in the specific
  reference frame the scalar potential perturbations
  can be defined as ${\cal N}^{-1}=1+\overline{\nu}$ and
 $\widetilde{\psi}=e^{\overline{\mu}}=1+\overline{\mu}+...$, where
 $\overline{\mu},\overline{\nu}$ are given in the class of
 functions distinguished by the projection operator
 $\overline{F}=F-\langle F\rangle$
 ($\langle \overline{F}\rangle \equiv 0$).

 The explicit dependence of the metric simplex and the energy
 tensor
 $\widetilde{T}_{\rm d}$ on $\widetilde{\psi}$
  can be given in terms of the scale-invariant Lichnerowicz variables \cite{lich}
  introduced in Appendix C (\ref{1-21}) and
 \bea\label{adm-2}
 \omega^{(L)}_{(0)}&=&\widetilde{\psi}^4{\cal N}d\zeta,~~~~~
 \omega^{(L)}_{(b)}={\bf e}_{(b)k}[dx^k +{\cal N}^kd\zeta],
 \\
\label{La-2}
 \widetilde{T}_{\rm d}&=& \widetilde{\psi}^{7}\hat \triangle
 \widetilde{\psi}+
  \sum\limits_{I} \widetilde{\psi}^I a^{\frac{I}{2}-2}{\cal T}_I,
  ~~~~~~{\cal T}_I\equiv\langle{\cal T}_I\rangle+\overline{{\cal T}_I},
 \eea
where $\hat \triangle
 \widetilde{\psi}\equiv\dfrac{4\varphi^2}{3}\partial_{(b)}
 \partial_{(b)}\widetilde{\psi}$ is the
 Laplace operator and  ${\cal T}_I$ is partial energy density
  marked by the index $I$ running a set of values
   $I=0$ (stiff), 4 (radiation), 6 (mass), and 8 (curvature)
 in correspondence with a type of matter field contributions
 considered in Appendix C (\ref{h32}) -- (\ref{h35})
 (except of the $\Lambda$-term, $I=12$).
 The negative contribution $-({16}/{\vh^2})\overline{p_{\psi}}^2$ of the
 spatial determinant momentum  in the energy
 density ${\cal T}_{I=0}$
can be removed by the Dirac constraint \cite{dir} of the
 zeroth velocity of the spatial volume element (\ref{rule-I})
 \be\label{La-6}
 \overline{p_{\psi}}=
 -8\vh^2\frac{\partial_{\zeta}\widetilde{\psi}^6-\partial_l
 [\widetilde{\psi}^6{\cal N}^l]}{\widetilde{\psi}^6{\cal N}}=0.
 \ee
 The diffeo-invariant part of the lapse function ${N}_{\rm int}$
 is determined by the local part (\ref{3-29})
 of the energy constraint (\ref{3-22}) that can be
 written as
 \be\label{ec1}
 \widetilde{T}_{\rm d}={\cal N}^{-2}\rho_{(0)}, ~~~~\to
 ~~~~{\cal N}^{-1}=\sqrt{\widetilde{T}_{\rm d}}\,\rho_{(0)}^{-1/2},
 \ee
where $\rho_{(0)}=\left\langle \sqrt{\widetilde{T}_{\rm
d}}\right\rangle^2$\!\!.
 In the class of functions $\overline{F}=F-\langle F\rangle$, the classical
 equation ${\delta S}/{\delta \log \widetilde{\psi}}=0$ takes the
 form
 \be\nonumber
 \widetilde{\psi}\dfrac{\delta S}{\delta \widetilde{\psi}}=-\widetilde{T}_{\psi}=\widetilde{N_d}\widetilde{\psi}
 \frac{\partial \widetilde{T}_{\rm d}}{\partial \widetilde{\psi}}+
 \widetilde{\psi}\triangle \left[
 \frac{\partial \widetilde{T}_{\rm d}}{\partial
 \triangle\widetilde{\psi}}\widetilde{N_d}\right]=0.
 \ee
 Using the property of the deviation projection operator
 $\delta S/\delta\overline{\mu}=\overline{D}=D-\langle D\rangle$,
 where $\overline{\mu}=\log \widetilde{\psi}$, we got
 the following equation
\be\label{e2}
 7{{\cal N}\widetilde{\psi}^7\hat \triangle\widetilde{\psi}}
 \!+\!{\widetilde{\psi}\hat \triangle[{\cal N}\widetilde{\psi}^7]}
 \!+\!{{\cal N}}\sum\limits_{I} I{\widetilde{\psi}^I
 a^{\frac{I}{2}-2}{\cal T}_I}=\rho_{(1)},
 \ee
 where $\rho_{(1)}=\left\langle 7{{\cal N}\widetilde{\psi}^7\hat \triangle\widetilde{\psi}}
 \!+\!{\widetilde{\psi}\hat \triangle[{\cal N}\widetilde{\psi}^7]}
 \!+\!\sum\limits_{I} I{\widetilde{\psi}^I
 a^{\frac{I}{2}-2}{\cal T}_I}\right\rangle$. Using (\ref{ec1})
 we can write for $\widetilde{\psi}$ a nonlinear equation
 \be\nonumber\label{nl}
 ({{\widetilde{T}_{\rm d}})^{-1/2}
 \left[7\widetilde{\psi}^7\hat \triangle\widetilde{\psi}
 \!+\sum\limits_{I} I\widetilde{\psi}^I
 a^{\frac{I}{2}-2}{\cal T}_I \right]+
 \widetilde{\psi}\hat \triangle[({\widetilde{T}_{\rm d}})^{-1/2}\widetilde{\psi}^7]}
 =\rho_{(1)}\rho_{(0)}^{-1/2}.
 \ee

 In the infinite volume limit $\rho_{(n)}=0,~a=1$
 Eqs.  (\ref{ec1}) and (\ref{e2})  coincide with the equations
 of the diffeo-variant formulation of GR
 $T_{\rm d}=0$ and (\ref{1-37ab}) considered in Section 2.3.

 For the small deviations $N^{-1}_{\rm int}=1+\overline{\nu}$ and
 $\widetilde{\psi}=e^{\overline{\mu}}=1+\overline{\mu}+...$ the
 first orders of Eqs.  (\ref{ec1}) and (\ref{e2}) take the form
 \bea\label{1e1-2}
   (-\hat \triangle-\rho_{(1)})\overline{\mu}~~~~ +&
   2\rho_{(0)}\overline{\nu}&=~~~\overline{{\cal T}}_{(0)},
 \\\label{1ec1-2}
 (14\hat \triangle+\rho_{(2)})\overline{\mu}~~
 -&~~~~~(\hat
 \triangle+\rho_{(1)})\overline{\nu}&=-~\overline{{\cal T}}_{(1)},
 \eea
 where
 \bea\label{ec1-3}
 \rho_{(n)}=\langle{\cal T}_{(n)}\rangle\equiv\sum_II^na^{\frac{I}{2}-2}\langle{\cal T}_{I}\rangle\\
\label{ec1-4} {\cal T}_{(n)}=\sum_II^na^{\frac{I}{2}-2}{\cal T}_{I}.
 \eea

 The set of Eqs. (\ref{e1-2}) and (\ref{ec1-2})
 gives $\overline{\nu}$ and $\overline{\mu}$ in the form of a sum
  \bea\nonumber\label{2-17}
 {\overline{\mu}}=\frac{1}{14\beta}\int d^3y\left[D_{(+)}(x,y) \overline{T_{(+)}}(y)-
 D_{(-)}(x,y) \overline{T_{(-)}}(y)\right],\\\nonumber\label{2-18}
 {\overline{\nu}}=\frac{1}{2\beta}\int d^3y\left[(1+\beta)D_{(+)}(x,y) \overline{T_{(+)}}(y)-
 (1-\beta)D_{(-)}(x,y) \overline{T_{(-)}}(y)\right],
  \eea
 where
 \be\label{beta}
 \beta=\sqrt{1+[\langle {\cal T}_{(2)}\rangle-14\langle
{\cal T}_{(1)}\rangle]/(98\langle {\cal T}_{(0)}\rangle)}, \ee
 \be\label{cur1}
 \overline{T}_{(\pm)}=(7\overline{{\cal T}}_{(0)}-
 \overline{{\cal T}}_{(1)})~\pm~ 7\beta\overline{{\cal T}}_{(0)}
 \ee
 are the local currents, $D_{(\pm)}(x,y)$ are the Green functions satisfying
 the equations
 \bea\label{2-19}
 [\pm \hat m^2_{(\pm)}-\hat \triangle
 ]D_{(\pm)}(x,y)=\delta^3(x-y),
 \eea
 where $\hat m^2_{(\pm)}= 14 (\beta\pm 1)\langle {\cal T}_{(0)}\rangle \mp
\langle {\cal T}_{(1)}\rangle$.

   In the case of point mass distribution in a finite volume $V_0$ with the zeroth pressure
  and  the  density
  $\overline{{\cal T}}_{(1)}=\dfrac{\overline{{\cal T}}_{(2)}}{6}
  \equiv \sum\limits_{J} M_J\left[\delta^3(x-y_J)-\dfrac{1}{V_0}\right]$,
 solutions   (\ref{2-17}),  (\ref{2-18}) take
 a very important form
 \bea\label{2-21}
  \overline{\mu}(x)&=\sum\limits_{J}
  \dfrac{r_{gJ}}{4r_{J}}\left[{\gamma_1}e^{-m_{(+)}(z)
 r_{J}}+ (1-\gamma_1)\cos{m_{(-)}(z)
 r_{J}}\right],\\\label{2-22}
 \overline{\nu}(x)&=\sum\limits_{J}
 \dfrac{2r_{gJ}}{r_{J}}\left[(1-\gamma_2)e^{-m_{(+)}(z)
 r_{J}}+ {\gamma_2}\cos{m_{(-)}(z)
 r_{J}}\right],
 \eea
 where
 $$
  {\gamma_1}=\frac{1+7\beta}{14\beta},~~~
 {\gamma_2}=\frac{(1-\beta)(7\beta-1)}{16\beta},~~
 $$
 $$r_{gJ}=\frac{3M_J}{4\pi\vh^2},~~
 r_{J}=|x-y_J|,~~~~m^2_{(\pm)}=\hat m^2_{(\pm)}\frac{3}{4\vh^2}.
 $$
 The minimal surface (\ref{rule-I})
  $\partial_i[\overline{\psi}^6{\cal N}^i]-(\overline{\psi}^6)'=0$
 gives the shift of the coordinate
  origin in the process of evolution
 \be \label{2-23}
{\cal
 N}^i=\left(\frac{x^i}{r}\right)\left(\frac{\partial_\zeta V}{\partial_r V}\right),~~~
 ~~~V(\zeta,r)=\int\limits_{0}^{r}d\widetilde{r}
 ~\widetilde{r}^2\widetilde{\psi}^6(\zeta,\widetilde{r}).
  \ee
In the infinite volume limit $\langle {\cal T}_{(n)}\rangle=0$ these
 solutions take the standard Newtonian form:
 $\overline{\mu}=D\cdot {\cal T}_{(0)}$, $\overline{\nu}=D\cdot [14{\cal T}_{(0)}-{\cal T}_{(1)}]$,
 ${\cal N}^i=0$
 (where $\hat \triangle D(x)=-\delta^3(x)$).

 \subsection{Generalization of the Schwarzschild solution}
One can see that another choice of variables for
 scalar potentials rearranges the perturbation series and leads
 to another result. In order to demonstrate this fact, let us
 change the lapse function
 as
 ${{\cal N}}\widetilde{\psi}^{7}
 =1-\overline{\nu_1}$ and keep
 $\widetilde{\psi}=1+\overline{\mu_1}$.
In order to simplify equations of the scalar potentials
  ${\cal N},\widetilde{\psi}$, one can introduce new
 table of symbols:
 \bea\label{nts}
 N_{\rm s}=\psi^7 {\cal N},\\T(\widetilde{\psi})=
 \sum\limits_{I} \widetilde{\psi}^{(I-7)}
 a^{\frac{I}{2}-2}{\cal T}_I,\\\label{4-1r}\rho_{(0)}=\left\langle
\sqrt{\widetilde{T}_{\rm d}}\right\rangle^2=\vh'^2.
 \eea
 In terms of these symbols the action (\ref{Asv110}) can be presented
 as a generating functional of equations of the local scalar potentials
 $N_{\rm s},\widetilde{\psi}$ and  field variables $F$ in terms of
 diffeo-invariant time $\zeta$:
 \be\label{gf}
 S[\varphi_0]\!\!=\!\!\int d\zeta\int d^3x \left[\sum_F P_F\partial_\zeta
 F\!\!-\!\!{N}_{\rm s}\left(\hat \triangle
 \widetilde{\psi}\!+\!{T}(\widetilde{\psi})\right)\!\!-\!\!\frac{\widetilde{\psi}^7
 \rho_{(0)}}{N_{\rm s}}\right].
 \ee

 The variations of this action with respect to
 $N_{\rm s},\widetilde{\psi}$ lead to equations
 \bea\label{4-1}
\hat \triangle
 \widetilde{\psi}+{T}(\widetilde{\psi})&=&\frac{\widetilde{\psi}^7
 \rho_{(0)}}{N^2_{\rm s}},
 \\\label{4-2}
\widetilde{\psi}\hat \triangle{N}_{\rm s}
 +{N}_{\rm s}\widetilde{\psi}\partial_{\widetilde{\psi}}{T}
 +7\frac{\widetilde{\psi}^7
 \rho_{(0)}}{N_{\rm s}}&=&\rho_{(1)},
 \eea
 respectively, we have used here the constraint  (\ref{La-6}) and
  the property of the deviation projection operator
 $\delta S/\delta\overline{\mu}=\overline{D}=D-\langle D\rangle$,
  according
 to which $\rho_{(1)}=\langle\widetilde{\psi}\hat \triangle{N}_{\rm s}
 +{N}_{\rm s}\widetilde{\psi}\partial_{\widetilde{\psi}}{T}
 +7{\widetilde{\psi}^7
 \rho_{(0)}}/{N_{\rm s}}\rangle$.

One can see that in the infinite volume limit
$\rho_{(n)}=\langle{\cal T}_I\rangle=0$ Eqs. (\ref{4-1}) and
(\ref{4-2}) reduce to the  equations of the conventional GR with the
Schwarzschild solutions
  ${\overline{\psi}}=1+\dfrac{r_g}{4r};~~
  N_{\rm s}
  =1-\dfrac{r_g}{4r}$ in empty space, where Eqs. (\ref{4-1}) and
(\ref{4-2}) become
  $\hat \triangle {\overline{\psi}}=0, ~\hat \triangle N_{\rm
  s}=0$.

 For the small deviations $N_{\rm s}=1-{\nu}_1$ and
 $\widetilde{\psi}=1+{\mu}_1$ the
 first orders of Eqs.  (\ref{4-1}) and
(\ref{4-2}) take the form
 \bea\nonumber\label{e1-2}
   [-\hat{\triangle}+14\rho_{(0)}-\rho_{(1)}]\mu_{1} +
   2\rho_{(0)}\nu_1=\overline{{\cal T}}_{(0)}
 \\\nonumber\label{ec1-2}
 [7\cdot 14\rho_{(0)}-14\rho_{(1)}+\rho_{(2)}]\mu_1
 +[-\hat{\triangle}+
14\rho_{(0)}-\rho_{(1)}]\nu_1=7\overline{{\cal
T}}_{(0)}-\overline{{\cal T}}_{(1)},
 \eea
where
 \bea\label{1ec1-3}
 \rho_{(n)}=\langle{\cal T}_{(n)}\rangle
 \equiv\sum_II^na^{\frac{I}{2}-2}\langle{\cal T}_{I}\rangle.
 \eea

 This choice of variables
 determines $\overline{\mu_1}$ and $\overline{\nu_1}$ in the form of a sum
  \bea\nonumber\label{12-17}
 \widetilde{\psi}
 &=&1+\frac{1}{2}\int d^3y\left[D_{(+)}(x,y)
\overline{T}_{(+)}^{(\mu)}(y)+
 D_{(-)}(x,y) \overline{T}^{(\mu)}_{(-)}(y)\right],\\\nonumber\label{12-18}
 {\cal N}\widetilde{\psi}^7
 &=&1-\frac{1}{2}\int d^3y\left[D_{(+)}(x,y)
\overline{T}^{(\nu)}_{(+)}(y)+
 D_{(-)}(x,y) \overline{T}^{(\nu)}_{(-)}(y)\right],
  \eea
 where $\beta$ are given by Eqs. (\ref{beta})
 \bea\label{1cur1}\overline{T}^{(\mu)}_{(\pm)}
 &=&\overline{{\cal T}}_{(0)}\mp7\beta
  [7\overline{{\cal T}}_{(0)}-\overline{{\cal T}}_{(1)}],\\
  ~\overline{T}^{(\nu)}_{(\pm)}&=&[7\overline{\cal T}_{(0)}
 -\overline{\cal T}_{(1)}]
 \pm(14\beta)^{-1}\overline{\cal T}_{(0)}
 \eea
 are the local currents, $D_{(\pm)}(x,y)$ are the Green functions satisfying
 the equations (\ref{2-19})
 where $\hat m^2_{(\pm)}= 14 (\beta\pm 1)\langle {\cal T}_{(0)}\rangle \mp
\langle {\cal T}_{(1)}\rangle$. In the finite volume limit these
solutions for $\widetilde{\psi}, {\cal N}$ coincide with solutions
(\ref{2-17}) and (\ref{2-18}), where
$\overline{\nu_1}=\overline{\nu} -7\overline{\mu}$ and
$\overline{\mu_1}=\overline{\mu}$.

  In the case of point mass distribution in a finite volume $V_0$ with the
zeroth pressure
  and  the  density
 \be\overline{{\cal T}}_{(0)}(x)=\dfrac{\overline{{\cal T}}_{(1)}(x)}{6}
  \equiv  M\left[\delta^3(x-y)-\dfrac{1}{V_0}\right],\ee
 solutions   (\ref{12-17}),  (\ref{12-18}) take
 a  form
 \bea\label{12-21}
  \widetilde{\psi}&=1+
  \dfrac{r_{g}}{4r}\left[{\gamma_1}e^{-m_{(+)}(z)
 r}+ (1-\gamma_1)\cos{m_{(-)}(z)
 r}\right],\\\label{12-22}
 {\cal N}\widetilde{\psi}^{7}&=1-
 \dfrac{r_{g}}{4r}\left[(1-\gamma_2)e^{-m_{(+)}(z)
 r}+ {\gamma_2}\cos{m_{(-)}(z)
 r}\right],
 \eea
 where
 ${\gamma_1}=\dfrac{1+7\beta}{2}$,
 ${\gamma_2}=\dfrac{14\beta-1}{28\beta}$,
 $r_{g}=\dfrac{3M}{4\pi\vh^2}$, $r=|x-y|$.
 Both choices of variables  (\ref{2-21}),  (\ref{2-22}) and
(\ref{12-21}),  (\ref{12-22}) have spatial oscillations and the
nonzero shift of the coordinate
  origin of the type of (\ref{2-23}).

In the infinite volume limit $\langle {\cal T}_{(n)}\rangle=0,~a=1$
 solutions (\ref{12-21}) and  (\ref{12-22}) coincide with
 the isotropic version of  the Schwarzschild solutions:
 $\widetilde{\psi}=1+\dfrac{r_g}{4r}$,~
 ${N_{\rm inv}}\widetilde{\psi}^{7}=1-\dfrac{r_g}{4r}$,~$N^k=0$.
 It is of interest  to find  an exact solution of Eq. (\ref{nl}) for
 different equations of state.

 \subsection{Investigation of CMB fluctuations}
\subsubsection{CMB fluctuation problem}
The investigation of CMB fluctuations is one of the highlights of
present-day cosmology with far-reaching implications and more
precise observations are planned for the near future. Therefore, the
detailed investigation of any possible flaw of the standard theory
deserves attention and public discussion.

``CMBR anisotropy'' in the inflationary model is described in
\cite{lif,MFB,KS}
 by the decomposition of the metric interval
\bea\label{1.4.1} &&ds^2=g_{\mu\nu}dx^\mu dx^\nu=\\\nonumber
&=&\label{1.4.1h}a^2(\eta)\left[(1+2\Phi)d\eta^2-2{\cal
N}_kdx^kd\eta -(1-2\Psi)dx^2-dx^idx^j(h_{ij})\right]  \eea
 associated with the Lifshits cosmological perturbation theory
 \cite{lif}. The comparison of this interval with
 the exact interval (\ref{3L+1G-1}) gives for scalar components
 the relations
 \bea \label{1.4.1hp}\widetilde{\psi}&=&1-\frac{\Psi}{2},
 \\\label{1.4.1hn}{\cal N}\widetilde{\psi}^6&=&1+{\Phi}.
 \eea

  The final expression for the temperature fluctuations  induced by scalar
fluctuations of the metric components (known as the Sachs-Wolfe (SW)
effect) can be written as \cite{MG} \bea\nonumber \label{scalSW}
&&\biggl(\frac{\Delta T}{T}\biggr)_{\rm s} = \biggr[ \frac{\delta
\rho_{\rm r}}{4\rho_{r}} + \Phi + n_{i} v^{i}_{\rm
b}\biggr]_{\eta_{i}} \!\!+\!\! \int_{\eta_{i}}^{\eta_{0}}d\eta(
\Psi' +\Phi')=\\\label{scalSW-h}\!\!\!&=&\!\!\!\biggr[ \frac{\delta
\rho_{\rm r}}{4\rho_{r}} \!+\! ({\cal N}\widetilde{\psi}^6\!-\!1)
\!+\! n_{i} v^{i}_{\rm b}\biggr]_{\eta_{i}} \!\!+\!\!
\int_{\eta_{i}}^{\eta_{0}}d\eta\left( \frac{\widetilde{\psi}'}{2}
\!+\!({\cal N}\widetilde{\psi}^6)'\!\!\right)\!. \eea Equation
(\ref{scalSW}) is integrated once\footnote{Notice
 that integration by parts is necessary in order to integrate the term $2 \partial_{i}\phi n^{i}$. Recall, in fact
 that $d\Phi / d\eta = \Phi' + \partial_{i} \phi n^{i}$.}
with respect to $\eta$ between the time $\eta_{i}$ (coinciding with
the decoupling time) and the time $\eta_{0}$ (coinciding with the
present time). Equation  (\ref{scalSW}) has three contribution
\begin{itemize}
\item{} the ordinary SW effect given by the first two terms at the right hand side of Eq. (\ref{scalSW})
i.e. ${\delta \rho_{\rm r}}/{(4\rho_{r})}$ and $\phi$;
\item{} the Doppler term (third term in Eq, (\ref{scalSW}));
\item{} the integrated SW effect (last term in Eq, (\ref{scalSW})).
\end{itemize}
The ordinary SW effect is due both to the intrinsic temperature
inhomogeneities on the last scattering surface and to the
inhomogeneities of the metric.  On large angular scales the ordinary
SW contribution dominates. The Doppler term arises thanks to the
relative velocity of the emitter and of the receiver. At large
angular scales its contribution is subleading but it becomes
important at smaller scales, i.e. in multipole space, for $\ell \sim
200$ corresponding to the first peak. The induced temperature
fluctuations induced by the vector modes of the geometry can be
written as
\begin{equation}
\biggl(\frac{\Delta T}{T}\biggr)_{\rm v} = [ -\vec{{\cal V}}\cdot
\vec{n}]_{{\cal T} _{i}}^{\eta_{f}} + \frac{1}{2}
\int_{\eta_{i}}^{\eta_{f}} (\partial_{i} {\cal N}_{j} + \partial_{j}
{\cal N}_{i}) n^{i} n^{j} d\eta.
\end{equation}
 where ${\cal V}^{i}_{\rm b}$ is the rotational component of the
 baryonic peculiar velocity.

 \subsubsection{Canonical Cosmological Perturbations Theory versus Lifshitz's one}
 We shell use the definition of the
 scalar components of the energy momentum tensor
 (\ref{La-2}) and (\ref{e2}). The energy momentum tensor components
 are
 \bea \label{3L+1G-3h}
 \widetilde{T}_{\rm d}&=&\dfrac{4\vh_0^2a^2}{3}{\widetilde{\psi}}^{7}
 \triangle
{\widetilde{\psi}}+\\\nonumber
  &+&\sum\limits_{I=0,4,6,8,12} a^{I/2-2}{\widetilde{\psi}}^I{\cal T}_I
  =2(T_{00}-T_{kk}),
  \\\label{3L+1G-8h}
   \widetilde{T}_{\psi}&=&\dfrac{4\vh_0^2a^2}{3}
   \left\{7{\cal N}{\widetilde{\psi}}^{7}
  \triangle {\widetilde{\psi}}+{\widetilde{\psi}} \triangle
\left[{\cal N}{\widetilde{\psi}}^{7}\right]\right\}+\\\nonumber
  &+&{\cal N}\sum\limits_{I=0,4,6,8,12}I a^{I/2-2}{\widetilde{\psi}}^I{\cal
  T}_I=12 T_{kk}
\eea Let us compare the equations of the canonical perturbation
theory  (\ref{4-1r}),  (\ref{4-1}) and (\ref{4-2}) with the
Lifshits cosmological perturbation theory for the scalar
components (\ref{1.4.1hp}) and (\ref{1.4.1hn})
  \bea\label{1.4.2}4\pi G a^2T_{00}&=&
 -3{\cal H}({\cal H}\Phi+\Psi')+\triangle\Psi  \\
 \label{1.4.3}\nonumber
 4\pi G a^2T_{kk}&=&
 3[(2{\cal H}'+{\cal H}^2)\Phi+{\cal H}\Phi'+\Psi''+2{\cal
 H}\Psi']+\triangle(\Phi-\Psi), \eea
 here     ${\cal H}=a'/a$
 in the case of the zeroth vector and tensor components
 \be\label{1.4.4}\nonumber N_k=0,~~~h_{ij}=0. \ee
One can see that
\begin{enumerate}
\item The $\Lambda$CDM Model omits the decomposition of the
potential energy
\be\label{li-1}\sum\limits_{I=0,4,6,8,12}^{}a^{I/2-2}(1-\Psi/2)^I{\cal
T}_I,~~{\cal T}_I=\langle{\cal T}_I\rangle+\overline{{\cal
T}}_I\ee with respect $\Psi$ that leads to the effective mass
terms in the Hamiltonian linear equations (\ref{2-19}) (this
effective mass is absent in Eq. (\ref{1.4.2})).
 \item The
$\Lambda$CDM Model chooses the gauge $N_k=0$ instead of the Dirac
minimal surface $p_{\psi}=0, N_k\not =0$ consistent with the
vacuum postulate in the Hamiltonian approach.
 \item The action
principle of GR in \cite{MFB} (see the second formula in Eq.
(10.7) p. 261) contains the double counting of the zeroth
Fourier-harmonics of the spatial metrics determinant presented by
two variables: $a$ and $\int d^3x\Psi(\eta,x_i)\not =0$ instead of
one. In other words, the $\Lambda$CDM Model uses doubling of the
zeroth Fourier harmonic of the scalar metric component
$\widetilde{\psi}=1-\Psi/2$, $\int d^3x\Psi \not =0$ in the action
\cite{MFB}, that destroys the the Hamiltonian approach. 
Nevertheless,   the linear equations (\ref{1.4.2}) in  the
$\Lambda$CDM Model  satisfy the opposite conditions $\int d^3x\Psi
=0$ and $\int d^3x\Phi =0$? This means that the description of the
``primordial power spectrum'' by  the inflationary model is
contradictable.
\end{enumerate}
 If we impose the constraints
 \be\label{1}
 \int d^3x p_\psi(\eta,x_i)=0,~~~~\int d^3x\log \widetilde{\psi}=0
 \ee
 in order to remove the ``double counting'',
 we  shall return back to the Einstein theory, where
 the equations of $\Psi$ and $\Phi$  will  not contain
 the time derivatives
 that are responsible for the ``primordial power spectrum'' in
  the inflationary model.

 In the contrast to standard cosmological perturbation theory
  \cite{lif,MFB}
   the diffeo-invariant version of the perturbation theory
 do not contain time derivatives that are responsible for
the CMB ``primordial power spectrum'' in the $\Lambda$CDM Model
\cite{linde}. However, the diffeo-invariant version of the Dirac
Hamiltonian approach to GR gives another possibility to explain
the CMB radiation spectrum and other topical problems of cosmology
by cosmological creation of the vector bosons in the Standard
Model \cite{bpzz}.

\newpage
\renewcommand{\theequation}{3.\arabic{equation}}
\section{Unification of GR and SM} \setcounter{equation}{0}
\subsection{The Unification}

The action of the SM in the electroweak sector, with presence of the
conformally coupled Higgs field can be write in the form
 \be\label{2-sm}
 S_{\rm SM}=
\int d^4x\sqrt{-g}
 \left[\frac{\phi^2}{6}{}^{(4)}R(g)+{\cal L}_{\rm Inv}
 +{\cal L}_{\rm Higgs}
 \right],
 \ee
that differs from  (\ref{1-sm}) by the curvature term.

 The acceptable
 unification of the General Relativity and the Standard Model
 is considered as the direct algebraical sum of GR (\ref{1-1})
 and SM (\ref{1-sm}) actions \be\label{t-1}
 S_{\rm GR\&SM}=S_{\rm GR}+S_{\rm SM}.
 \ee
 in the Riemannian manifold.

\subsection{The Newton's law in the GR\&SM theory}

The General Relativity and the Standard Model  reflect almost all
physical effects and phenomena
 revealed by measurements and observations,
 however, it does not means that the direct sum
 of the actions of GR an SM lies in agreement with all these effects and
 phenomena.
 One can see that the conformal coupling Higgs field $\phi$
 with conformal weight $n=-1$
 distorts the Newton coupling constant in the Hilbert
  action (\ref{1-1})
 \be\label{22-1}
 S_{\rm GR+Higgs}=\int d^4x\sqrt{-g}
 \left[-\left(1\!-\!\frac{\phi^2}{\vh_0^2}\right)\frac{\vh_0^2}{6}R(g)
 +g^{\mu\nu}\partial_\mu\phi\partial_\nu\phi
 \right]
 \ee
 due to
 the additional curvature term in the Higgs Lagrangian (\ref{22-1})
 $1\!-\!{\phi^2}/\vh_0^2$.
 This distortion
 changes
 the Einstein equations
 and their standard solutions
  of the Schwarzschild type and other \cite{B74,PPG,PS}.

   The coefficient $1\!-\!{\phi^2}/{\vh_0^2}$
 restricts  region, where the Higgs   field   is given,
  by the
 condition $\phi^2 < {\vh_0^2}$,
  because in other region $\phi^2 > {\vh_0^2}$
  the sign before the 4-dimensional curvature is changed in the
  Hilbert action  (\ref{1-1}).

In order to keep the Einstein theory  (\ref{1-1}),  one needs to
consider only the field configuration such that $\phi^2 <
{\vh_0^2}$.
  For this case one can introduce  new  variables by the Bekenstein--Wagoner transformation \cite{B74}
\bea\label{9-h11}
 g_{\mu\nu}&=&g_{\mu\nu}^{\rm (B)}\cosh^2Q\simeq g_{\mu\nu}^{\rm (B)},
  \\\label{9-h6}
 \phi^2&=&\vh_0^2\sinh^2Q\simeq\vh_0^2 Q^2, \\\label{s9-h6}
 s_{\rm (B)}&=&(\cosh{Q})^{-3/2} s
  \eea
 considered in \cite{PPG,PS}.
 These
 variables restore the initial Einstein--Hilbert action  (\ref{22-1})
 with the standard Newton law in the following way
 \be\label{22-2}
 S_{\rm GR+Scalar}=\vh_0^2\int d^4x\sqrt{-g_{\rm (B)}}
 \left[-\frac{R(g_{\rm (B)})}{6}+g_{\rm (B)}^{\mu\nu}
 \partial_\mu Q \partial_\nu Q\right].
 \ee
 Now it is clear that \emph{the  Bekenstein--Wagoner (BW)
 transformation converts the ''conformal  coupling'' Higgs field
 with the  weight $n=-1$ into
 the  ''minimal coupling''} scalar field $Q$ - an angle
 of the scalar -- scale mixing
 that looks like a {\it scalar graviton} with the conformal weight
 $n=0$.

 The Planck mass became one more  parameter
  of the Higgs Lagrangian,  so that
    the lowest order of the Lagrangian
    after  the separation of the zeroth Fourier harmonic $\langle Q\rangle$
 \be\label{hig-1}
 Q=\langle Q\rangle+\frac{h}{\vh_0\sqrt{2}}
 \ee
 over small  $\langle Q\rangle\ll 1$
   reproduces  the acceptable Standard Model action  (\ref{0-Ma})
\begin{eqnarray}\nonumber
\label{t-2small} {\mathcal{L}^{\rm \lambda}_{\rm{Higgs}}}(\langle
Q\rangle)&=&\vh_0^2g_{\rm (B)}^{00}
 \partial_0 \langle Q\rangle \partial_0 \langle Q\rangle-\langle Q\rangle\vh_0\sum_s f_s \bar s_{\rm (B)} s_{\rm
(B)} +\\\nonumber&+& \frac{\langle Q\rangle^2\vh_0^2}{4}\sum_{\rm v}
g^2_{\rm v}V^2-4{\lambda}  \langle Q\rangle^2\vh^2_0h^2.
\end{eqnarray}

\subsection{The GR\&SM   cosmology}

\subsubsection{Diffeo-invariant cosmological dynamics} Finally, we got the unified GR\&SM  theory
 \be\label{3-sm}
 S_{\rm GR\&SM}
 =\int d^4x\sqrt{-g_{\rm (B)}}
 \left[-\vh_0^2\frac{R(g_{\rm (B)})}{6}+{\cal L}_{\rm Inv}(F)
 +{{\mathcal{L}_{\rm{Higgs}}}}\right],
 \ee
where $F=g_{\rm (B)},W,Z,s$ and Lagrangians are given by Eqs.
(\ref{M}), (\ref{Ma}), (\ref{1-6a}),  (\ref{w1-6a}), and
\begin{eqnarray}
\label{3t-4} {{\mathcal{L}_{\rm{Higgs}}}}=\vh_0^2\partial_\mu
Q\partial_\nu Qg_{\rm (B)}^{\mu\nu}
  -\Phi\!{\sum_s f_s\bar ss}
+\frac{\Phi^2 }{4}\sum_{\rm v}g^2_{\rm v}V^2,
\end{eqnarray}
where $\Phi=\Phi(Q)=\varphi_0\sinh{Q}$. This Lagrangian is depend on
a one dimensional parameter $\vh_0$ only, that is given by Eq.
(\ref{1-2}).

The next step is to clear up the cosmological consequences of the
unified theory.
 The simplest way to made this step is the
 extraction of the cosmological scale factor $a(x^0)$
 by  scale transformations of all field variables
  obtained by the WB
 transformation  (\ref{9-h11}), (\ref{9-h6}), and (\ref{s9-h6})
 \bea
 g_{\rm (B)}&=&a^2\widetilde{g},\\
 W,Z&=&\widetilde{W},\widetilde{Z},\\
 s_{\rm (B)}&=&a^{-3/2}\widetilde{s},\\
 \vh&=&a\vh_0.
 \eea
 In particular,
 the
   curvature
 $\sqrt{-g_{\rm (B)}}\,\,{}^{(4)}R(g_{\rm
(B)})=a^2\sqrt{-{\widetilde{g}}}\,\,{}^{(4)}R({\widetilde{g}})-6a
 \partial_0\left[{\partial_0a}\sqrt{-{\widetilde{g}}}~
 {\widetilde{g}}^{00}\right]$
  can be expressed in terms of
   the new lapse
 function
 \be \label{lfsda}
 {\widetilde{N}}_d=[\sqrt{-{\widetilde{g}}}~{\widetilde{g}}^{00}]^{-1}
 \ee
and spatial metric determinant $|{\widetilde{g}^{(3)}}|$. In this
case, one can repeat the diffeo-invariant Hamiltonian formulation of
the GR presented in the previous Section 2 \cite{242,242a}, where
$\log {a}$ is identified with a zeroth mode as
 the  spatial volume ``averaging''
 \be\label{2non1}
 \log {a}=\frac{1}{6 V_0}\int
 d^3x\log|{{g}_{\rm (B)}^{(3)}}|\equiv\frac{1}{6}
 \langle \log|{{g}_{\rm (B)}^{(3)}}|\rangle,
\ee here
 the finite Lichnerowicz \cite{lich} diffeo-invariant volume $V_0=\int d^3x$ is
 introduced\footnote{One should emphasize
that modern cosmological models
 \cite{lif}
 are considered in the finite space and ``finite time-interval''
 in a reference frame
 identified with the frame of the Cosmic Background Microwave (CMB) radiation.}.
 In this case,
 \be\label{3non1}
 \log|{\widetilde{g}^{(3)}}|=\log|{{g}_{\rm (B)}^{(3)}}|-6\log {a}\ee
 is identified with
  the nonzero Fourier harmonics that satisfy  the  constraint
 \be\label{4non1}
 \langle\log|{\widetilde{g}^{(3)}}|\rangle \equiv 0.
\ee

 A scalar field can be also presented as a sum of a zeroth Fourier
  harmonics and nonzero ones
 \bea\label{z-s1}
 Q= \langle Q\rangle+\overline{Q}; ~~~~
 \langle\overline{Q}\rangle=0.
 \eea
 Finally, the action (\ref{3-sm}) takes the form of the sum of
 nonzero and zeroth-mode-contributions
\be \label{6-6}
 S_{\rm GR\&SM}[\vh_0|F,Q]= S_{\rm GR\&SM}[\vh|\widetilde{F},\overline{Q}]+
 S_{\rm zm}[\vh|\overline{Q}];
 \ee
here the first action
 repeats action $S_{\rm GR\&SM}[\vh_0|F,Q]$ (\ref{3-sm}), where $[\vh_0|F,Q]$
 are replaced by
 $[\vh|\widetilde{F},\overline{Q}]$, and the second
\be \label{zm-1}
  S_{\rm zm}[\vh|\overline{Q}]\Big|_{N_0\not =1}=
 \underbrace{V_0\!\int\! dx^0  \!
 \frac{1}{{N}_0}\left[ \vh^2\left(\frac{d \langle
 Q\rangle}{dx^0}\right)^2-\left(\frac{d
 \vh}{dx^0}\right)^2\right]}_{zeroth-mode~contribution}\equiv\int\! dx^0
 L_{\rm zm}
 \ee
is the action of the zeroth modes  $\vh,\langle Q\rangle$; here
 \be \label{6-5}
 \frac{1}{N_{0}}=\frac{1}{V_{0}}\int\frac{{d^3x}}{\widetilde{N}_{d}}\equiv
 \left\langle \frac{1}{\widetilde{N}_{d}}\right\rangle
 \ee
 is the homogeneous component of the lapse function.
The action of the local variables in  (\ref{6-6}) determines
 the correspondent   densities for the local variables
 \bea \label{6-9e}\widetilde{T}_{\rm d}&=&-\frac{\delta
 S_{\rm GR\&SM}[\vh_0|{\widetilde{F}},{\overline{Q}}]} {\delta
 \widetilde{N}_{\rm d}},\\\nonumber\widetilde{T}_{\psi}-
 \langle\widetilde{T}_{\psi}\rangle&=&-\widetilde{\psi}\,\frac{\delta
 S_{\rm GR\&SM}[\vh_0|{\widetilde{F}},{\overline{Q}}]} {\delta
 \widetilde{\psi}}=0,
 \eea
 where $\widetilde{\psi}=\sqrt[6]{|\widetilde{g}^{(3)}|}$ is
 the Dirac notation of the spatial metric determinant \cite{dir}.

The action (\ref{6-6}) coincides with the action of the relativistic
mechanics, where the dimension cosmological scale factor plays the
role of the external evolution parameter in the  field ``space of
events'' $[\vh|{\widetilde{F}},{\overline{Q}},\langle{Q}\rangle]$,
where $\vh$ is the time-like variable in this ``space'', and
${\widetilde{F}},{\overline{Q}},\langle{Q}\rangle$ are the
space-like ones.

 The action principle for the $S[\vh_0|{F},{Q}]$
 with respect to the lapse function $\widetilde{N}_{\rm d}$
 gives  the energy constraints equation
 \be\label{6-7}
\frac{1}{\mathcal{N}^2(\zeta,x^k)}
\left[\left(\frac{d\vh(\zeta)}{d\zeta}\right)^2-\vh^2(\zeta)
\left(\frac{d{\langle
Q\rangle}(\zeta)}{d\zeta}\right)^2\right]-\widetilde{T}_{\rm
d}(\zeta,x^k)=0,
 \ee
 where $\widetilde{T}_{\rm d}$ is given by Eq. (\ref{6-9e}) and
 \be \label{6-9}
 \zeta=\int{dx^0}N_{0}
 \ee
  is the ``diffeo-invariant homogeneous time-interval'' with its
derivative  and  \be \label{6-8}
 \mathcal{N}(\zeta,x^k)={\widetilde{N}_{\rm
 d}(\zeta,x^k)}{\langle{\widetilde{N}_{\rm d}^{-1}}\rangle(\zeta)},\ee
 is difeo-invariant part of the local lapse function with the
 unit constraint
 \be \label{6-8a}
 \langle\mathcal{N}^{-1}\rangle\equiv\frac{1}{V_0}\int d^3x\mathcal{N}^{-1}=1
\ee
 following from the definition of
  the homogeneous component of the lapse function $N_0$ given by Eq.
  (\ref{6-5}).
 This equation is the algebraic one with respect to
 the diffeo-invariant lapse function ${\cal N}$ and has solution
 satisfying the constraint  (\ref{6-8})
 \be\label{6-10}
 {\cal N}=
 {\langle\widetilde{T}_{\rm
 d}^{1/2}\rangle}{\widetilde{T}_{\rm d}^{-1/2}}.
 \ee
 The substitution of this solution into the energy constraint
 (\ref{6-7}) leads to the cosmological type equation
\be\label{6-11}
 \vh'^2=\rho_{\rm tot}(\vh)=\rho_{\rm loc}(\vh)+\rho_{\rm zm}(\vh)
 ;
 \ee
 here
 the total energy density $\rho_{\rm tot}(\vh)$ is
 split on the sum of  the energy density of local fields (loc)
  and the zeroth mode (zm) one defined as
 \be\label{6-12}
 \rho_{\rm loc}(\vh)={\langle(\widetilde{T}_{\rm
 d})^{1/2} \rangle}^2,~~~~~~
 \rho_{\rm zm}(\vh)=\vh^2 {\langle Q\rangle'}^2=\frac{P_{\langle
 Q\rangle}^2}{4V_0^2\vh^2}
 \ee
 where
\be\label{6-12a}P_{\langle Q\rangle}=\frac{\partial L_{\rm
zm}}{\partial(\partial_0 {\langle Q\rangle})}=
2V^0\vh^2\frac{d{\langle Q\rangle}}{d\zeta}\equiv2V^0{\vh^2\langle
Q\rangle'}
 \ee
  is the scalar field zeroth mode momentum that is
an integral of motion of the considered model because
 the action does not depend on $\langle Q\rangle$.
  The constraint-shell value
 of the momentum of  external time $\vh$
 \be\label{6-13} P_\vh=\frac{\partial L_{\rm
zm}}{\partial(\partial_0 \vh)}=2V_0\vh'= \pm 2V_0\sqrt{\rho_{\rm
tot}(\vh)}\equiv\mp E_\vh\ee
   can be considered as the Hamiltonian generator of evolution
   of all field variables with respect to $\vh$ in the
   ``space of events'' $[\vh|\widetilde{F},\overline{Q},\langle Q\rangle]$.
 The value of  the momentum $P_\vh=\pm E_\vh$ onto
  solutions of the motion equations
 can be  considered as  an ``energy of the universe'', in accord with the
 analogy with relativistic mechanics.
 We can see also that a solution of Eq. (\ref{6-13}) with respect
 to diffeo-invariant time-interval $\zeta$ (\ref{6-9})
 \be\label{6-14} \zeta_{\pm}=  \pm \int\limits_{\vh_I}^{\vh} \frac{d\widetilde{\vh}}{
 \sqrt{{\rho_{\rm tot}(\widetilde{\vh})}}}
 \ee
 is the Hubble law in the exact theory, that includes the initial
  datum $\vh_I=\vh(\zeta=0)$.

\subsubsection{Zeroth mode sector of GR\&SM theory as a ``cosmological model''}

Let us consider solutions of
 Eqs. (\ref{6-11}), (\ref{6-12}),  (\ref{6-13}), and  (\ref{6-14}) in the case
 of the
 zeroth mode sector in the action (\ref{6-7}), i.e. for $\rho_{\rm
 loc}=0$.
 The zeroth mode sector $[\vh|\langle Q\rangle]$ in the action (\ref{6-7})
 \be \label{zm-1a}
  S_{\rm zm}=
 \underbrace{V_0\!\int\! dx^0  \!
 \frac{1}{{N}_0}\left[ \vh^2\left(\frac{d \langle
 Q\rangle}{dx^0}\right)^2-\left(\frac{d
 \vh}{dx^0}\right)^2\right]}_{zeroth-mode~contribution}
 \ee
 is most important at the beginning of the Universe, when
 all particle like excitations are absent. One can say that at the beginning
 there were only the  ``beginning data''
 $\vh_I,\vh'_I,\langle Q\rangle_I, \langle Q\rangle'_I,$.
 \bea\label{zm-2}
 ds^2&=& ds^2_{\rm WDW}=a^2(x^0)[(N_0(x^0)dx^0)^2 -(dx^idx^i)],\\\label{zm-3}
  a&=&\vh/\vh_0,\\\label{zm-4}
  \eta&=&\int dx^0 N_0(x^0).
 \eea
The conformal vacuum Higgs effect considered in Section 1,
 in the  cosmological
 approximation, is described by the action
 \be \label{zm-1ab}
  S_{\rm vac}=
 V_0\!\int\! dx^0  \!
 \left[ \frac{\vh^2}{{N}_0}\left(\frac{d \langle
 Q\rangle}{dx^0}\right)^2-\frac{1}{{N}_0}\left(\frac{d
 \vh}{dx^0}\right)^2-{N}_0\textsf{V}^{\rm
 conf}_{eff}\right],
 \ee
 where $\textsf{V}^{\rm conf}_{eff}$
 is the Coleman--Weinberg effective potential and is given by the
 formula (\ref{eff-1}). These action and interval keep the symmetry
  with respect to
  reparametrizations of the  coordinate evolution
 parameter $x^0~\to~\overline{x}^0=\overline{x}^0(x^0)$.
 Therefore, the cosmological model  (\ref{zm-1ab}) can be considered
 by analogy with
 a model of a relativistic particle in the Special Relativity (SR)
 including the Hamiltonian approach to this theory.
 The canonical conjugate momenta of the theory (\ref{zm-1ab}) are
 \be\label{pp-1} P_\vh=2\vh'V_0,
~~~P_{\langle{Q}\rangle}=2\vh^2\langle{Q}\rangle'V_0\ee where
$f'=\dfrac{df}{d\eta}$. The Hamiltonian  action has a form
\be\label{hama}
 S_{\rm vac}=\int
 dx^0\left[P_{\langle Q\rangle}\frac{d \langle
 Q\rangle}{dx^0}-P_\vh \frac{d\vh}{dx^0}-\frac{N_0}{4V_0}\left(-P^2_\vh+E^2_\varphi\right)\right],
 \ee
 where
 \be\label{ev1}
 E_\varphi=2V_0\left[\frac{P_{\langle{Q}\rangle}^2}{4V_0^2\vh^2}+\textsf{V}^{\rm
 conf}_{eff}\right]^{1/2}
 \ee
 is treated as the {\it"energy of a universe"}.

 The classical energy constraint in the model (\ref{hama}) is
 \bea\label{2.3-11}
 P_\vh^2-E_\vh^2=0
\eea and repeat completely the cosmological equations
  in the case of
 the rigid state equation $\Omega_{\rm rigid}=1$ because
 due to the unit vacuum-vacuum transition amplitude $V^{\rm
 conf}_{eff}(\vh \langle\,Q_I\rangle)=0$
\bea\label{2.3-12}
 \vh_0^2 a'^2=\frac{P_{Q_I}^2}{4V_0^2\vh^2}
 =H_0^2 \frac{\Omega_{\rm rigid}}{a^2},
 \eea
  where $P_{Q_I}$ is a constant of the motion
 \be\label{2.3-14a}
  P_{Q_I}'=0,
  \ee
 because in the equation of motion
\be \label{2.3-14} P\,'_{\langle{Q}\rangle}+\frac{d\textsf{V}^{\rm
 conf}_{eff}(\vh \langle{Q}\rangle)}{d\langle{Q}\rangle}=0, \ee
 the last term is equal zeroth
  $\dfrac{d\textsf{V}^{\rm
 conf}_{eff}(\vh \langle{Q}\rangle)}{d\langle{Q}\rangle}=0$
if all masses satisfy the Gell-Mann--Oakes--Renner type relation
(\ref{31-t}).

 The solution of these equations take the form
\bea\label{2.3-16}
 \vh(\eta)=\vh_I\sqrt{1+2{\cal H}_I\eta}, ~~~~
 \langle{Q}\rangle(\eta)=Q_I+\log {\sqrt{1+2{\cal H}_I\eta}},
\eea
 where
\bea\label{2.3-17}
 \vh_I&=&\vh(\eta=0),~~~~~{\cal
 H}_I\equiv\frac{\vh'(\eta=0)}{\vh(\eta=0)}=
 \frac{P_{\langle{Q}\rangle}}{2V_0\vh_I^2},\\\label{2.3-18}
  Q_I&=&{\langle{Q}\rangle}(\eta=0),~~~~P_{\langle{Q}\rangle}={\rm const}
\eea
 are the ordinary ``free'' initial data of the equation of
 the motion.
 Besides of the Higgs field $Q_H$ can be one more
  (massless) scalar field $Q_A$ (of the type of axion (A)).
 In this case
\bea\label{2.3-16a}
 \vh(\eta)&=&\vh_I\sqrt{1+2{\cal H}_I\eta}, \\
  Q_A(\eta)&=&Q_{AI}+\frac{{\cal H}_A}{2{\cal H}_I}\log {(1+2{\cal
  H}_I\eta)},\\
Q_H(\eta)&=&Q_{HI}+\frac{{\cal H}_H}{2{\cal H}_I}\log {(1+2{\cal
  H}_I\eta)},
\eea where $\vh_I,{\cal H}_I=\sqrt{{\cal H}_A^2+{\cal H}_H^2}$ and
$Q_{AI},{\cal H}_A=
 \dfrac{P_A}{2V_0\vh_I^2}$ and $Q_{HI},{\cal H}_H=
 \dfrac{P_H}{2V_0\vh_I^2}$ are free initial data in the CMB frame of
 reference, in the contrast to the Inflationary model, where
 $\vh_I={\cal H}(\eta=\eta_0)$. One can see that this main hypothesis
 of the Inflationary model contradicts to
 the diffeo-invariant constraint-shell dynamics of the GR, in
 particular the cosmological model (\ref{hama}), where
 the  constraint-shell Hamiltonian  action takes a form
\be\label{ham1}
 S^{\rm constraint~shell}_{\rm vac}=\int\limits_{\vh_I}^{\vh_0}
 d\vh\left[P_{\langle Q\rangle}\frac{d \langle
 Q\rangle}{d\vh}\mp E_\vh
 \right].
 \ee
 As it was shown \cite{242,242a,114:a,gpk} there are initial data of
quantum creation of matter at $z_I+1= 10^{15}/3$, and a value of the
Higgs-metric mixing ``angle'' $Q_0\simeq3 \times 10^{-17}$ is in
agreement with the present-day energy budget of the Universe.

\subsubsection{Quantum universes versus classical ones}
 The standard  pathway from  SR to QFT of particles
  shows us the similar  pathway
 to ``QFT'' of universes \cite{WDW,Bog} that  include
 the following steps.
\begin{enumerate}
\item The Hamiltonian approach: (\ref{hama}),
\item Resolution of the energy constraint:
$P^2_{\varphi}-E^2_\varphi=0$ with respect to $P_\vh=\pm E_\vh$
(\ref{6-13}),
\item Reduction as  substitution of these solutions
 into action (\ref{hama})  gives us the ``reduced action''
\be\label{ham-2}
 S=\int\limits_{\vh_I}^{\vh_0}d\widetilde{\vh}\left[P_{\langle
Q\rangle}\frac{d \langle
 Q\rangle}{d\widetilde{\vh}}\mp E_\vh
 \right]\ee
and the time-interval ($\eta$) -- time-variable ($\vh$) relation
(\ref{6-14})
  \be\label{6-14zm} \eta_{\pm}=  \pm \int\limits_{\vh_I}^{\vh}
\frac{d\widetilde{\vh}}{
 \sqrt{{\rho_{\rm zm}(\widetilde{\vh})}}}=V_0
 \left|\frac{\vh^2_0-\vh^2_I}{P_{\langle
Q\rangle}}\right|
 \ee
that is treated  in cosmology as the Hubble law (and in SR, as the
Lorentz transformation),
\item Primary quantization of the energy constraint:
$[\hat{P}^2_{\varphi}-E^2_\varphi]\Psi=0$, here
$\hat{P}_{\varphi}=-i\dfrac{d}{d\varphi}$,
\item Secondary quantization  of the energy constraint:
 $\Psi=\dfrac{A^++A^-}{\sqrt{2E_\vh}}$,
\item The Bogoliubov transformation:
 $ A^+=\alpha
 B^+\!+\!\beta^*B^-$,
\item The Bogoliubov vacuum : $B^-|0>_{U}=0$, and
\item Cosmological creation of $N_U={}_{U}<0|A^+A^-|0>_{U}$
 {\it universes} from the Bogoliubov vacuum $|0>_{U}$ at $\eta=0$
 (see in detail Appendices A and B \cite{Bog}).
\end{enumerate}
 The arrow of the time-interval $\eta \geq 0$ arises at the step 5.)
 of the decomposition of the wave function onto the sum of
  the creation operator of a universe going
  forward $\vh \geq \vh_I$ with positive energy $P_\vh\geq 0$,
   and
  the annihilation operator of a universe going backward $\vh \leq \vh_I$
  with  energy $P_\vh\leq 0$.
 This ``eightfold pathway'' shows us that
 two quantizations 4.) and 5.) are needed, in order to remove
 the negative energy and provide
 the stable system \cite{Bog}.

The eight principles  are the basis of
 the fundamental
operator quantization as the result of the QFT experience in the
twentieth century.  The main principle providing   this quantization
is
  the ``coordinate time reparametrization symmetry of the action''
 leading to the concepts of energy constraint, space of events,
   time-event -- time-interval relation,
particle, quasiparticle, vacuum, and quantum creation from vacuum as
a QFT mathematical model of ``Big-Bang'' considered above. In the
model of rigid state, where  $E_\vh=P_{\langle Q\rangle}/\vh$,
  we  have an exact solution  for
 number of created universes (see Eqs. (\ref{usv3}))
\be\label{11cu1}
 N_{\rm U}=\frac{1}{4P_{\langle Q\rangle}^2-1}
 \sin^2\left[\sqrt{P_{\langle Q\rangle}^2-\frac{1}{4}}~~
 \ln\frac{\vh}{\vh_I}\right]\not
 =0,
\ee
 where
 \be\label{cc1}
 \vh=\vh_I\sqrt{1+2H_I\eta}
 \ee
  and
$\vh_I,H_I=\vh'_I/\vh_I=P_{\langle Q\rangle}/(2V_0\vh_I^2)$ are the
initial data.

There is another version of  GR
 accepted in $\Lambda$CDM cosmological perturbation theory \cite{MFB,M}.
 For simplicity, one can compare this version using as example
 the zeroth mode sector (\ref{zm-1}).
 Instead of  the SR type  theory (\ref{zm-1})
  in \cite{MFB,M} one uses its version
  obtained by the substitution of  the $N_0=1$,  $x^0=\eta$ gauge into
  the action (\ref{zm-1})
 \be \label{zm-11}
  S_{\rm zm}[\vh|\overline{Q}]\Big|_{N_0 =1}=
 \underbrace{V_0\!\int\! dx^0  \!
 \left[ \vh^2\left(\frac{d \langle
 Q\rangle}{dx^0}\right)^2-\left(\frac{d
 \vh}{dx^0}\right)^2\right]}_{zeroth-mode~contribution}\Big|_{x^0=\eta}.
 \ee
 In this case,  the reparametrization
 symmetry is postulated on the level of
 classical equations \cite{bardeen} so that  the measurable
 conformal time becomes an object
 of  reparametrizations
  $\eta \to \widetilde{\eta}=\widetilde{\eta}(\eta)$  in the contrast
  to the Dirac
 definition of measurable quantities as  diffeo-invariants.
 The quantum  theory  (\ref{zm-11}) does not contain a vacuum
 as a state with the minimal energy
 because the corresponding Hamiltonian
 is not  restricted from bottom; therefore, this theory is not stable
 in contrast to the initial theory (\ref{zm-1}), where the primary
 and secondary quantizations determine the vacuum as state with
 the minimal energy of the constraint-shell Universe motion
 in its space of events.

 However, very the problem of unification GR
  with SM based on the fundamental
 quantization  of relativistic QFT supposes the consideration of SM
 and GR on equal footing. Therefore, instead of  the theory (\ref{zm-11})
 without the vacuum postulate
  accepted in $\Lambda$CDM model \cite{MFB},
 we shall consider the theory (\ref{zm-1}) with vacuum postulate
   accepted in SR and GR. The theory (\ref{zm-1}) has
 a  particular  quantum solution (\ref{11cu1}),  (\ref{cc1})
   predicting an inevitable
  vacuum creation
  of a number of ``universes''
    determined by the ``free'' initial
  data $\vh_I,{\cal H}_I$, and $\langle Q\rangle_I=Q_0$
  including $P'_{\langle Q\rangle}=0$ (\ref{2.3-14}) at the moment $\eta=0$.

In this case,
 the corresponding equation of motion in SM
 admits an arbitrary value
 of the initial datum $\langle Q\rangle_I=Q_0$.
These initial data
 determine the mass spectrum of the SM particles
 (vector, bosons and fermion) in  the SM Lagrangian.

\subsection{Hamiltonian GR\&SM}

\subsubsection{GR\&SM theory in the $3L+1G$ Hamiltonian approach}
 As we have seen above in Subsection 1.5, that the diffeo-invariant
  $3L+1G$ version
 \bea \label{3L+1G-1}ds^2&\equiv&\omega_{(\alpha)}\omega_{(\alpha)}=
 a^2\widetilde{\omega}_{(\alpha)}\widetilde{\omega}_{(\alpha)}=
 a^2\widetilde{\psi}^4\,\,\omega^{(L)}_{(\alpha)}\omega^{(L)}_{(\alpha)},
 \\\label{3L+1G-4}\omega^{(L)}_{(0)}&=&\widetilde{\psi}^4{\cal N}d\zeta,
 \\\label{3L+1G-5}
 \omega^{(L)}_{(b)}&=& {\bf e}_{(b)i}dx^i+
 {\cal N}_{(b)} d\zeta,
\\\label{3L+1G-7}
 -\frac{\delta S}{\delta \widetilde{N}_{\rm
 d}}&=&0~~~ \Rightarrow~~~{\cal N}=
 {\langle\widetilde{T}_{\rm
 d}^{1/2}\rangle}{\widetilde{T}_{\rm d}^{-1/2}},\\\label{3L+1G-8}
 \widetilde{T}_{\rm d}&=&\dfrac{4\vh_0^2a^2}{3}{\widetilde{\psi}}^{7}
 \triangle
{\widetilde{\psi}}+
  \sum\limits_{I=0,4,6,8,12} a^{I/2-2}{\widetilde{\psi}}^I{\cal T}_I,
 \\\label{3L+1G-3}
 -\widetilde{\psi}\frac{\delta S}{\delta \widetilde{\psi}}&=&0
 ~~~ \Rightarrow~~~
 \widetilde{T}_{\psi}-\langle \widetilde{T}_{\psi}\rangle=0,\\\label{3L+1G-8d}
   \widetilde{T}_{\psi}&=&\dfrac{4\vh_0^2a^2}{3}
   \left\{7{\cal N}{\widetilde{\psi}}^{7}
  \triangle {\widetilde{\psi}}+{\widetilde{\psi}} \triangle
\left[{\cal N}{\widetilde{\psi}}^{7}\right]\right\}+\\\nonumber &+&
  {\cal N}\sum\limits_{I=0,4,6,8,12}I a^{I/2-2}{\widetilde{\psi}}^I{\cal T}_I
 \eea
  is more adequate to finite volume
  and finite time of the cosmological dynamics of the
  Universe as the whole, with the Hubble law (\ref{6-14})
   than the
  Dirac -- ADM $4L$ version
\bea
 \label{4L-4}\omega_{(0)}&=&\psi^6N_{\rm d}dx^0,
 \\\label{4L-5}
 \omega_{(b)}&=&\psi^2 {\bf e}_{(b)i}(dx^i+N^i dx^0),
 \\\label{4L-2}-\frac{\delta S}{\delta N_{\rm d}}&\equiv& T_{\rm d}=
 \dfrac{4\varphi_0^2}{3}{\psi}^{7}
 \triangle
{\psi}+
  \sum\limits_{I=0,4,6,8,12} {\psi}^I{\cal T}_I=0,
 \\\label{4L-3}-\psi\frac{\delta S}{\delta \psi}&\equiv& T_{\psi}=
  \dfrac{4\varphi_0^2}{3}\left\{7N_d{\psi}^{7}\triangle {\psi}+{\psi}
  \triangle
\left[N_d{\psi}^{7}\right]\right\}+\\\nonumber&+&
  N_d\sum\limits_{I=0,4,6,8,12}I {\psi}^I{\cal T}_I=0.
 \eea
  Moreover, the  diffeo-invariant
  $3L+1G$ Hamiltonian  approach can be considered as
  the finite volume generalization of
   the acceptable Dirac -- ADM one with
   $4L$ constraints. Both these approaches coincide
  in the infinite volume limit
  $a=1,\langle {\widetilde{T}_{\rm d}}^{1/2} \rangle \to 0$.
 However, the $4L$ version loses  reparametrization time
 symmetry principle
 and its direct consequences, such as the evolution parameter
$\vh_0a=\vh$ in the ``field space of events'' and the ``energy of
event''
 that arises in
the Hamiltonian  constraint-shell action (\ref{3-sm})
 \bea\label{h-21} S_{\rm GR\&SM}\big|_{{\rm constraint-shell}}\!\!\!&=&\!\!
 \int\! dx^0\!\int d^3x
 \sum\limits_{{F}={\psi},e,\,{Q}}P_{{F}}\partial_0F.
 \eea
  The kinemetric subgroup (\ref{1-8}) essentially
  simplifies the solution of the energy constraint (\ref{6-7}),
  if the homogeneous variable is extracted from the
  the determinant
  $\psi^2(x^0,x^k)=a(\zeta)\widetilde{\psi}^2(\zeta,x^k)$
  with the additional constraints
 \bea\label{3-20a}
 \int d^3x \log\widetilde{\psi}
 &\equiv& 0,~~~~~\\
 \frac{1}{V_0}\int d^3x\frac{1}{{\cal N}}&\equiv&\left\langle
 \frac{1}{{\cal N}}\right\rangle=1,\\
 |{\bf
 e}_{(b)i}|&=&1,~~\partial_k{\bf
 e}^k_{(b)}=0,\\
 ({\widetilde{\psi}}^6)'
 &=&\partial_{(b)}({\widetilde{\psi}}^6 {\cal N}_{(b)})
\\\label{3L+1G-2}
  \zeta_{\pm}=  &\pm& \int\limits_{\vh_I}^{\vh}
\frac{d\widetilde{\vh}}{
 \sqrt{{\langle(\widetilde{T}_{\rm
 d})^{1/2} \rangle^2}+P^2_{\langle Q\rangle}/(2V_0\vh_0 a)^2}}
 .
 \eea

 The action (\ref{h-21}) after the separation of
 the zeroth modes (\ref{2non1}) and (\ref{z-s1}) takes the form
\bea\label{h-22}\nonumber
  (3.69)\!\!\!&=&\!\!\!\int dx^0
 \left\{\left[\int\limits_{V_0}
d^3x\!\!\!\sum\limits_{\widetilde{F}=\widetilde{\psi},e,\,
\overline{Q}}
 P_{\widetilde{F}}\partial_0 \widetilde{F}\right]
 +P_{\langle Q\rangle}\frac{d\langle
 Q\rangle}{dx^0}
 -P_\vh\frac{d\vh}{dx^0}\right\}
\\\label{100}
\!\!\!&=&\!\!\!\int\limits_{\vh_I}^{\vh_0}d{\vh}\left\{\left[\int\limits_{V_0}
d^3x\!\!\!\sum\limits_{\widetilde{F}=\widetilde{\psi},e,\,
\overline{Q} }
 P_{\widetilde{F}}\partial_\vh \widetilde{F}\right]
 +P_{\langle Q\rangle}\frac{d\langle
 Q\rangle}{d\vh}
 \pm E_\vh\right\}.
 \eea
 where
 \bea\label{cs-1} P_\vh&=&\pm E_\vh
 =\pm2 V_0\sqrt{{\langle(\widetilde{T}_{\rm
 d})^{1/2} \rangle^2}+P^2_{\langle Q\rangle}/(2V_0\vh_0 a)^2}
 \eea
 is the constraint-shell Hamiltonian in the ``space of events'' given
 by the resolving the energy constraint (\ref{6-13}).

\subsection{Correspondence principle}

 The physical meaning of this constraint-shell Hamiltonian
 can be revealed
  in the limit
 of the tremendous contribution of
 the homogeneous energy density
 \bea\nonumber&&2V_0\langle(\widetilde{T}_{\rm
 d})^{1/2}\rangle^2\!=\!2V_0\langle(\rho_{\rm s}\!+\!{T}_{\rm
 sm})^{1/2}\rangle^2\!=\!2V_0\rho_{\rm s}\!+\!\int d^3x{T}_{\rm sm}
,\\\label{cs-4} &&2V_0\rho_{\rm s}\simeq 10^{79} \mbox{\rm GeV}\gg
 \int d^3x{T}_{\rm sm}\simeq 10^{2} \mbox{\rm GeV}.
  \eea
 In this case the constraint-shell Hamiltonian takes the form
 \bea\nonumber\label{cs-5} P_\vh&=&\pm E_\vh
 =\pm2 V_0\sqrt{{\langle(\widetilde{T}_{\rm
 d})^{1/2} \rangle^2}+P^2_{\langle Q\rangle}/(2V_0\vh_0 a)^2}=\\\label{cs-6}
&=&\pm\left[2V_0\sqrt{\rho_{\rm cr}}+\frac{1}{\sqrt{\rho_{\rm
cr}}}\int\limits_{V_0}d^3x T_{\rm sm}+...\right],\\\label{100-2z}
 \rho_{\rm cr} &=&\rho_{\rm s}+\rho_{\rm zm}, ~~~~
 ~~~~\rho_{\rm zm}=\frac{P^2_{\langle Q\rangle}}{(2V_0\vh_0 a)^2}.
 \eea
 Using the standard definition of the
 conformal time in cosmology $d\vh=d\eta\sqrt{\rho_{\rm cr}} $
 one can see that the constraint-shell action (\ref{100})
\bea\label{100-1} S&=&\int\limits_{\vh_I}^{\vh_0}d{\vh}\left\{
 +P_{\langle Q\rangle}\frac{d\langle
 Q\rangle}{d\vh}
 \pm 2V_0\sqrt{\rho_{\rm cr}}\right\}+\\\nonumber &+&
 \int\limits_{0}^{\eta_0}d\eta \left[\int\limits_{V_0}
d^3x\sum\limits_{\widetilde{F}=\widetilde{\psi},e,\, \overline{Q} }
 P_{\widetilde{F}}\partial_\vh \widetilde{F}\pm T_{\rm sm}\right].
 \eea
is the sum of action of homogeneous cosmology and the action of the
local field theory  with the SM Hamiltonian,
  where all masses
  are determined by the Higgs-metric ``angle''
 $\langle Q\rangle$ and the cosmological scale factor
 $a(\eta)$.

 The cosmological dynamics in the form of
 the Hubble law is
 \bea \label{3L+1G-3x}
  \eta_{\pm}&=&  \pm \vh_0\int\limits_{\vh_0a_I}^{\vh_0a}
 \frac{d\widetilde{a}}{\sqrt{\rho_{\rm cr}(\widetilde{a})}},
 \eea
is a one of consequences of the time reparametrization invariance
 principle. In the homogeneous approximation
 \bea\nonumber
 \label{100-2}
 \rho_{\rm cr} &=&\rho_{\rm s}+\rho_{\rm zm}=\\\label{100-3}
 &=&\rho_{0\rm cr}\sum\limits_{I=0,4,6,8,12}^{}\Omega_I
 a^{I/2-2}(\eta),\\\nonumber
 ~&&\sum\limits_{I=0,4,6,8,12}^{}\Omega_I=1,
 \eea
 where $\Omega_I$ is partial energy density
  marked by the index $I$ running a collection of values
   $I=0$ (rigid), 4 (radiation), 6 (mass), 8 (curvature), 12
   ($\Lambda$-term)
in accordance with a type of matter field contributions.

 The equation of $\langle Q\rangle$ take the standard form
 \be
 P'_{\langle Q\rangle}-\frac{d L}{d \langle Q\rangle}=0.
 \ee
In the case of  $\dfrac{d L}{d \langle Q\rangle}=0$,
 $P_{\langle Q\rangle}/(2V_0)=\langle Q\rangle' \vh^2_0 a^2$ is an
 integral of motion.
  Therefore
  \be\label{1-q}
 \langle Q\rangle(\eta)=Q_0+\frac{P_{\langle Q\rangle}}{2\vh^2_0V_0}\int\limits_{0}^{\eta}
 \frac{d\widetilde{\eta}}{a^2(\widetilde{\eta})}.
  \ee
 The accepted $\Lambda$CDM cosmological model arises in the case
 if $P_{\langle Q\rangle}\simeq 0$, when the rigid state is suppressed
 by the $\Lambda$ -term $\Omega_{I=12}$ in the total density
 (\ref{100-2})
\be
\label{gg-1}\Omega_{I=12}a^{4}(\eta)\gg\frac{\Omega_{I=0}}{a^{2}(\eta)}.
 \ee
 At the Planck epoch of the primordial inflation
 \be\label{gg-2}
 \vh_I=\vh_0 a_0\simeq H_0\simeq 10^{-61}\vh_0
 \ee
 this means that $\Lambda$ -term is greater than the rigid, if
\be \label{gg-1a}\Omega_{\Lambda (I=12)}\geq \frac{\Omega_{\rm
rigid(I=0)}}{a^{6}_I}=10^{366}\Omega_{\rm rigid(I=0)}.
 \ee
Here a question arises: What is a reason of this strong dominance of
the $\Lambda$-term, if its  contribution is suppressed  in
$10^{366}$ times in comparison with the rigid state?
\newpage
\renewcommand{\theequation}{4.\arabic{equation}}
\section{GR\&SM theory as a conformal brane}
 \setcounter{equation}{0}

\subsection{The Lichnerowicz variables and {\it relative units}
of the dilaton gravitation}

 One can say that
 the manifest dependence on the energy density $T_{\rm d}$
 on the spacial determinant $\psi$ in the expression (\ref{h-1})
 is equivalent to a choice the Lichnerowicz (L)-coordinates (\ref{1-12})
  $\omega^{(L)}_{(\mu)}$
 and L-variables
 (\ref{1-13}),  (\ref{1-14}) as observable ones.
 The L-observables are physically equivalent with the case
 when the field
  with the mass $m=m_0\psi^2$  is contained in
 space-time
 distinguished by the unit
 spatial metric determinant and the volume element
 \be \label{1-15}
 dV^{(L)}=\omega^{(L)}_{(1)}\wedge \omega^{(L)}_{(2)}\wedge
 \omega^{(L)}_{(3)} =d^3x.
 \ee
 In terms of the L-variables and L-coordinates $\vh_0\psi^2=w$
 the Hilbert action of classical theory of gravitation
  (\ref{1-1}) is formally the same as the action of
  the dilaton gravitation (DG) \cite{pct}
\bea\label{dg-1a}
  S_{DG}[\hat g^w]&=&-\int d^4x\frac{\sqrt{-\hat g^w}}{6}~R(\hat g^w)
  \equiv\\\nonumber&\equiv&
  -\int d^4x\left[\frac{\sqrt{- g}w^2}{6}~R( g)-w
  \partial_\mu(\sqrt{- g}\partial_\nu g^{\mu\nu})\right],
 \eea
 where $\hat g^w=w^2g$ and
   $w$ is the dilaton scalar field. This action is invariant with
 respect to the scale transformations
 \be\label{dg2a}
 {F^{(n)\Omega}}=\Omega^{n}F^{(n)}, ~~{g^{\Omega}}=\Omega^{2}g,
 ~~{w^{\Omega}}=\Omega^{-1}w.
 \ee
 One can see that there is a transformation \be\label{ct1a}
 \Omega=\dfrac{w}{\vh_0}\ee
 converting the dilaton action  (\ref{dg-1}) into the Hilbert one
   (\ref{1-1}).
In this manner, the CMB frame reveals the possibility to
 choose the units of measurements in the canonical  GR.
The dependence on the energy momentum tensors
    on the
   spatial determinant {\it potential} $\psi$ is completely determined by
   the Lichnerowicz (L) transformation to the
   conformal variables (\ref{1-12}), (\ref{1-13}), and (\ref{1-14}).
 The manifest dependence on the energy density $T_{\rm d}$
 on the spacial determinant $\psi$ in the expression (\ref{h-1})
 is equivalent to a choice the L-coordinates (\ref{1-12}) $\omega^{(L)}_{(\mu)}$
 and L-variables
 (\ref{1-13}),  (\ref{1-14}) as observable ones.
 The L-observables are physically equivalent with the case
 when the field
  with the mass $m=m_0\psi^2$  is contained in
 space-time
 distinguished by the unit
 spatial metric determinant and the volume element (\ref{1-15}).

 In terms of the L-variables and L-coordinates
 \be\label{21-15}
 \vh_0\psi^2=X_{(0)}
 \ee
 the Hilbert action of classical theory of gravitation
  (\ref{1-1}) is formally the same as the action of
  the dilaton gravitation (DG) \cite{pct}
\bea\label{dg-1}&&S_{DG}=\\\nonumber
  &-&\int d^4x\left[\frac{\sqrt{- g_{(L)}}X_{(0)}^2}{6}~R( g_{(L)})-X_{(0)}
  \partial_\mu(\sqrt{- g_{(L)}}\partial_\nu g_{(L)}^{\mu\nu}X_{(0)})\right],
 \eea
 where $g_{(L)\mu\nu}=|g^{(3)}|^{-1/3}g_{\mu\nu}$ and
   $X_{(0)}$ is the dilaton scalar field.
   This action supplemented by  conformally coupling scalar $\Phi=X_{(1)}$
   field takes the form of a relativistic brane
 \bea\label{brane-m}
&&S_{\mathrm{brane}}^{(D=4/N=2)}[X_{(0)},X_{(1)}]
\!=-\!\int\!d^4x\!\Bigg[\sqrt{-g}\!
\frac{X_{(0)}^2-X_{(1)}^2}{6}\,{}^{(4)}
\!R(g)-\nonumber\\&-&X_{(0)}\partial_\mu(\sqrt{-g}g^{\mu\nu}\partial_\nu{X}_{(0)})
+X_{(1)}\partial_\mu(\sqrt{-g}g^{\mu\nu}\partial_\nu{X}_{(1)})
\Bigg],\eea
 with two external ``coordinates''  defined as
 \be\label{ct-1}
 X_{(0)}=\vh_{0}\psi^2, ~~~~~~~~~~X_{(1)}=\Phi
 \ee
in accord with the standard definition of
 the general action for brane in $D/N$ dimensions
given in \cite{bn} by
 \bea S^{(D/N)}_{\mathrm{brane}}&=&-\int\!
d^Dx\!\sum_{A,B=1}^N\eta^{AB}\!\Bigg[\!\sqrt{-g}\frac{X_A
X_B}{(D-2)(D-1)}{}^{(D)}\!R(g)\!-\nonumber\\&&\!X_A\partial_\mu(\sqrt{-g}
 g^{\mu\nu} \!\partial_\nu X_B)\Bigg]. \label{braneDN}
 \eea
 In this case, for $D=4$,
 $N=2$ we have $\eta^{AB}=\mathrm{diag}\{1,-1\}$.
  The hidden
 conformal invariance of the theory  (\ref{brane-m}) admits to replace
   the Einstein definition of a measurable
   interval  (\ref{1-2}) in GR    by
  its conformal invariant version
 as a Weyl-type ratio
  \be \label{1-10a}
 ds_{\rm (L)}^2=\frac{ds^2}{ds_{\rm units}^2},
 \ee
 where  $ds_{\rm units}^2$ is an interval of the units
 that is defined like
 the Einstein definition of a measurable  interval  (\ref{1-2}) in GR.

   This action is invariant with
 respect to the scale transformations
 \be\label{dg2}
 {F^{(n)\Omega}}=\Omega^{n}F^{(n)}, ~~{g^{\Omega}}=\Omega^{2}g,
 ~~{X_{(0)}^{\Omega}}=\Omega^{-1}X_{(0)}.
 \ee
 One can see that there is a transformation \be\label{ct1}
 \Omega={X_{(0)}}{\vh^{-1}_0}\ee
 converting the dilaton action  (\ref{dg-1}) into the Hilbert one
   (\ref{1-1}).
In this manner, the CMB frame reveals the possibility to
 choose the units of measurements in the canonical  GR.

\subsection{"Coordinates" in brane "superspace of events"}

 The analogy of the conformally coupling scalar field in GR
  with  a relativistic brane (\ref{brane-m})
   allows us to formulate the choice of variables (\ref{9-h11}) and (\ref{9-h6})
   as a choice of the ``frame'' in the brane ``superspace of events''
 \bea\label{br-1}
 \widetilde{X}_{(0)}&=&\sqrt{X^2_{(0)}-X^2_{(1)}}, \\
 Q&=&\rm{arc}\coth \frac{X_{(0)}}{X_{(1)}}
 \eea
As we have seen above the argument in favor of the choice of these
variables is the definition of the measurable value of the Newton
constant
 \be\label{nc-1}
  G=\frac{8\pi}{3}\widetilde{X}_{(0)}^{-2}\Big|_{\rm present-day}=
  \frac{8\pi}{3}\vh^{-2}_0\ee
    as the present-day value
 of  the ``coordinate'' $\widetilde{X}_{(0)}=\vh_0$.

 In the case the action (\ref{brane-m})  takes the form
\bea\nonumber &&
S_{\mathrm{brane}}^{(D=4/N=2)}[X_{(0)},X_{(1)}]\!=\\\nonumber
&&\!\int\! d^4x\!\Bigg[\sqrt{-g_{(L)}}\widetilde{X}_{(0)}^2\!
\left(-\frac{{}^{(4)}\!R(g_{(L)})}{6}+g_{(L)}^{\mu\nu}\partial_\mu
Q\partial_\nu Q\right)\,+\\&&\label{brane-m3}
\widetilde{X}_{(0)}\partial_\mu\left(\sqrt{-g_{(L)}}
g_{(L)}^{\mu\nu}\partial_\nu \widetilde{X}_{(0)}\right)\Bigg].\eea
 This form is  the brane
 generalization of the relativistic conformal mechanics
\bea\label{confm-1}
S_{\mathrm{particle}}^{(D=1/N=2)}[X_{0},Q_{0}]\!&=& \!\int\!
ds\!\left[{X}_{0}^2\! \left(\frac{d
Q_{0}}{ds}\right)^2-\left(\frac{d {X}_{0}}{ds}\right)^2
\right];~~~\\\label{confm-2} &&ds = dx^0e(x^0)\eea  considered as a
simple example in the Section 3.

 The relativistic mechanics  (\ref{confm-1}) has two diffeo-invariant
 measurable times.
  They are
 the geometrical interval (\ref{confm-2}) and the time-like
 variable  $X_{0}$ in the external ``superspace of events''.
 The relation between these two ``times'' $X_{0}(s)$
 are conventionally treated as  a relativistic
 transformation.
 The main problem is to point out similar
  two measurable time-like diffeo-invariant
 quantities in both GR  and a brane  (\ref{brane-m3}).

 The brane/GR correspondence (\ref{brane-m})
 and special relativity (\ref{confm-1}) allows us to treat an external time
  as homogeneous component of the time-like external ``coordinate''
  $\widetilde{X}_{(0)}(x^0,x^k)$
  identifying this homogeneous component with
 the cosmological scale factor $a$ (\ref{d-t1})
  \be\label{nc-2}
  \widetilde{X}_{(0)}(x^0,x^k)\to\vh(x^0)=\vh_0a(x^0)
  \ee
  because  this factor is
 introduced in the cosmological perturbation theory \cite{lif}
 by the scale transformation of the metrics (\ref{ct-1}) too.

 The question
 arises:
 What is the value of this initial datum $\vh_I=\vh_0a_I$, if
 cosmological factor is treated as one of  ``degrees of freedom''?

\subsection{Free initial data  versus ``Planck's epoch''}

 The ``degrees of freedom''  means that
their initial data are beyond  equations of motion (i.e. ``free''
from any theoretical explanation)
   and they can be defined by only fitting
  of diffeo-invariant observational data.
 This ``freedom'' is main difference of a ``theory''
 from  Inflationary Model \cite{linde}, where these data are determined by
 fundamental parameters of equations of motion of the type of the Planck mass.
  In particular, in the Inflationary Model (and $\Lambda$CDM Model too),
  the initial datum $a_I$ is explained by
  the constraint
  \be\label{im-1}
 a_I=a'_0/\vh_0\equiv{\cal H}_0/\vh_0
  \ee
 called the Planck epoch. This constraint is
 justified by  the fundamental status of the Planck mass
 parameter
 in the initial Einstein -- Hilbert action.

 However, as
 we have seen in both the exact  GR (\ref{t-1}), (\ref{6-6})
   and its cosmological
 approximation (\ref{zm-1}),
  diffeo-invariant solutions
 of Einstein equations in GR
 contain the Planck mass as a multiplier
 of the cosmological scale factor $\vh=\vh_0a$, so that
 the Planck mass arises in the diffeo-invariant
 reduced action (\ref{ham-2})  as a present-day datum $\vh_0=\vh(\eta=\eta_0)$.

 The present-day status of the Planck mass in GR  is one
 of consequences of the diffeo-invariant reduction of theory
 to the  constraint-shell action (\ref{3-sm}).

 Therefore, the justification of the Planck epoch, in Inflationary
 Model,
 in the form of  the constraint (\ref{im-1}) looks
 as artefact of the diffeo-non-invariant consideration
 that is not compatible (as we have seen above) with the practice of SR
 and the Dirac definition
 of observables as diffeo-invariants.

 Moreover, the Planck's constraint (\ref{im-1})
 is not compatible with the causality principle
 of the diffeo-invariant action, in accord to which,
  in the sum
 \be\label{sm-2}
 \int\limits_{\vh_I}^{\vh_0}d\widetilde{\vh} E_{\widetilde{\vh}}=
 \int\limits_{\vh_I}^{\vh}d\widetilde{\vh} E_{\widetilde{\vh}}
 +\int\limits_{\vh}^{\vh_0}d\widetilde{\vh} E_{\widetilde{\vh}},
 \ee
 the initial datum $\vh_I$ in the first integral
 does not depend on the present-day data $\vh_0,\vh'_0={\cal H}_0\vh_0$
 in the second integral in contrast with the acceptable
 treatment  (\ref{im-1})
 of the ``Planck epoch'' as the Early Universe one $\vh_I={\cal H}_0$.

\section{Observational tests}

\subsection{Test I. The supernova Ia data}

 \begin{figure}[h!]\label{fig1}
\centering\includegraphics[scale=0.45]{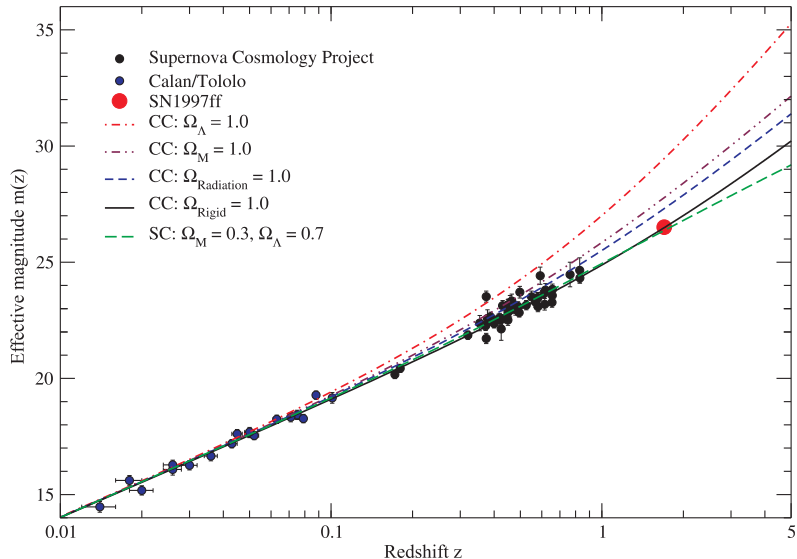} \caption{\small{The
Hubble diagram in cases of the {\it absolute} units of standard
cosmology (SC) and the {\it relative} units of conformal cosmology
(CC) \cite{039a, Danilo, zakhy}.
 The points include 42 high-redshift Type Ia
 supernovae \cite{snov} and the reported
 farthest supernova SN1997ff \cite{SN}. The best
fit to these data  requires a cosmological constant
$\Omega_{\Lambda}=0.7$, $\Omega_{\rm CDM}=0.3$ in the case of SC,
whereas in CC
 these data are consistent with  the dominance of the rigid (stiff)
 state. The Hubble Scope Space Telescope team analyzed 186 SNe Ia \cite{riess1} to test the CC
 \cite{zakhy}.}
 }
\end{figure}

Since the end of the XX century supernovae data has widespread
tested for all theoretical cosmological models. The main reason of
this is the fact that supernovae "standard candles" are still
unknown or absent \cite{Panagia_05}. Moreover, the first
observational conclusion about accelerating Universe and existence
of non-vanishing the $\Lambda$-term was done with the cosmological
SNe Ia data. Therefore, typically  standard (and alternative)
cosmological approaches are checked with the test.

Models of Conformal Cosmology are also discussed among other
possibilities \cite{zpz,Bog,bpzz,242,242a,114:a,039,Danilo,CC-2}.
 Conformal Cosmology is an alternative description of the Supernovae data without
the $\Lambda$-term as evidence for  the Weyl geometry \cite{we} with
the relative units interval $ds^2_{\rm Weyl}=ds^2_{\rm
Einstein}/ds^2_{\rm Einstein~Units}$ where all measurable quantities
and their  units are considered on equal footing.
 There is the scalar version of the Weyl geometry
  described by the conformal-invariant action of
a massless scalar field \cite{pct} with the negative sign
 that is mathematically equivalent to the Hilbert action of the  General
 Relativity where the role of the scalar field $\phi$ is played by the
  parameter of the scale
transformation $g^{\Omega}=\Omega^2 g$ multiplied by the Planck mass
$\phi=
 \Omega M_{\rm Planck}\sqrt{{3}/{(8\pi)}}$ \cite{kl}.

 We have seen above that
 the correspondence principle \cite{pp}
 as the low-energy
 expansion of the {\it``reduced action''} (\ref{2ha2}) over the
 field density
shows that the Hamiltonian approach to the General
 Theory of Relativity
 in terms of the Lichnerowicz scale-invariant variables
 (\ref{adm-2}) identifies the ``conformal quantities''
  with the observable ones including the conformal time $d\eta$,
  instead of $dt=a(\eta)d\eta$, the coordinate
 distance $r$, instead of the Friedmann one $R=a(\eta)r$, and \emph{the conformal
 temperature $T_c=Ta(\eta)$, instead of the standard one $T$}.
 Therefore,
 the scale-invariant variables  distinguish the conformal
 cosmology (CC) \cite{039, Narlikar},
  instead of the standard cosmology (SC).
 In this case,
 the
  red shift of the wave lengths of the photons
  emitted at the time $\eta_0-r$ by atoms on a cosmic object in the comparison
  with the Earth ones emitted at emitted at the time $\eta_0$,
  where $r$ is the distance between the Earth and the object:
\be \frac{\lambda_0}{\lambda_{\rm
cosmic}(\eta_0-r)}=\frac{a(\eta_0-r)}{a (\eta_0)}\equiv a(\eta_0-r)
=\frac{1}{1+z}. \ee
 This red shift can be  explained by the running masses
 $m=a(\eta)m_0$ in action (\ref{12h5}). In this case, the
 Schr\"odinger wave equation \bea\label{schL}
 \left[\frac{\hat p^2_r}{2 a(\eta)m_0}-\frac{\alpha}{r}\right]\Psi_L(\eta,r)=
 \frac{d}{id\eta}\Psi_L(\eta,r)
 \eea
 can be converted
 by the substitution $r=\dfrac{R}{a(\eta)}$, $p_r=P_{R}a(\eta)$, $a(\eta)d\eta=dt$,
 $a(\eta)\Psi_L(\eta,r)=\Psi_{0}(t,R)$ into the
 standard Schr\"odinger wave equation with the constant mass
\bea\label{sch0}
 \left[\frac{\hat P^2_R}{2 m_0}-\frac{\alpha}{R}\right]\Psi_0(t,R)=
 \frac{d}{i dt}\Psi_0(t,R).
 \eea
 Returning back to the Lichnerowicz variables $\eta,r$
 we obtain the spectral decomposition of the wave
 function of an atom with the running mass
 \bea\label{1sch0}
 \Psi_L(\eta,r)=\frac{1}{a(\eta)}\sum\limits_{k=1}^{\infty}
 e^{-i\varepsilon_0^{(k)}
 \int\limits_{\eta}^{\eta_0}d\widetilde{\eta}a(\widetilde{\eta})}\Psi^{(k)}_0(a(\eta)r)=\sum\limits_{k=1}^{\infty}\Psi^{(k)}_L(\eta,r).
 \eea
 Where $\varepsilon_0^{(k)}=\alpha^2m_0/k^2$ is a set of eigenvalues
 of the Schr\"odinger wave equation in the Coulomb potential. We got the
 equidistant spectrum $-i(d/d\eta)\Psi^{(k)}_L(\eta,r)=
 \varepsilon_0^{(k)}\Psi^{(k)}_L(\eta,r)
 $ for any wave lengths of cosmic photons
 remembering the size of the atom at the moment of their emission.

 The conformal observable distance  $r$ loses the factor $a$, in
 comparison with the nonconformal one \mbox{$R=ar$}. Therefore, in the
 case of CC, the redshift --
  coordinate-distance relation $\eta=\int{d\vh}(\sqrt{\rho_0(\vh)})^{-1}$
  corresponds to a different
  equation
  of state than in the case of SC \cite{039}.
 The best fit to the data,  including
  Type Ia supernovae~\protect \cite{snov, SN},
 requires a cosmological constant $\Omega_{\Lambda}=0.7$,
$\Omega_{\rm CDM}=0.3$ in the case of the ``scale-variant
quantities`` of standard cosmology. In the case of ``conformal
 quantities'' in CC, the Supernova data \cite{snov, SN} are
consistent with the dominance of the stiff (rigid) state,
$\Omega_{\rm Rigid}\simeq 0.85 \pm 0.15$, $\Omega_{\rm Matter}=0.15
\pm 0.15$ \cite{039, 039a, Danilo}. If $\Omega_{\rm Rigid}=1$, we
have the square root dependence of the scale factor on conformal
time $a(\eta)=\sqrt{1+2H_0(\eta-\eta_0)}$. Just this time dependence
of the scale factor on
 the measurable time (here -- conformal one) is used for description of
 the primordial nucleosynthesis \cite{Danilo, three}.

 This stiff state is formed by a free scalar field
 when $E_\vh=2V_0\sqrt{\rho_0}=\dfrac{Q}{\vh}$. In this case there is an exact
solution of the Bogoliubov equations  of the number of universes
created from a vacuum with the initial data
$\vh(\eta=0)=\vh_I,H(\eta=0)=H_I$ \cite{origin}.

\subsection{Test II. Particle creation and the present-day energy budget}

 These initial data $\vh_I$ and $H_I$ are determined by the
 parameters of matter cosmologically created from the Bogoliubov
 vacuum  at the beginning of a universe $\eta\simeq 0$.
\begin{figure}[ht!]
\begin{center}
 \includegraphics[scale=0.22,angle=-90]{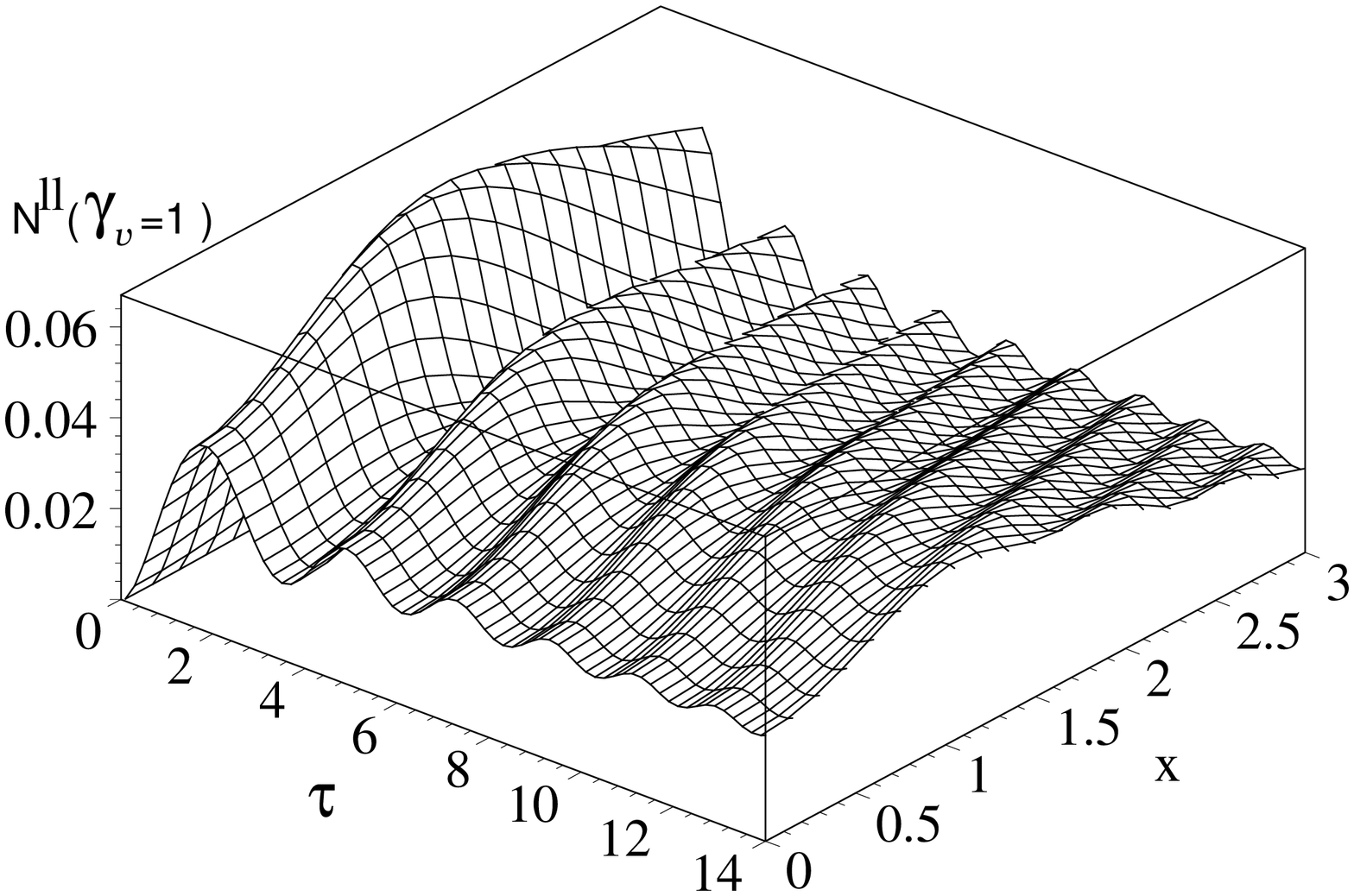}\hspace{-5mm}
 \includegraphics[scale=0.22,angle=-90]{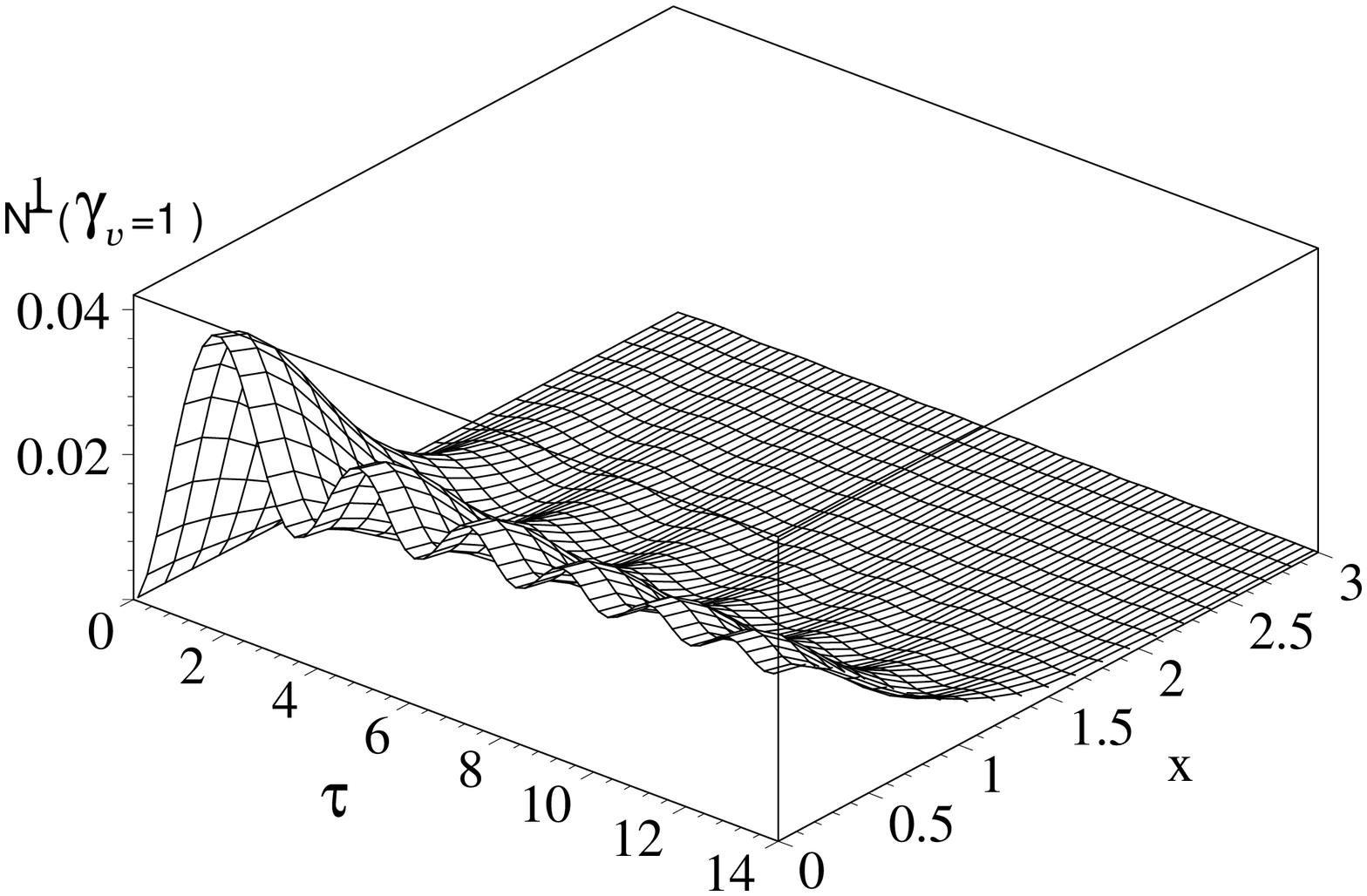}
\parbox{10cm}{  \caption{\small{
Dependence of longitudinal ${N^\|}$ and transverse ${N^\bot }$
components of the distribution function of vector bosons  on time
$\tau=2h_i\eta$ and momentum $(x=q/m_i)$. Their momentum
distributions in  units of the primordial mass $ x = q/M_I $ show
the large contribution of longitudinal bosons. Values of the initial
data $M_I = H_I$  follow from the uncertainty principle and give the
temperature of relativistic bosons $T\sim H_I=(M_0^2H_0)^{1/3}=2.7 K
$ \cite{114:a}}}}
\end{center}
\end{figure}
 The Standard
 Model (SM) density ${T}_{{\rm s}}$ in action (\ref{12h5})
  shows
 us that W-, Z- vector bosons have maximal probability of this
 cosmological creation
 due to their mass singularity \cite{114:a}. One can introduce the notion of
\emph{a particle in a universe} if the Compton length of a particle
 defined by its inverse mass
 \mbox{$M^{-1}_{\rm I}=(a_{\rm I} M_{\rm W})^{-1}$} is less than the
 universe horizon defined by the inverse Hubble parameter
 $H_{\rm I}^{-1}=a^2_{\rm I} (H_{0})^{-1}$ in the
 stiff state. Equating these quantities $M_{\rm I}=H_{\rm I}$
 one can estimate the initial data of the scale factor
 \mbox{$a_{\rm I}^2=(H_0/M_{\rm W})^{2/3}=10^{-29}$} and the primordial
 Hubble parameter
 \mbox{$H_{\rm I}=10^{29}H_0\sim 1~{\rm mm}^{-1}\sim 3K$}.
 Just at this moment there is  an effect of intensive
  cosmological creation of the vector bosons described in paper
  \cite{114:a} (see Fig. 2);
 in particular, the distribution functions of the longitudinal vector bosons
demonstrate clearly a large contribution of relativistic momenta.
 Their conformal (i.e. observable) temperature $T_c$
 (appearing as  a consequence of
 collision and scattering of these bosons) can be estimated
from the equation in the kinetic theory for the time of
establishment of this temperature \mbox{$\eta^{-1}_{relaxation}\sim
n(T_c)\times\sigma\sim H$}, where $n(T_c)\sim T_c^3$ and $\sigma
\sim 1/M^2$ is the cross-section. This kinetic equation and values
of the initial data $M_{\rm I} = H_{\rm I}$ give the temperature of
relativistic bosons \be\label{t}
 T_c\sim (M_{\rm I}^2H_{\rm I})^{1/3}=(M_0^2H_0)^{1/3}\sim 3 K
\ee as a conserved number of cosmic evolution compatible with the
Supernova data \cite{039,SN,snov}.
 We can see that
this  value is surprisingly close to the observed temperature of the
CMB radiation
 $ T_c=T_{\rm CMB}= 2.73~{\rm K}$.

 The primordial mesons before
 their decays polarize the Dirac fermion vacuum
 (\emph{as the origin of axial anomaly} \cite{riv,ilieva,gip,j})
 and give the
 baryon asymmetry frozen by the CP -- violation.
The
 value of the baryon--antibaryon asymmetry
of the universe following from this axial anomaly was estimated in
paper \cite{114:a} in terms of the coupling constant of the
superweak-interaction
 \be\label{a1}
 n_b/n_\gamma\sim X_{CP}= 10^{-9}.
 \ee
The boson life-times     $\tau_W=2H_I\eta_W\simeq
\left({2}/{\alpha_W}\right)^{2/3}\simeq 16,~ \tau_Z\sim
2^{2/3}\tau_W\sim 25
 $ determine the present-day visible
baryon density
\be\label{b}\Omega_b\sim\alpha_W=\dfrac{\alpha_{QED}}{\sin^2\theta_W}\sim0.03.\ee
All these results (\ref{t}) -- (\ref{b})
 testify to that all  visible matter can be a product of
 decays of primordial bosons, and the observational data on CMB
 radiation
 can reflect  parameters of the primordial bosons, but not the
 matter at the time of \emph{recombination}. In particular,
 the length of  the semi-circle on the surface of  the last emission of
photons at the life-time
  of W-bosons
  in terms of the length of an emitter
 (i.e.
 $M^{-1}_W(\eta_L)=(\alpha_W/2)^{1/3}(T_c)^{-1}$) is
 $\pi \cdot 2/\alpha_W$.
 It is close to $l_{min}\sim  210 $ of CMB radiation,
 whereas $(\bigtriangleup T/T)$ is proportional to the inverse number of
emitters~
 $(\alpha_W)^3 \sim    10^{-5}$.

 The temperature history of the expanding universe
 copied in the ``conformal quantities'' looks like the
 history of evolution of masses of elementary particles in the cold
 universe with the constant conformal temperature $T_c=a(\eta)T=2.73~ {\rm K}$
 of the Cosmic Microwave Background radiation.

 Equations of the
 vector bosons
 in SM are very close  to
 the  equations of the $\Lambda$CDM model
  with  the inflationary scenario used
  for description  of  the CMB ``power primordial spectrum''.

 \subsection{Test III: The Newton potential and the Large-scale structure}
The equations describing the longitudinal vector bosons
 in SM, in this case, are close to
 the equations that follow from
 the Lifshits perturbation theory and are  used, in  the inflationary model, for
 description of the ``power primordial spectrum'' of the CMB radiation.

 The next differences are a nonzero shift vector and  spatial oscillations of
 the scalar potentials determined by $\hat m^2_{(-)}$.
 In the conformal  cosmology model \cite{039}, the
  SN data corresponds to the dominance of rigid state $\Omega_{\rm rigid}\sim
  1$.
  The rigid state determines the parameter
  of spatial oscillations
  \begin{equation}\hat m^2_{(-)}=\dfrac{6}{7}H_0^2[\Omega_{\rm R}(z+1)^2+\dfrac{9}{2}\Omega_{\rm
  Mass}(z+1)].\end{equation}
The redshifts  in the recombination
  epoch $z_r\sim 1100$ and the clustering parameter \cite{kl}
\begin{equation}
r_{\rm clustering}=\dfrac{\pi}{\hat m_{(-)} }\sim \dfrac{\pi}{
 H_0\Omega_R^{1/2} (1+z_r)} \sim 130\, {\rm Mpc}
 \end{equation}
  recently
 discovered in the researches of a large scale periodicity in redshift
 distribution \cite{a1,a2}
 lead to a reasonable value of the radiation-type density
  $10^{-4}<\Omega_R\sim 3\cdot 10^{-3}<5\cdot 10^{-2}$ at the time of this
  epoch.
\newpage
\section{Summary}

%
%
We have proposed
\begin{itemize}
\item to make the Bekenstein -- Wagoner transformation in the
unified GR\&SM theory, in order to restore the Newton law, (this
restoration converts  the Higgs field  into dimensionless "angle"
of the metric-Higgs mixing),

\item to convert the fundamental constant in the Higgs potential
onto the zeroth Fourier harmonic of the Higgs field, that predicts
mass of  the Higgs field by initial date of the zeroth Fourier
harmonic,

\item to attach the cosmological scale factor to the Planck mass
as the single  dimension parameter of the theory,

\item to choose the Conformal Cosmology description
\cite{zpz,Bog,bpzz,242,242a,114:a,039,Danilo,zakhy,CC-2} where the
Rigid state is an equivalent of the Quintessence state in the
Standard Cosmology and gives satisfactory explanation of the last
Supernovae Ia data for the luminosity distance-redshift relation
without a cosmological constant,

\item to choose the free initial data for cosmological scale
factor instead of the Early Unverse Planck epoch where initial
data of the scale factor is determined by its present-day velocity
(\emph{i.e.} the Hubble parameter). In the class of the
diffeo-invariant solutions of the Einstein field equations, the
Planck epoch can be unambiguously treated as the present-day one,
whereas the opposite case contradicts to the Causality Principle.
\end{itemize}

In the Section 4 it was shown that there are  free  initial data
of the Electro-Weak epoch for both the zeroth modes (scale factor
and "angle" of the metric-Higgs  mixing) that initiated  the
intensive vacuum  creation of the SM particles that can be treated
as the "Big-Bang".

 In this case the Hilbert action-interval principle
 with two times-- the time of events as the cosmological scale factor, and
the time-interval gives responses on almost all problems of the
Inflationary  Model, if we replace its Early Universe Planck epoch
by the free initial data of the Electro-Weak epoch.

\begin{itemize}
\item The Hubble law is the time of events -- time-interval  relation;
\item Energy of events is the scale factor canonical momentum;
\item Homogeneity is consequence of the zeroth mode metric excitation obtained by
averaging of the spacial metric determinant logarithm over the
volume;
\item Horizon is given by the Weyl
definition of the measurable interval that attaches the cosmological
scale factor to all  masses so that we get  running masses instead
of  the expanded Unverse;
\item Flatness is given by the initial data;
\item Planck era is the present-day one;
\item Singularity  is absent in Quantum Universe with stable vacuum;
\item Arrow of time-interval is consequence of the causal quantization and primary and secondary
quantization of the energy constraint, in order to obtain the QFT
Bogoliubov vacuum,
 as a state with the minimal energy of events;
\item Cosmological vacuum creation  treated in the modern  literature as "Big-Bang" is consequence of the running masses;
\item Origins of temperature (~2.7 K) are consequence of scattering and
collisions of primordial particle created from vacuum.
\end{itemize}

 The unified theory keeps the Newton gravity,
 provided that the Higgs scalar field mixes with
 the scalar metric component
 that convert the Higgs field into the "metric-Higgs mixing angle" $Q$.

 Then, in order to adapt the Standard Model (SM) to cosmology, we
consider the SM version, where the fundamental dimensional
parameter $C$ in the Higgs potential $\lambda (\Phi^2-C^2)^2$ is
replaced by zeroth Fourier harmonic of the Higgs field $\Phi$, so
that all masses in SM  are determined by initial data of the
potential free (i.e. {\it inertial}) equation of this harmonic.
Such the replacement of the  fundamental  parameter by an initial
datum
 immediately predicts mass of Higgs field
$\sim 250$ GeV that follows from the extremum of the quantum
Coleman -- Weinberg effective potential obtained from the unit
vacuum--vacuum transition amplitude in the {\it inertial} motion
regime. We consider a cosmological model, where  the difference
between
 the Higgs field and  any additional scalar field forming the density
 of the Early Universe can be   explained  by  initial data of
 {\it inertial}  motions of these scalar fields.  The inertial motion
 of a scalar field corresponds to
 the rigid state equation. The dominance of the rigid state
 is consistent with
 the best fit to the Hubble diagram of  high-redshift Type Ia supernovae and
SN1997ff \cite{riess1} in the Conformal Cosmology
\cite{zpz,Bog,bpzz,242,242a,114:a,039,Danilo,zakhy,CC-2}, whereas
in Standard Cosmology  these data requires  a cosmological
constant $\Omega_{\Lambda}=0.7$, $\Omega_{\rm Cold Dark
Matter}=0.3$ \cite{mukh}.

The Conformal Cosmology is the generalization of the
 Copernican relativity of  positions and velocities to
 the Weyl relativity of values of  intervals.
The Conformal Cosmology is based on the Weyl definition of the
measurable interval as the ratio of the Einstein one and the units
determined by the standard mass $m_0$. If we joint the
cosmological scale factor to the length, we get the expanded
Universe in the Standard Cosmology. If we joint the cosmological
scale factor to the units (i.e. masses including $m_0$), we get
the Conformal Cosmology as the {\it collapsed units of an
observer}, just as Copernicus  explained the Planet motion
  by a {\it relative position of an observer}  in the
Heliocentric system. In both the cases, the {\it Copernican
relativity of positions and velocities} and
 the {\it Weyl relativity of values} of  intervals mean free initial data
 of equations of motion,  in particular, the free initial data
 of the cosmological scale factor in contrast to the Planck epoch
 of the Inflationary Model \cite{linde}.
All masses in Conformal Cosmology depend on  the  initial data of
a solution of the equation of motions in the CMB frame, like a
trajectory of a particle in the Newton mechanics depends on
initial data fitted from observations in a ``frame of reference to
initial data''.

 One more difference from the Inflationary Model \cite{linde} is
the {\it Einstein -- Hilbert  relativity of times}. The {\it
relativity of times} allows us to treat cosmological scale factor
as the time-variable in space of events, cosmological equations --
as the energy constraint in space of events, and Big-Bang as an
inevitable consequence of the primary and secondary quantization
of this energy constraint in the form of vacuum creation of the
universe and matter with the arrow of time-interval as an quantum
anomaly. This anomaly is the consequence of the  vacuum postulate
about a quantum state with the minimal energy in the space of
events.

All relativity principles are consistent with  the Dirac
Hamiltonian approach based on  the classification of all field
components onto the Laplace-like {\it potentials} with the
boundary condition and
 the d'Alambert-like {\it degrees of freedom} with initial data.

 The CMB frame fixing and the finite space-time
 suppose a formulation of GR with the complete separation of the frame
  transformations
  from the general coordinate ones identified with
  the diffeomorphisms \cite{origin,242}.
 Such the separation states the following questions.
\begin{enumerate}
\item What is a status of the conformal time in
 the Hubble redshift -- luminosity
 distance relation? Is the conformal time an
 object of Bardeen's gauge transformations \cite{bardeen}, or it
 is gauge-invariant measurable quantity in CMB frame in accord with
 the Dirac definition of observable quantities as invariants?
\item What is a status of the cosmological scale factor? Is it an
additional variable in Lifshitz's cosmological perturbation theory
\cite{MFB}, or it is a zeroth mode of the scalar metric component?
\item What are initial data of the cosmological scale factor? Are they
the Planck epoch data as the origin of the numerous problems in the
Inflationary Model \cite{linde}, or they are free from equations of
motion in accord with the standard causality principle for any
dynamic systems?
\item What is the ``vacuum'' of cosmological models?
\item What is the origin of arrow of time?
\end{enumerate}
 Responses to these questions can be given by the mentioned above
 the Hamiltonian tool \cite{dir,242,242a,ADM}.

  Just  the complete separation of the frame
  transformations
  from the general coordinate ones
is the main difference of the Hamiltonian approach
 to GR from the naive Bardeen's approach \cite{bardeen}
 to the cosmological perturbation theory accepted in the Inflationary Model.
 The  Bardeen's gauge transformations of the measurable
 time \mbox{$\eta\to
 \widetilde{\eta}=\widetilde{\eta}(\eta)$} identifies  the
 unmeasurable coordinate time  $x^0$ as an object of
 the reparametrizations with the diffeo-invariant
 measurable interval
 $\eta=x^0$. This confusing $x^0=\eta$ is an obstacle for
  the consistent description of the dynamics in the
 reduce phase space  of events $[\vh|Q]$, where the
 scale factor $\vh$ is the single independent ``evolution parameter''.
 This confusing two times $x^0=\eta$ also prevents
 to understand the
 relativistic status of the Hubble law, space  of events $[\vh|Q]$,
 the energy of events, the vacuum of events, the vacuum creation
 of the Universe,  and the arrow of time in the cosmological
 models \cite{Bog}.

  We show that there are the free initial data
 $a_I=3 \times 10^{-15}, Q_0=3 \times 10^{-17}$ that
 give the similar inevitable vacuum creation
 of particle in agreement with the present day
 energy budget of the Universe.

\section*{Acknowledgements}

Authors are grateful to A.B. Arbuzov, B.M. Barbashov, A. Beckwith,
D.B. Blaschke, A. Borowiec, K.A. Bronnikov, \v{C}. Burdik, P.
Flin, D.V. Ivashchuk, D.I. Kazakov, E. Kapu\'scik, E.A. Kuraev, P.
Leach, \mbox{V.N. Melnikov}, A.G. Nikitin, M.R. Pi\c atek,
\mbox{V.B. Prieezzhev}, Yu.P. Rybakov, S.A. Shuvalov, V.V. Skokov,
L.M. Soko\l owski, \mbox{I.A. Tkachev}, M. Vittot, A. Woszczyna,
and A.F. Zakharov for interesting and critical discussions.

\newpage
\appendix
\section{Hilbert's QFT Foundations}

\renewcommand{\theequation}{A.\arabic{equation}}

\setcounter{equation}{0}

\subsection{Hilbert's formulation of Special Relativity}

  The Hilbert geometric formulation  of a relativistic particle
  \cite{H,pp,bpp}
  is based on  the action:
 \be\label{2} S^{\mbox{\tiny SR}}_{1915}
 =-\frac{m}{2}\int\limits_{\tau_1}^{\tau_2}d\tau
 \left[\frac{(\dot X_\alpha)^2}{e(\tau)}+e(\tau)
 \right],
 \ee
 and an {\it geometric interval}
  \be\label{3}
 ds=e(\tau)d\tau~~~~~\longmapsto~~~~~s(\tau)=
 \int\limits_{0}^{\tau}d\overline{\tau}e(\overline{\tau}),
 \ee
 where $\tau$ is the {\it coordinate evolution parameter}
 given in a one-dimensional Riemannian manifold with
 a single component of the metrics $e(\tau)$ and
 the variables $X_\alpha$ form the Minkowskian {\it space of
 events}, where $\left(X_\alpha\right)^2=X_0^2-X_i^2$.

 The action (\ref{2}) and interval (\ref{3}) are
 invariant with respect to reparametrizations of
 the {\it coordinate evolution parameter}
 $\tau ~\to~\widetilde{\tau}
 =\widetilde{\tau}(\tau);$
 therefore, the theory given by (\ref{2}) and  (\ref{3})
 can be considered as the simplest model of GR.
 A single component of the metrics $e(\tau)$
  plays the role
 of the Lagrange multiplier in the Hamiltonian form of
 the action (\ref{2}):
 \be \label{4}
 S^{\mbox{\tiny SR}}_{1915}=\int\limits_{\tau_1}^{\tau_2}d\tau
 \left[-P_\alpha\partial_\tau X^\alpha+
 \frac{e(\tau)}{2m}\left(P_\alpha^2-m^2\right)\right].
 \ee
 Varying action  (\ref{4}) over lapse-function $e(\tau)$  defines the
 {\it``constraint''}:
 \be\label{co}
 (P_\alpha)^2-m^2=0.
 \ee
 Varying action  (\ref{4})  over
 dynamical variables $(P_{\alpha}, X_{\alpha})$
 gives the equations of motion: $P_\alpha=m{dX_\alpha}/{ds}$,
  ${dP_\alpha}/{ds}=0$,
 taking into
 consideration $ds=e(\tau)d\tau$.
 Solutions of these equations  in terms of gauge-invariant
   geometric interval (\ref{3}) take the form
 \be  \label{5}
 X_\alpha(s)=X_\alpha(0)+
 \frac{P_\alpha(0)}{m}s.
 \ee

  The physical meaning of this solution is revealed  in
  a  specific {\it``frame of reference''}. In particular,
   solutions of energy constraint (\ref{co}) with respect to
  a temporal component
  $P_{0}$ of momentum $P_{\alpha}$
 \be  \label{6}
 {P_0}_{\pm}=\pm \sqrt{m^2+P_i^2}
 \ee
   are considered as the {\it``reduced
  Hamiltonian''} in the {\it``reduced phase space''} $\left\{X_i,P_j\right\}$
  that becomes the energy $E(P)=\sqrt{m^2+P_i^2}$ onto a trajectory
  \cite{poi,ein}. The time component of  solution
 (\ref{5})
 \be  \label{7}
 s=\frac{m}{{P_0}_{\pm}}[X_0(s)-X_0(0)] \ee
 shows us that the {\it``time-like variable''} $X_0$ is identified with the
 time measured in the rest frame of reference, whereas an interval $s$
 is the time measured in the comoving frame.

 The dynamic version of SR  \cite{poi,ein}
 can be obtained as values of the geometric action (\ref{4})
 onto solutions of the constraint (\ref{6})
 \be \label{8} S^{\mbox{\tiny SR}}_{1915}
|_{P_0={P_0}_{\pm}}=S^{\mbox{\tiny SR}}_{1905}
=\int\limits_{X_{0I}}^{X_0}d\overline{X}_0
 \left[P_i\frac{d{X}_i}{d\overline{X}_0}-{P_0}_{\pm}\right].
 \ee
  Just the values of the {\it`` geometric interval''} (\ref{7})
  and action (\ref{8}) onto  resolutions (\ref{6})
  of constraint (\ref{co}) in
  the specific frame of reference will be called
 the {\it ``Hamiltonian reduction''} of Hilbert's geometric
 formulation of SR given by Eqs. (\ref{2}) and  (\ref{3}) (see \cite{pp,dir}).

\subsection{Dynamic Special Relativity of 1905}

 The {\it ``Hamiltonian reduction''} leads to action (\ref{8}) of
 the dynamic theory of a relativistic particle of ``1905''
 \cite{poi,ein} that
  establishes a correspondence with the classical
 mechanic action by the low-energy decomposition
  \be \label{9}
 E(P)=\sqrt{m^2+P_i^2}=m+\frac{P_i^2}{2m}+...
 \ee
 It gives us the very important  concept of particle ``energy'' $E(0)=mc^2$.
We can see that
 relativistic  relation (\ref{7})
between the ``time as the variable'' and the ``time as the
interval''
 appears in the geometric version  of ``1915'' \cite{H}
 as a consequence of the variational equations  (\ref{6}), whereas
 in the dynamic
 version of  ``1905'' \cite{poi,ein} the same relativistic
 relation in the form of a kinematic Lorenz
 relativistic transformation is supplemented to variational equations following
 from the dynamic action (\ref{8}).

\subsection{Quantum geometry of a relativistic particle}

 The next step forward to QFT is the primary quantization of
 particle variables: $i[\hat P_\mu,X_\nu]=\delta_{\mu\nu}$, that
 leads to
 the quantum version of the energy constraint (\ref{co})
 $[\square+m^2]\psi(X_0,X_i)=0$ known  as the Klein -- Gordon equation
 of the wave function.
  The general solution of this
 equation
 \be\label{kg}
 \partial^2_0\Psi_p+E_p^2\Psi_p=0
 \ee
 for a single p-Fourier harmonics
 $\Psi_p(X_0)=\int d^3X \exp{iP_jX^j}\psi(X_0,X_i)$
 takes the form of the sum of two terms
 \be\nonumber
 \Psi_p=\frac{1}{\sqrt{2E_p}}\{a_{p}^{(+)}(X_0)+a_p^{(-)}(X_0)\},
 \ee
 where $a_{p}^{(+)}(X_0),a_p^{(-)}(X_0)$ are  solutions of the
 equations
\be \label{1pa} (i\partial_0+E_p)a_{p}^{(+)}=0,~~~~~~~~
(i\partial_0-E_p)a_p^{(-)}=0.
  \ee
 They are treated as the Shr\"odinger equations of the
 dynamic
 theory (\ref{8}) for the case of positive and negative
 particle ``energies'' (\ref{6}) revealed
 by resolving energy constraint (\ref{co}).

  QFT is formulated as the secondary quantization of a relativistic
 particle $[a_p^{(-)},a_{p}^{(+)}]=1$ \cite{Logunov}. In order to remove
 the negative ``energy'' $-E_p$ and to provide the quantum system with
 stability, the field $a_{p}^{(+)}$ is considered as the operator of
 creation of a particle and $a_{p}^{(-)}$  as the operator of
 annihilation of a particle, both with positive ``energy''.
 The initial datum $X_{I(0)}$ is treated as a point of this
 creation or annihilation. This interpretation means
 postulating vacuum as a state with minimal ``energy'' $a_p^{(-)} |0\rangle =0$,
 \label{pos} and it restricts the motion of a particle in
 the space of events, so that a particle with $P_{0+}$ moves
 forward and with $P_{0-}$ backward.
 \be \label{b12sr}
 P_{0+}~~~~ \to ~~~~X_{I{(0)}}\leq {X_{(0)}};~~~~~P_{0-}~~~~ \to ~~~~X_{I{(0)}}\geq {X_{(0)}}.
  \ee
  As a result of such a restriction the interval (\ref{7})
  becomes
 \be  \label{7+}
 s_{(P_{0+})}=\frac{m}{{E_p}}[X_0(s)-X_0(0)];~~~~~~X_{I{(0)}}\leq {X_{(0)}},
 \ee
 \be  \label{7-}
 s_{(P_{0-})}=\frac{m}{{E_p}}[X_0(0)-X_0(s)];~~~~~~X_{I{(0)}}\geq {X_{(0)}}.
 \ee
 One can see that  in both cases the geometric interval is
 positive. In other words, the
 stability of quantum theory and the vacuum postulate as its consequence
 lead to the absolute reference point  of this interval $s=0$
 and its positive arrow.
 The last  means  violation of the symmetry of classical theory
 with respect to the transformation $s \to $ $-s$. Recall that the
 violation of the symmetry of classical theory by their
 quantization is called the quantum anomaly \cite{riv,ilieva,gip}.
 The quantum anomaly as the consequence of the vacuum postulate
 was firstly discovered by Jordan \cite{j} and then
 rediscovered by a lot of authors (see \cite{riv}).

 \subsection{Creation of particles}

  Creation of particles is described by QFT obtained by
   quantization of  classical fields with masses depending on time $m=m(X_0)$.
  Classical field equation (\ref{kg}) can be got by varying the action
  \be\label{ft1}
 S_{\rm p}=\int dX_0 \left\{P_p\partial_0\Psi_{p}-H_{\rm
 p}\right\},
 \ee
  where $H_{\rm p}=\frac{1}{2}\left[P_p^2+E_p^2(X_0)\Psi_p^2\right]$ is the
  field Hamiltonian,
   here we kept only one p-harmonics.

 \noindent The holomorphic representation of the fields \cite{ps1,skok}
 \be\label{g1}
 \Psi_p=\frac{1}{\sqrt{2E_p(X_0)}}\left\{a_{p}^{(+)}(X_0)+a_p^{(-)}(X_0)\right\},
 \ee
 \be\label{g2}
 P_p=i\sqrt{\frac{E_p(X_0)}{2}}\left\{a_{p}^{(+)}(X_0)-a_p^{(-)}(X_0)\right\}.
 \ee
 allows us
  to express the field Hamiltonian in action  (\ref{ft1})
 in terms of observable quantities --- the one-particle energy $E_p(X_0)$
 and  ``number'' of particles
 $N_p(X_0)=[a_{p}^+a_p^-]$:
 \be\label{ufh1z1}
 H_{\rm p}=\frac{1}{2}\left[P_p^2+E_p^2(X_0)\Psi_p^2\right]=
 E(X_0)\left[N_p(X_0)+\frac{1}{2}\right].
 \ee
   While the canonical structure
 $P_p \partial_0 \Psi$ in (\ref{ft1}) takes the form:
 \be\nonumber
   P_p \partial_0 \Psi_p=
  \left[\frac{i}{2}(a_p^+\partial_0 a_p^--a_p^+
 \partial_0 a^-)-
 \frac{i}{2}(a_p^+a_p^+- a_p^-a^-_p)\frac{\partial_0 E(X_0)}{2E(X_0)}\right].
 \ee
 The one-particle energy and the number of particle are not
 conserved. In order to find a set of  integrals of
 motion, we can use the Bogoliubov transformations
  \be\label{bogo}
  a_p^+=\alpha b_p^+\!+\!\beta^*b_p^-
  ~~~~~~~~~~~~~(\alpha=e^{i\theta}\cosh r,~~~
  ~\beta=e^{-i\theta}\sinh{r}),
  \ee
 so that  the equations of $b_p^+,b_p^-$ become diagonal
\be\label{bogo1}
 (i\partial_0+E_B)b_p^+=0,~~~~~~~~
 (i\partial_0-E_B)b_p^-=0,
 \ee
 and the conserved vacuum is defined by the postulate:
  \be\label{bogo2}
   b_p^-|0>_{\rm p}=0.
  \ee
 The  corresponding Bogoliubov equations of
 diagonalization expressed in terms  of the distribution function
 of {\it``particles''} $N_{\rm p}(X_0)$ and the rotation function
 $R_{\rm p}(X_0)$
 \bea\nonumber
 N_{\rm p}(X_0)&=&|\beta|^2\equiv {}_{\rm p}<a_p^+a_p^- >_{\rm
 p}\equiv\sinh{r}^2,\\\nonumber
 R_{\rm
 p}(X_0)&=&i(\alpha\beta^*-\alpha^*\beta)\equiv-\sin(2\theta)\sinh{2r}
 \eea
 take the form \cite{ps1}
 \be\label{1p}
 \left\{\begin{aligned}
 \frac{dN_{\rm p}}{dX_0}&=&\frac{\partial_0 E(X_0)}{2E(X_0)}
 \sqrt{4N_{\rm p}(N_{\rm p}+1)-R_{\rm p}^2},  \\
 \frac{dR_{\rm p}}{dX_0}&=&-{2E(X_0)}
 \sqrt{4N_{\rm p}(N_{\rm p}+1)-R_{\rm p}^2},
 \end{aligned}\right.
 \ee
 \be \label{6pa}
 E_B(X_0)=\frac{E_p(X_0)-\partial_0\theta}{\cosh 2r}.
 \ee
 These equations supplemented by the quantum geometric interval
 (\ref{7+}) and (\ref{7-}) are the complete set of equations  for description
 of the phenomenon of particle creation.

 Thus, the direct way from
 Hilbert's geometric formulation of any relativistic theory to the
 corresponding
 ``quantum field theory''
 goes through  Dirac's Hamiltonian reduction and Bogoliubov's
 transformations. As a result, we have
 the description of creation of a relativistic particle in the space of events
 at the absolute reference point
 of geometric interval $s$ of this particle. The physical meaning of this
 interval is revealed in the Quantum Cosmology considered below.

\section{Quantum universes}

\renewcommand{\theequation}{B.\arabic{equation}}

\setcounter{equation}{0}

 \subsection{QFT of universes}

 After the primary  quantization of the cosmological scale factor $\vh$:
 $i[P_\vh,\vh]=1$ the energy constraint $P^2_\vh=E_\vh^2$
 transforms
 to the WDW equation
\be\label{wdw}
 \partial^2_\vh\Psi+E_\vh^2\Psi=0.
 \ee
 This equation can be obtained in the corresponding classical
 WDW field theory for universes of the type of the Klein -- Gordon
 one:
 \be\label{uf}
 S_{\rm U}=\int d\vh \frac{1}{2}
 \left[(\partial_\vh\Psi)^2-E_\vh^2\Psi^2\right]\equiv \int d\vh
  L_{\rm U}.
 \ee
 Introducing the canonical momentum
$P_\Psi=\partial L_{\rm U}/\partial(\partial_\vh\Psi)$, one can
obtain the Hamiltonian form of this theory
 \be\label{ufh}
 S_{\rm U}=\int d\vh \left\{P_\Psi\partial_\vh\Psi-H_{\rm
 U}\right\},
 \ee
 where
\be\label{ufh1}
 H_{\rm U}=\frac{1}{2}\left[P_\Psi^2+E_\vh^2\Psi^2\right].
 \ee
 is the Hamiltonian. The concept of the one-universe ``energy'' $E_\vh$
 gives us the opportunity to present this Hamiltonian $H_{\rm U}$
 in the standard forms of the product of this ``energy''  $E_\vh$ and
 the ``number'' of universes
 \be\label{AA}
 N_U=A^+A^-,
 \ee
 \be\label{ufh1z12}
 H_{\rm U}=E_\vh\frac{1}{2}\left[A^+A^-+A^-A^+\right]=E_\vh[N_U+\frac{1}{2}]
 \ee
 by  means of the transition to the holomorphic variables
 \be\label{g}
 \Psi=\frac{1}{\sqrt{2E_\vh}}\{A^{+}+A^{-}\},~~~~~~~~
 P_\Psi=i\sqrt{\frac{E_\vh}{2}}\{A^{+}-A^{-}\}.
 \ee
 The  dependence of $E_\vh$  on $\vh$ leads to the additional term
 in the action expressed in terms the holomorphic variables
 \be\label{und3}
 P_\Psi \partial_\vh \Psi\!=\!
  \frac{i}{2}(A^+\partial_\vh A^-\!-\!A^+
 \partial_\vh A^-)\!-\!
 \frac{i}{2}(A^+A^+\!-\!A^-A^-)\triangle (\vh),
\ee
 where
 \be\label{tri}
 \triangle(\vh)=\frac{\partial_\vh E_\vh}{2E_\vh}.
 \ee
  The last term in (\ref{und3}) is responsible for
  the cosmological creation of  ``universes'' from ``vacuum''.

 \subsection{Bogoliubov transformation. Creation of universes}

 In order to define stationary physical states, including
 a ``vacuum'', and a set of integrals of motion, one usually uses
 the Bogoliubov transformations \cite{B,ps1,skok} of the
 variables of universes
 $(A^+,A^-)$:
 \be \label{u17} A^+=\alpha
 B^+\!+\!\beta^*B^-,~~\;\;A^-=\alpha^*
 B^-\!+\!\beta B^+~~(|\alpha|^2-|\beta|^2=1),
  \ee
  so that the classical equations of the field theory  in terms of universes
 \be \label{1un} (i\partial_\vh+E_\vh)A^+=iA^-\triangle(\vh),~~~~~~~~
(i\partial_\vh-E_\vh)A^-=iA^+\triangle(\vh),
  \ee
  take a diagonal form in terms of  {\it quasiuniverses}
  $B^+,B^-$:
 \be \label{2un} (i\partial_\vh+E_B(\vh))B^+=0,~~~~~~~~
 (i\partial_\vh-E_B(\vh))B^-=0.
  \ee
  The diagonal form is possible, if the  Bogoliubov coefficients $\alpha,\beta$ in
  Eqs. (\ref{u17})
  satisfy to equations
 \be \label{3un} (i\partial_\vh+E_\vh)\alpha=i\beta\triangle(\vh),~~~~~~~~
(i\partial_\vh-E_\vh)\beta^*=i\alpha^*\triangle(\vh).
  \ee
 For the parametrization
 \be \label{4un} \alpha=e^{i\theta(\vh)}\cosh r(\vh),~~~~~~~~
 \beta^*=e^{i\theta(\vh)}\sinh{r}(\vh),
  \ee
  where $r(\vh),\theta(\vh)$ are the parameters of ``squeezing''
  and ``rotation'', respectively, Eqs. (\ref{3un}) become
 \bea \label{5un}\nonumber
 (i\partial_\vh\theta-E_\vh)\sinh 2r(\vh)&=&-\triangle(\vh)\cosh 2r(\vh)\sin
 2\theta(\vh),\\\partial_\vh r(\vh) &=&\triangle(\vh)\cos
 2\theta(\vh),
 \eea
  while ``energy'' of  {\it quasiuniverses} in Eqs. (\ref{2un})
  is defined by expression
   \be \label{6un}
  E_B(\vh)=\frac{E_\vh-\partial_\vh\theta(\vh)}{\cosh 2r(\vh)}.
  \ee
 Due to Eqs. (\ref{2un}) the ``number'' of {\it quasiuniverses}
  ${\cal N}_B=B^+B^-$ is conserved
  \be \frac{d{\cal N}_B}{d\vh}\equiv
  \frac{d(B^+B^-)}{d\vh}=0.
  \ee
  Therefore, we can introduce the ``vacuum''
 as a state without {\it quasiuniverses}:
 \be \label{sv}
 B^-|0>_{\rm U}=0.
 \ee
 A number of created {\it universes} from this Bogoliubov vacuum
 is equal to the expectation value of the operator
 of the  {\it number of universes } (\ref{AA}) over the Bogoliubov
 vacuum
 \be\label{usv1}
 N_{\rm U}(\vh)={}_{\rm U}<A^+A^-
  >_{\rm U}\equiv |\beta|^2=\sinh^2r(\vh),
 \ee
 where $\beta$ is the coefficient in the Bogoliubov transformation
 (\ref{u17}), and $N_{\rm U}(\vh)$ is called the ``distribution
 function''. Introducing the Bogoliubov ``condensate''
\be\label{usv2}
 R_{\rm U}(\vh)=i(\alpha\beta^*-\alpha^*\beta)
 \equiv {}_{\rm U}<P_\Psi\Psi >_{\rm U}
 =\frac{i}{2}~\,{}_{\rm U}\!<[A^+A^+-A^-A^-]>_{\rm U},
 \ee
 one can rewrite the Bogoliubov equations of the diagonalization (\ref{3un})
\be\label{usv3}
 \left\{\begin{aligned}
 \frac{dN_{\rm U}}{d\vh}&=\triangle(\vh)
 \sqrt{4N_{\rm U}(N_{\rm U}+1)-R_{\rm U}^2},  \\
 \frac{dR_{\rm U}}{d\vh}&=-{2E_\vh}
 \sqrt{4N_{\rm U}(N_{\rm U}+1)-R_{\rm U}^2}.
 \end{aligned}\right.
 \ee
  It is natural to propose that at the moment of creation of
  the universe $\vh(\eta=0)=\vh_I$ both these functions are equal to zeroth
   $N_{\rm U}(\vh=\vh_I)=R_{\rm U}(\vh=\vh_I)=0$.
  This moment of the conformal time
  $\eta=0$
   is distinguished by the vacuum postulate (\ref{sv}) as the beginning
   of a universe.

\subsection{Quantum anomaly of conformal time}

 As it was shown in the case of a particle in QFT \cite{Logunov},
 the postulate of a vacuum as a state with minimal ``energy''
 restricts the motion of a ``universe'' in
 the space of events, so that a ``universe'' with $P_{\vh+}$ moves
 forward and with $P_{\vh-}$ backward.
 \be \label{1b12sr}
 P_{\vh+}~~~~ \to ~~~~\vh_{I}\leq {\vh_{0}};~~~~~~~~~P_{\vh-}~~~~ \to ~~~~\vh_{I}\geq {\vh_{0}}.
  \ee
  If we substitute this restriction into the interval (\ref{zm-4})
 \bea  \label{u7+}
 \eta_{(P_{vh+})}&=&\int\limits_{\vh_I}^{\vh_0}\frac{d\vh}{\sqrt{{\rho_0(\vh)}}}
 ;~~~~~~~~~~~~~~~~~~\vh_{I}\leq {\vh_{0}},\\\label{u7-}
 \eta_{(P_{\vh-})}&=&\int\limits_{\vh_0}^{\vh_I}\frac{d\vh}{\sqrt{\rho_0(\vh)}}
 ;~~~~~~~~~~~~~~~~~~\vh_{I}\geq {\vh_{0}},
 \eea
 one can see that the geometric interval in both cases is
 positive. In other words, the stability of quantum theory
 as the vacuum postulate leads to the absolute point of
 reference of this interval $\eta=0$  and its positive arrow.
   In QFT the initial datum $\vh_I$ is considered  as a point of
 creation or annihilation of universe.
 One can propose that the singular point $\vh=0$ belongs to
 antiuniverse. In this  case, a universe with a positive
 energy goes out of the singular point   $\vh =0$.

 In the model of rigid state $\rho=p$, where  $E_\vh=Q/\vh$
 Eqs. (\ref{usv3}) have an exact solution
\be\label{11cu}
 N_{\rm U}=\frac{1}{4Q^2-1}
 \sin^2\left[\sqrt{Q^2-\frac{1}{4}}~~\ln\frac{\vh}{\vh_I}\right]\not
 =0,
\ee
 where
 \be\label{cc}
 \vh=\vh_I\sqrt{1+2H_I\eta}
 \ee
  and
$\vh_I,H_I=\vh'_I/\vh_I=Q/(2V_0\vh_I^2)$ are the initial data.

 We see that there are results of the type of the arrow of
 time and absence of the cosmological singularity (\ref{u7+}),
 which can be understood only on the level of
  quantum theory, where symmetry $\eta~\to~-\eta$ is broken \cite{riv,ilieva}.

\section{Massive electrodynamics in GR}

\renewcommand{\theequation}{C.\arabic{equation}}
\setcounter{equation}{0}

As the model of the matter let us consider massive electrodynamics
in GR
 \be \label{c1-1}
 S=\int d^4x\sqrt{-g}\left[-\frac{\vh_0^2}{6}R(g)
 +{\cal L}_{\rm m}\right]\equiv\int d^4x\sqrt{-g}\mathcal{L},
 \ee
 where ${\cal L}_{\rm m}$ is the Lagrangian of the massive vector and spinor
 fields
\be \label{c1-2}\nonumber
 {\cal L}_{\rm
 m}=-\frac{1}{4}F_{\mu\nu}
 F_{\alpha\beta}g^{\mu\alpha}g^{\nu\beta}-M_0^2 A_\mu A_\nu
 g^{\mu\nu}-\widetilde{\psi}i\gamma^\sigma
(D_{\sigma}-ieA_\sigma)\Psi-m_0\widetilde{\psi}\hat \Psi
 \ee
 $F_{\mu\nu}=\partial_\mu A_\nu-\partial_\nu A_\mu$ is the stress tensor,
 $
 D_{\delta}=\partial_{\delta}-i\frac{1}{2}[\gamma_{(\alpha)}
 \gamma_{(\beta)}]\sigma_{\delta(\alpha)(\beta)},
 $
is the Fock covariant derivative\cite{fock29},
 $\gamma_{(\beta)}=\gamma^\mu e_{(\beta)\mu}$ are the Dirac $\gamma$-matrices,
 summed with tetrads $e_{(\beta)\nu}$, and
 $\sigma_{\sigma(\alpha)(\beta)}=e^\nu_{(\beta)}(\nabla_\mu
 e_{(\alpha)\nu})$ are
  coefficients of spin-connection \cite{fock29,ll}.

 The Lagrangian of the massive fields (\ref{c1-2})  can be
 rewritten
 in terms of the Lichnerowicz variables
 \be\label{1-21}
 {A^{L}}_{\mu}={A}_{\mu},~~~~~~~~~~ \Psi^{L}=a^{3/2}\psi^{3}\Psi,~
 \ee
 that lead to fields with masses depending on the scale factor $a\psi^{2}$
 \be\label{1-22}
  m_{(\rm L)}=m_0a\psi^{2}=m\psi^{2},~~~~~M_{(\rm
  L)}=M_0a\psi^{2}=M\psi^{2}.
  \ee
  These fields are in the space defined by the component of the frame
 \bea \label{1-23}
 \omega^{(\rm L)}_{(0)}&=&\widetilde{\psi}^4~\widetilde{N}_ddx^0,\\\label{1-24}
 \omega^{(\rm L)}_{(a)}&=&{\bf e}_{(a)i}(dx^i+N^i dx^0).
 \eea
  with the unit metric determinant $|{\bf e}|=1$.

  As the result, the Lagrangian of the matter fields (\ref{c1-2}) takes the form
  \bea\nonumber
  &&\sqrt{-g}{\cal L}_{\rm m}(A,\widetilde{\psi},\Psi)=
 \frac{1}{i}\widetilde{\psi}^L\gamma^{0}\left(\partial_0-
 N^k\partial_k+\frac{1}{2}\partial_lN^l-
 ie A_0\right)\Psi^L \\\nonumber
 \!\!\!\!\!&-&\!\!\!\!\!
 \widetilde{N}_d\left[\widetilde{\psi}^6 m\widetilde{\psi}^L\Psi^L+{\cal
 H}_\Psi\right]\!\!-\!\!\left[\widetilde{N}_d\frac{\pi_0^2}{M^2}+\widetilde{\psi}^8{M^2}A^2_{(b)}\right]\!\!-\!
 \pi_0[{N^iA_i-A_0}]
 \\\nonumber
 \label{1-25}&+&\widetilde{N}_d\left[-J_{5(c)}v_{[ab]}\varepsilon_{(c)(a)(b)}
 +\frac{\widetilde{\psi}^4}{2}\left(v_{i(\rm A)}v^i_{(\rm A)}-
 \frac{1}{2}F_{ij}F^{ij}\right)
 \right],
 \eea
 where the Legendre transformation
 $A_0^2/(2\widetilde{N}_d)=\pi_0A_0-\widetilde{N}_d\pi^2_0/2$ with the subsidiary
 field $\pi_0$
 is used for linearizing the massive term;
  \be\label{1-26} {\cal
 H}_\Psi=
 -\widetilde{\psi}^4[i\widetilde{\psi}^L\gamma_{(b)}D_{(b)}\Psi^L +J^0_5 \sigma-
 \partial_kJ^k]
 \ee
 is the Hamiltonian density of the fermions,
 \bea\label{1-27}
 v_{[ab]}&=&\frac{1}{2}\left({\bf
e}_{(a)i}v^i_{(b)}-{\bf
 e}_{(b)i}v^i_{(a)}\right),\\\label{1-28}
 D_{(b)}\Psi^L&=&[\partial_{(b)}-\frac{1}{2}
 \partial_k {\bf e}^k_{(b)}-ieA_{(b)}]\Psi^L,\\\label{1-29}
 v_{i(\rm A)}&=&\frac{1}{\psi^4\widetilde{N}_d}[\partial_0A_i-\partial_iA_0+F_{ij}N^j]
 \eea
 are the field  velocities, and
 \bea\label{1-30}
 J_{5(c)}&=&\frac{i}{2}(\widetilde{\psi}^L\gamma_5
 \gamma_{(c)}\Psi^L),\\
 J^0_5&=&\frac{i}{2}(\bar\Psi^L\gamma_5\gamma^0\Psi^L),\\
 J_{k}&=&\frac{i}{2} \bar\Psi^L\gamma_{k}\Psi^L
 \eea
 are the currents, $\sigma=\sigma_{(a)(b)|(c)}
 \varepsilon_{(a)(b)(c)}$, where $\varepsilon_{(a)(b)(c)}$ denotes the Levi-Civita tensor.

 The  canonical conjugated momenta take the  form
\bea\label{1-31}
 P_\vh&=&-2V_0\frac{\partial_0\vh}{N_0}=-2V_0\frac{d\vh}{d\zeta}\equiv-2V_0
 \vh'\\\label{1-32}\overline{p_{\psi}}&=&\frac{\partial[\sqrt{-g}\mathcal{L}]}{\partial
 (\partial_0\ln{{\widetilde{\psi}}})}=-8\vh^2{\overline{v}},
 \\\label{1-33}
 p^i_{(b)}&=&\frac{\partial [\sqrt{-g}{\cal L}]}{\partial(\partial_0{\bf e}_{(a)i})}
 ={\bf e}^i_{(a)}\left[\frac{\vh^2}{3} v_{(a b)}-J_{5(c)}
 \varepsilon_{(c)(a)(b)}\right],\\\label{1-34}
 P^i_{\rm (A)}&=&\frac{\partial [\sqrt{-g}\mathcal{L}]}{\partial(\partial_0{ A}_{i})}=\widetilde{\psi}^4v^i_{\rm
 (A)},\\\label{1-35}
 P_{\rm (\Psi)}&=&\frac{\partial [\sqrt{-g}\mathcal{L}]}
 {{\partial(\partial_0{\Psi^L })}}=\frac{1}{i}\widetilde{\psi}^L\gamma^{0}.
 \eea
 Then, the action  (\ref{c1-1}) one can be represented in the Hamiltonian form
 \be\label{1-36}\nonumber
 S\!=\!\int dx^0\left[-P_{\vh}\partial_0\vh+
 N_0\frac{P^2_\vh}{4V_0}+\int d^3x
 \left(\sum\limits_{{F}
 } P_{F}\partial_0F
 +{\cal C}-\widetilde{N}_d T_{0t}^0\right)\right],
 \ee
 where $P_{F}$ is a set of the field momenta (\ref{1-32}) -- (\ref{1-35}),
\be\label{1-37}
 T_{0t}^0= \widetilde{\psi}^{7}\hat \triangle
\widetilde{\psi}+
  \sum\limits_{I=0,4,6,8} \widetilde{\psi}^I\tau_I,
 \ee
 is the sum of the Hamiltonian densities including the gravity density
  \bea\label{h31}
 \widetilde{\psi}^{7}\hat \triangle
 \widetilde{\psi}&\equiv&\widetilde{\psi}^{7}\dfrac{4\varphi^2}{3}
 \partial_{(b)}\partial_{(b)}\widetilde{\psi},\\\label{h32aa}
 \tau_{I=0}&=&\dfrac{6\widetilde{p}_{(ab)}\widetilde{p}_{(ab)}}{\vh^2}
 -\dfrac{16}{\vh^2}\overline{p_{\psi}}^2+\frac{\pi_0^2}{2a^2M^2},\\\label{h33}
 \tau_{I=4}\!\!\!&=&\!\!\!\frac{P^2_{(\rm A)}}{2}\!+\!
 \frac{F_{ij}F^{ij}}{4}\!-\!i\widetilde{\psi}^L\gamma_{(b)}D_{(b)}\Psi^L\!-\!J^0_5
 \sigma\!+\!
 \partial_kJ^k\!,\!\\\label{h34}
 \tau_{I=6}&=&m \widetilde{\psi}^L\Psi^L,\\\label{h35}
 \tau_{I=8}&=&\dfrac{\varphi^2}
  {6}R^{(3)}({\bf e})+\frac{M^2A^2_{(b)}}{2},
\eea
  here $\widetilde{p}_{(ab)}=\frac{1}{2}({\bf e}^i_{(a)}\widetilde{p}_{(b)i}+
  {\bf e}^i_{(b)}\widetilde{p}_{(a)i})$, $\widetilde{p}_{(b)i}={p}_{(b)i}+{\bf e}^i_{(a)}
  \varepsilon_{(c)(a)(b)}J_{(c)}$,
 \be\label{1-41}
 {\cal C}=A_0[\partial_iP^i_{(A)}+eJ_0-\pi_0]+N_{(b)}
  {T}^0_{(b)t} +\lambda_0\overline{p_\psi}+ \lambda_{(a)}\partial_k{\bf e}^k_{(a)}
 \ee
 denotes the sum of the constraints, where
 $J_0=\widetilde{\psi}^L\gamma_0\Psi^L$ is the zeroth component of the current; $A_0, N_d, N^i,
 \lambda_0,
\lambda_{(a)}$
 are the Lagrange multipliers  including the Dirac condition
 of the minimal 3-dimensional hyper-surface \cite{dir}
\be\label{1-42}
 p_{\widetilde{\psi}}=\overline{v}=0 \to
 (\partial_0-N^l\partial_l)\log{
 \widetilde{\psi}}=\frac16\partial_lN^l,
 \ee
 that gives a positive value of the Hamiltonian density
 (\ref{h32aa}), and
 \bea\label{1-43}\nonumber
 {T^0_{(a)}}_t&=&\!\!\!-\overline{p_{\psi}}\partial_{(a)}
 \widetilde{\psi}+\frac{1}{6}\partial_{(a)}
 (\overline{p_{\psi}}{\overline\psi}) +
 2p_{(b)(c)}\gamma_{(b)|(a)(c)}-\partial_{(b)}p_{(b)(a)}
 \\\nonumber&&\!\!\!\!\!\!-\frac{1}{i}\bar\Psi^I\gamma^{0}\partial_{(a)}
\Psi^I\!-\!\frac{1}{2i}
\partial_{(a)}\left(\bar\Psi^I\gamma^{0} \Psi^I\right)\!-\!P^i_{(\rm
A)}F_{ik}{\bf e}^k_{(a)}\!-\!\pi_0A_{(a)} \eea are the components of
the energy-momentum tensor \mbox{${T}^0_{(a)}={T}^0_{i}{\bf
e}^i_{(a)}$}.

\section{Vacuum creation of  particles}

\renewcommand{\theequation}{D.\arabic{equation}}

\setcounter{equation}{0}

\subsection{Particle in Quantum Field
Theory}

In quantum field theory, the concept of a particle can be associated
only with those field variables that are characterized by a positive
probability and a positive energy. Negative energies are removed by
causal quantization, according to which the creation operator at a
negative energy is replaced by the annihilation operator at the
respective positive energy. All of the variables that are
characterized by a negative probability can be removed according to
the scheme of fundamental operator quantization \cite{sch2}. The
results obtained by applying the operator-quantization procedure to
massive vector fields in the case of the conformal flat metric are
given in \cite{hp,114:a}\footnote{ The vacuum creation of particles
in the conformal flat metric was  considered in \cite{grib,grib80}
and, in the time reparametrization-invariant models,  in
\cite{ps1}.}.

In order to determine the evolution law for all fields  $\bf \rm v$,
 it is convenient to use the Hamiltonian
 form of the action functional for their Fourier components
${\bf \rm v}_k^{I}=\int\limits_{} d^3xe^{\imath\bf k\cdot x}{\bf \rm
v}^{I}({\bf x})$; that is, \bea \label{grad}
S&=&\int\limits_{x^0_1}^{x^0_2 }dx^0 \Bigg\{ \sum\limits_{k}
 \left[{\bf p}_{k}^{\bot}\partial_0{\bf \rm v}_{k}^{\bot}
 + {\bf p}_{k}^{||}\partial_0{\bf \rm v}_{k}^{||}\right]
-  P_{a} \partial_0a\nonumber\\
&+&{N}_{0}\left[\frac{P_{a}^2}{4V_{0}\vh_0^2}- {V_{0}\rho_{\rm tot}}
\right]\Bigg\}, \eea where ${\bf p}_{k}^{\bot},{\bf p}_{k}^{||}$ are
the canonical momenta for, respectively, the transverse and the
longitudinal component of vector bosons and $\rho_{\rm tot}$ is the
sum of the conformal densities of the scalar field obeying the rigid
equation of state and the vector field,
\begin{eqnarray}
\label{totgrad}
\rho_{\rm tot}(a)&=&\frac{\vh_0^2 H_0^2}{a^2}+\rho_{\rm v}(a),~~~~\\
\rho_{\rm v}(a)  &=&V_{0}^{-1}(H^{\bot} + H^{||}), \label{totgrad1}
\end{eqnarray}
$H^{\bot}$ and $H^{{||}}$ being the Hamiltonians for a free
field,\footnote{In quantum field theory, observables that are
constructed from the above field variables form the Poincar\'e
algebra \cite{hp,sch2}. Therefore, such a formulation, which depends
on the reference frame used, does not contradict the general theory
of irreducible and unitary transformations of the relativistic group
  \cite{Logunov,Schweber}.
}
\begin{eqnarray}
&H^{\bot} = \sum\limits_{k} \dfrac{1}{2}\left[{\bf p}_k^{\bot}{}^2 +
\omega^2 {\bf \rm v}_k^{\bot}{}^2\right]~, \nonumber\\
[-8mm]&\label{grad1}
\\&\nonumber
H^{||} = \sum\limits_{k} \dfrac{1}{2}\left
[\left(\dfrac{\omega(a,k)}{M_{\rm v} a}\right)^2{\bf p}_{k}^{||}{}^2
+ (M_{\rm v} a)^2 {\bf \rm v}_{k}^{||}{}^2 \right].
\end{eqnarray}
Here, the dispersion relation has the form  $\omega = \sqrt{{\bf
k^2} + (M_{\rm v} a)^2}$; for the sake of brevity, we have also
introduced the notation \mbox{${\bf p}_{k}^{||2}\equiv {\bf
p}_{k}^{||}\cdot{\bf p}_{-k}^{||}$}.

Within the reparametrization-invariant models specified by action
functionals of the type in (\ref{grad}) with the Hamiltonians in
(\ref{grad1}), the concepts of an observable particle and of
cosmological particle creation were defined in \cite{ps1,skok}. We
will illustrate these definitions by considering the example of an
oscillator with a variable energy. Specifically, we take its
Lagrangian in the form
 \be \label{la1}
 {\cal L}=p_{\rm v}\partial_0 {\rm v}-N_0\frac{1}{2}[p^2_{\rm v}+\omega^2{\rm v}^2-\omega]+\rho_0(N_0-1)~.
 \ee
The quantity  $H_{\rm v}=[p^2_{\rm v}+\omega^2{\rm v}^2]/2$ has the
meaning of a ``conformal Hamiltonian'' as a generator of the
evolution of the fields
 $v$ and $p_v$
with respect to the conformal-time interval  $d\eta=N_0dx^0$, where
the shift function $N_0$ plays the role of a Lagrange multiplier.
The equation for  $N_0$  introduces the density $\rho_0=H_{\rm
v}-\omega/2$ in accordance with its definition adopted in the
general theory of relativity. In quantum field theory
\cite{grib,ps1,skok}, the diagonalization of precisely the conformal
Hamiltonian \be \label{2la2} H_{\rm v}=\frac{1}{2}[p^2_{\rm
v}+\omega^2{\rm v}^2]=\omega\left[\hat N_{\rm
part}+\frac{1}{2}\right]
 \ee
specifies   both   the   single-particle   energy $\omega=\sqrt{{\bf
k}^2+(M_va(\eta))^2}$ and the particle-number operator \be
\label{la2} \hat N_{\rm part} =\frac{1}{2\omega}[p^2_{\rm
v}+\omega^2{\rm v}^2]-\frac{1}{2}
 \ee
with the aid of the transition to the symmetric variables $p$ and
$q$ defined as \be \label{la3}
 p_{\rm v}= \sqrt{\omega}p=i\sqrt{\frac{\omega}{2}}(a^+-a),
 ~~~~~~~~{\rm v}=\sqrt{\frac{1}{\omega}}q=\sqrt{\frac{1}{2\omega}}(a^++a).
 \ee
In terms of the symmetric variables $p,q$ the particle-number
operator takes form  \be \label{la4}
 \hat N_{\rm part}= \frac{1}{2}[p^2+q^2]-\frac{1}{2}=a^+a.
 \ee
Upon going over to these variables in the Lagrangian in (\ref{la1}),
we arrive at  \be \label{la5}
 {\cal L}=p\partial_0q-pq\partial_0 \Delta^{\bot}-N_0\omega[\hat N_{\rm part}+1/2]~,
 \ee
where $\partial_0 \Delta^{\bot}=\partial_0\omega/2\omega$ and where
there appears sources of cosmic particle creation in the form
$pq=i[(a^+)^2-a^2]/2$. Here, we give a derivation of these sources
for transverse fields, whereas, for longitudinal fields [see
Eq.(\ref{grad1})], the analogous diagonalization of the Hamiltonian
leads to the factor $\partial_0
\Delta^{||}=\partial_0\vh/\vh-\partial_0\omega/2\omega$.

In order to diagonalize the equations of motion in terms of the
mentioned new variables, it is necessary to apply, to the phase
space, the rotation transformation  \be \label{la6}
 p=p_\theta \cos\theta + q_\theta \sin\theta,~~~~~
 q=q_\theta \cos\theta - p_\theta \sin\theta
 \ee
and the squeezing phase space transformation  \be \label{la7}
 p_\theta =\pi e^{-r},~~~~~~~~q_\theta =\xi e^{+r}.
 \ee
As a result, the Lagrangian in (\ref{la5}) assumes the form  \be
\label{la8}
 {\cal L}=\pi\partial_0\xi+\pi\xi[\partial_0r-\partial_0 \Delta\cos2\theta]+
 \ee
 $$ +\frac{\pi^2}{2}e^{-2r}[\partial_0\theta-N_0\omega-
\partial_0 \Delta\sin2\theta ]+
\frac{\xi^2}{2}e^{2r}[\partial_0\theta-N_0\omega+
\partial_0 \Delta\sin2\theta ].
$$
The equations of motion that are obtained from this Lagrangian, \be
\label{la9}
 \xi'+\xi[r'- \Delta'\cos2\theta]+
  {\pi}e^{-2r}[\partial_0\theta-N_0\omega-
\partial_0 \Delta\sin2\theta ]=0,
\ee \be \label{la10}
 \pi'-\pi[r'- \Delta'\cos2\theta]
-{\xi}{2}e^{2r}[\partial_0\theta-N_0\omega+
\partial_0 \Delta\sin2\theta ]=0,
 \ee
take a diagonal form, \be \label{la11}
 \xi'+\omega_b\pi=0,~~~~~~~~~-\pi'+\omega_b\xi=0,
 \ee
 if \be\label{be}
 \omega_b=e^{-2r}[\omega-\theta'-\Delta'\sin2\theta]=
 \frac{\omega-\theta'}{{\rm ch} 2r}
 \ee
  and
the rotation parameter  $\theta$ and the squeezing parameter $r$
satisfy the equations\footnote{These equations for transverse and
longitudinal bosons coincide completely with the equations for the
coefficients of the Bogolyubov transformation $b=\alpha a+\beta
a^+$,~~ $\alpha'-i\omega\alpha=\Delta'\beta$,
 derived by using the Wentzel-Kramers-Brillouin method in \cite{grib80},
{see Eqs. (9.68) and (9.69) in \cite{grib80} on page 185 in the
Russian edition of this monograph}, where it is necessary to make
the change of variables specified by the equations $
\Delta'=\frac{\omega^{(1)}}{2},~ \alpha^*= \exp[i\theta-i\int d\eta
\omega]{\rm ch} r,~ \beta= \exp[-i\theta+i\int d\eta \omega]{\rm sh}
r. $} \be \label{la12} [\theta'-\omega]{\rm sh}2r
=-\Delta'\sin2\theta {\rm ch}2r,~~~~~~r'=\ \Delta'\cos2\theta .
 \ee
By solving these equations, we can find the time dependence
 of the number of particles produced in cosmic evolution (\ref{la4})
 \be \label{la13} \hat
N_{\rm part}=\frac{{\rm ch}2r-1}{2}+{\rm ch}2r\hat N_{\rm q-part}+
{\rm sh}2r\frac{\pi^2-\xi^2}{2},
 \ee
where $\hat N_{\rm q-part}=[\pi^2+\xi^2-1]/{2}=b^+b$ is the number
of quasiparticles defined as variables that diagonalize the equation
of motion. Since the equation of motion is diagonal, the number of
quasiparticles is an integral of the motion, that is, a quantum
number that characterizes the quantum state of the Universe.  One of
these states is the physical vacuum state $|0\rangle_{\rm sq}$ of
quasiparticles (that is, the squeezed vacuum, which is labelled with
the subscript  ``sq'' in order to distinguish it from the vacuum of
ordinary particles), \be\label{vacuum} b_{\varsigma}|0\rangle_{\rm
sq} = 0~~~~~(b=\frac{1}{\sqrt{2}}[\xi+i\pi]). \ee In the
squeezed-vacuum state, the number of quasiparticles is equal to
zeroth \be \label{la14} {}_{\rm sq}\langle 0|\hat N_{\rm
q-part}|0\rangle_{\rm sq}=0.
 \ee
In this case, the expectation value of the particle-number operator
(\ref{la13}) in the squeezed-vacuum state is \be \label{la15}
{}_{\rm sq}\langle 0|\hat N_{\rm part}|0\rangle_{\rm sq}=\frac{{\rm
ch}(2r(\eta))-1}{2} ={\rm sh}^2r(\eta).
 \ee
The time dependence of this quantity is found by solving the
Bogolyubov equation  (\ref{la12}). The origin of the Universe is
defined as the conformal-time instant $\eta=0$,
 at which the number of particles and the number of
 quasiparticles are both equal to zeroth.
The resulting set of Eqs. (\ref{la12}) becomes closed upon
specifying the equation of state and initial data for the number of
particles. In just the same way, the number of particles
characterized by an arbitrary set of quantum numbers  $\varsigma$,
\be {\cal N}_{\varsigma}(\eta) = {}_{\rm sq} \langle 0|\hat
N_{\varsigma}|0\rangle_{\rm sq} = {\rm sh}^2
r_{\varsigma}(\eta),\nonumber \ee and produced from the ``squeezed''
vacuum by the time instant $\eta$ can be determined by solving an
equation of the type in (\ref{la12}).

Thus, just the conformal quantities of the theory, such as the
energy  $\omega_k=\sqrt{k^2+M^2_va^2}$, the number particles $\hat
N_{\rm part}$, the conformal density $$\rho_{\rm {\rm v}} = \sum_k
{}_{\rm sq}\langle 0|\hat N_{k ~\rm part}|0\rangle_{\rm
sq}\omega_k/V_{(r)}$$ that are associated with observables,
  in just the same way as the conformal time in
  observational cosmology is associated with
  the observed time \cite{039}.

\subsection{Physical implications}

\subsubsection{Calculation of the Distribution Function}

Let us consider the example where the above set of equations is
solved for the evolution law in the case of the rigid equation of
state,
 $$a(\eta)=a_I\sqrt{1+2H_I\eta}~~~~~(a^2_IH_I=H_0),$$
where $a_I=a(0)$ and $H_I$ are initial data at the matter-production
instant.

We introduce the dimensionless variables of time  $\tau$ and
momentum $x$ and the coefficient
 $\gamma_{\rm v}$ according to the formulas
\be\label{gas}
 \tau=2\eta H_I=\eta/\eta_I,~~~~~~~~ x=\frac{q}{M_{I}},
 ~~~~~~~~\gamma_{\rm v}=\frac{M_{I}}{H_I},
\ee where $M_{I}=M_{\rm v}(\eta=0)$  are initial data for the mass.
Now the single-particle energy has the form
 $\omega_{\rm v}=H_I\gamma_{\rm v}\sqrt{1+\tau+x^2}$.

The Bogolyubov equations~(\ref{la12}) can be represented as \bea
\tanh(2r^{||}_{\rm v}) &=&
-\dfrac{\dfrac{1}{2(1+\tau)}-\dfrac{1}{4\left[
(1+\tau)+x^2\right]}}{\dfrac{\gamma_{\rm v}}{2}\sqrt{(1+\tau)+x^2} -
\dfrac{d\theta^{||}_{\rm v}}{d\tau}}\sin(2\theta^{||}_{\rm v}),\\
\frac{dr^{||}_{\rm v}}{d\tau} &=&
\left[\frac{1}{2(1+\tau)}-\frac{1}{4\left[
(1+\tau)+x^2\right]}\right] \cos(2\theta^{||}_{\rm v}),\\\label{1g}
\tanh(2r^{\bot}_{\rm v}) &=& -\dfrac{\dfrac{1}{4\left[
(1+\tau)+x^2\right]}}{\dfrac{\gamma_{\rm v}}{2}\sqrt{(1+\tau)+x^2} -
\dfrac{d}{d\tau}\theta^{\bot}_{\rm v}} \sin(2\theta^{\bot}_{\rm
v}),\\
\frac{dr^{\bot}_{\rm v}}{d\tau}&=& \frac{1}{4\left[
(1+\tau)+x^2\right]}\cos(2\theta^{\bot}_{\rm v}). \eea We solved
these equations numerically at positive values of the momentum
$x=q/M_I$, considering that, for $\tau\to+0$, the asymptotic
behavior of the solutions is given by $r(\tau)\to{\rm
const}\cdot\tau$ and $\theta(\tau)=O(\tau)$.
 The distributions of
longitudinal ${\cal N}^{||}(x,\tau)$ and transverse ${\cal
N}^{\bot}(x,\tau)$ vector bosons are given in the Figure 1. for the
initial data $H_I=M_I~(\gamma_{\rm v}=1$).

From the Figure 1, it can be seen that, for $x>1$, the longitudinal
component of the boson distribution is everywhere much greater than
than the transverse component, this demonstrating a more copious
cosmological creation of longitudinal bosons in relation to
transverse bosons. A slow decrease in the longitudinal component as
a function of momentum leads to a divergence of the integral for the
density of product particles~\cite{114:a}: \be\label{nb}
 n_{\rm v}(\eta)=\frac{1}{2\pi^2}
\int\limits_{0 }^{\infty } dq q^2 \left[ {\cal N}^{||}(q,\eta) +
2{\cal N}^{\bot}(q,\eta)\right]\to\infty. \ee

\subsubsection{Thermalization of Bosons }

The divergence of the integral (\ref{nb}) stems from idealizing the
problem of the production of a pair of particles in a finite volume
for a system where there are simultaneous interactions associated
with the removal of fields having a negative probability and where
identical particles affect one another (so-called exchange effects).
In this case, it is well known~\cite{ll}, that one deals with the
production not a pair but a set of Bose -- particles, which
acquires, owing to the aforementioned interactions, the properties
of a statistical system.   As a model of such a statistical system,
we consider here a degenerate Bose-Einstein gas, whose distribution
function has the form  (we use the system of units where the
Boltzmann constant is
 $k_{\rm B}=1$) \be \label{bose} {\cal F}\left(T_{\rm v},q,M_{\rm v}(\eta),\eta\right)=
 \left\{\exp\left[\frac{\omega_{\rm v}(\eta)- M_{\rm v}(\eta)}{ T_{\rm v}}\right]
 -1\right\}^{-1}, \ee
 where $T_{\rm v}$ is the boson temperature.
 We set apart the problem of theoretically validating
 such a statistical system and its thermodynamic exchange,
 only assuming fulfillment of specific conditions ensuring
 its existence. In particular, we can introduce the notion of the temperature
  $T_{\rm v}$ only in an equilibrium system.
 A thermal equilibrium is thought to be stable if the time within which the
 vector-boson temperature
 $T_{\rm v}$ is established, that is, the relaxation time \cite{ber} \be \label{rel01}
 \eta_{\mbox{\small  rel}} =
 \left[{n(T_{\rm v})\sigma_{\mbox{\small scat}}}\right]^{-1}
 \ee (expressed in terms of their density  $n(T_{\rm v})$
 and the scattering cross section
 $\sigma_{\mbox{\small  scat.}} \sim 1/ M_{I}^2$), does not exceed the
 time of vector-boson-density formation owing to cosmological creation,
 the latter time being controlled by the primordial Hubble parameter
  $\eta_{\rm v}=1/ H_I$. From formula (\ref{rel01}) it follows,
  that the particle-number density is proportional to the product of
  the Hubble parameter and the mass squared, that is
  an integral of the motion in the present example:
  \be \label{bose11}
 n(T_{\rm v})=n(T_{\rm v},\eta_{\rm v})\simeq
  C_H H_IM_I^2,
\ee where $C_H$ is a constant. The expression for the density $
n(T_{\rm v},\eta)$ in Eq. (\ref{bose11}) assumes the form
\be\label{n1}
 n_{{\rm v}}(T_{\rm v},\eta)=\frac{1}{2\pi^2}
\int\limits_{0 }^{\infty } dq q^2{\cal F}\left(T_{\rm
v},q,M(\eta),\eta\right) \left[ {\cal N}^{||}(q,\eta) + 2{\cal
N}^{\bot}(q,\eta)\right]~. \ee Here, the probability of the
production of a longitudinal and a transverse boson with a specific
momentum in an ensemble featuring exchange interaction is given (in
accordance with the multiplication law for probabilities) by the
product of two probabilities, the probability of their cosmological
creation,  ${\cal N}^{||,\bot}$ and the probability of
 a single-particle state of vector bosons obeying
 the Bose-Einstein distribution~(\ref{bose}).

A dominant contribution to the integral (\ref{n1}) from the region
of high momenta (in the above idealized analysis disregarding the
Boltzmann factor, this resulted in a divergence)
 implies the relativistic temperature dependence of the density,
 \be \label{nc1}
 n(T_{\rm v},\eta_{\rm v}) = C_T T_{\rm v}^3,
 \ee
where $C_T$ is a coefficient. A numerical calculation of the
integral (\ref{n1}) for the values  $T_{\rm v}=M_I=H_I$, which
follow from the assumption about the choice of initial data
($C_T=C_H$), reveals that this integral (\ref{n1}) is weakly
dependent on time in the region  $\eta \geq \eta_{\rm v}=H_I^{-1}$
and, for the constant $C_T$, yields the value \be \label{nc} C_T =
\frac{ n_{\rm v}}{ T_{\rm v}^3} = \frac{1}{2\pi^2} \left\{
[1,877]^{||}+2 [0,277]^{\bot}=2,431 \right\}~, \ee where the
contributions of longitudinal and transverse bosons are labeled with
the superscripts  $(||,~ \bot)$, respectively.

 On the other hand, the lifetime $\eta_L$ of
 product bosons in the early Universe in dimensionless units
$\tau_L=\eta_L/\eta_I$, where $\eta_I=(2H_I)^{-1}$, can be estimated
by using the equation of state $a^2(\eta)=a_I^2(1+\tau_L)$ and the
$W$-boson lifetime within the Standard Model. Specifically, we have
\be \label{life} 1+\tau_L= \frac{2H_I\sin^2
\theta_{(W)}}{\alpha_{\rm QED} M_W(\eta_L)}= \frac{2\sin^2
\theta_{(W)}}{\alpha_{\rm QED}\gamma_{\rm v}\sqrt{1+\tau_L}}, \ee
where $\theta_{(W)}$ is the Weinberg angle,  $\alpha_{\rm
QED}=1/137$ is the fine-structure constant, and $\gamma_{\rm
v}=M_{I}/ H_I\geq 1$.

From the solution to Eq.~(\ref{life}), \be \label{lifes} \tau_L+1=
\left(\frac{2\sin^2\theta_{(W)}}{\gamma_{\rm v}\alpha_{\rm
QED}}\right)^{2/3} \simeq \frac{16}{\gamma_{\rm v}^{2/3}}~ \ee it
follows that, at $\gamma_{\rm v}=1$, the lifetime of product bosons
is an order of magnitude longer than the Universe relaxation time:
\be \label{lv} \tau_L =\frac{\eta_L}{\eta_I}\simeq
\frac{16}{\gamma_{\rm v}^{2/3}}-1=15. \ee

Therefore, we can introduce the notion of the vector-boson
temperature $T_{\rm v}$, which is inherited by the final vector
boson decay products (photons). According to currently prevalent
concepts, these photons form cosmic microwave background radiation
in the Universe. Indeed, suppose that one photon comes from the
annihilation of the products of $W^\pm$-boson decay and that the
other comes from  $Z$-bosons. In view of the fact that the volume of
the Universe is constant within the evolution model being
considered, it is then natural to expect that the photon density
coincides with the boson density \cite{114:a} \be \label{1nce}
n_\gamma={T_\gamma^3}\frac{1}{\pi^2} \left\{ 2.404 \right\} \simeq
n_{\rm v}. \ee

On the basis of  (\ref{bose11}), (\ref{nc1}), (\ref{nc}) and
(\ref{1nce}) we can estimate the temperature $T_{\gamma}$ of cosmic
microwave background radiation arising upon the annihilation and
decay of $W^+ $ and $Z$-bosons: \be \label{1nce1} T_\gamma\simeq
\left[\frac{ 2.431}{2.404 \cdot 2 }\right]^{1/3}T_{\rm v}=0.8 T_{\rm
v}, \ee   taking into account that the temperature of vector-bosons
$T_{\rm v}= [H_IM^{2}_I]^{1/3}$ is an invariant quantity in the
described model. This invariant can be estimated at \be T_{\rm v}=
[H_IM^{2}_I]^{1/3}=[H_0 M_W^2]^{1/3}=2.73/0.8K=3.41K \ee which is a
value that is astonishingly close to the observed temperature of
cosmic microwave background radiation. In the present case, this
directly follows, as is seen from the above analysis of our
numerical calculations, from the dominance of longitudinal vector
bosons with high momenta and from the fact that the relaxation time
is equal to the inverse Hubble parameter. The inclusion of physical
processes, like the heating of photons owing to electron-positron
annihilation
  $e^+~e^-$  \cite{ee}
amounts to multiplying the photon temperature (\ref{1nce1}) by
$(11/4)^{1/3}=1.4$ therefore, we have \be \label{1nce11}
T_\gamma(e^+~e^-)\simeq (11/4)^{1/3}0.8 T_{\rm v}=2.77 ~K~. \ee We
note that, in other models  \cite{mar}, the fluctuations of the
product-particle density are related to primary fluctuations of
cosmic microwave background radiation   \cite{33}.

 \subsubsection{Inverse Effect of Product
Particles on the Evolution of the Universe}

The equation of motion
 $\vh'^2(\eta)=\rho_{\rm tot}(\eta)$, with the
Hubble parameter defined as $H=\vh'/\vh$, means that, at any instant
of time, the energy density in the Universe is equal to the
so-called
 critical density; that is
$$
 \rho_{\rm tot}(\eta)= H^2(\eta)\vh^2(\eta)
\equiv \rho_{\rm cr}(\eta)~.
$$
The dominance of matter obeying the extremely rigid equation of
state implies the existence of an approximate integral of the motion
in the form
$$ H(\eta)\vh^2(\eta)=H_0\vh_0^2~.
$$

 On this basis, we can immediately find the ratio of the
 product-vector-boson energy, $ \rho_{\rm v}(\eta_I)\sim  T^4\sim
 H_I^4\sim M^4_{I}$, to the density of the Universe in the extremely
 rigid state,  $ \rho_{\rm tot}(\eta_I)=H_I^2\vh^2_{I}$,
 \be
 \label{Fas} \frac{\rho_{\rm v}(\eta_I)}{\rho_{\rm tot}(\eta_I)}=
 \frac{ M^2_{I}}{\vh_I^2} =\frac{M^2_{W}}{\vh_0^2}=y^2_{\rm
 v}=10^{-34}.
 \ee
 This value indicates that the inverse effect of product
particles on the evolution of the Universe is negligible.

{The primordial mesons before
 their decays polarize the Dirac fermion vacuum and give the
 baryon asymmetry frozen by the CP -- violation,
 so that $n_b/n_\gamma \sim X_{CP} \sim 10^{-9}$ and
 $\Omega_b \sim \alpha_{\rm qed}/\sin^2\theta_{(\rm W)}\sim
 0.03$.}

\subsubsection{\label{3.4}Baryon-antibaryon Asymmetry of Matter in
the Universe}

In each of the three generations of leptons
 (e,$\mu$,$\tau$) and color quarks, we have four fermion
 doublets-in all, there are $n_L=12$ of them. Each of 12 fermion
 doublets interacts with the triplet of non-Abelian fields
 $A^1=(W^{(-)}+W^{(+)})/\sqrt{2}$, $A^2=
 i(W^{(-)}-W^{(+)})/\sqrt{2}$, and $A^3=Z/\cos\theta_{(W)}$,
  the corresponding coupling constant being $g=e/\sin\theta_{(W)}.$

 It is well known that, because of a triangle anomaly, W- and
Z- boson interaction with lefthanded fermion doublets
 $\psi_L^{(i)}$, $i=1,2,...,n_L$, leads to
a nonconservation of the number of fermions of each type ${(i)}$
 ~\cite{bj,th,ufn},
 \bea \label{rub}
 \partial_\mu j^{(i)}_{L\mu}=\frac{1}{32\pi^2}
 {\rm Tr}\hat F_{\mu\nu}{}^*\!{\hat F_{\mu\nu}},
 \eea
 where $\hat
 F_{\mu\nu}=-iF^a_{\mu\nu}g_W\tau_a/2$ is the strength of the
 vector fields, $F^a_{\mu\nu}=
 \partial_\mu A_\nu^a-\partial_\nu
 A_\mu^a+g\epsilon^{abc}A_\mu^bA_\nu^c$.

 Taking the integral of the equality in (\ref{rub}) with respect to
 the four-dimensional variable $x$, we can find a relation between
 the change   $\Delta F^{(i)}=\int d^4x
 \partial_\mu j^{(i)}_{L\mu}$ the fermion number $ F^{(i)}=\int d^3x
 j_0^{(i)}$ and the Chern-Simons functional,
 $N_{CS}=\frac{1}{32\pi^2}\int d^4x {\rm Tr}\hat
 F_{\mu\nu}{}^*\!{\hat F_{\mu\nu}}$: \be \label{rub2}
 \Delta F^{(i)}= N_{CS} \not = 0, ~~~i=1,2,...,n_L.
 \ee
 The equality in (\ref{rub2}) is considered as a selection rule --
that is, the fermion number changes identically for all fermion
types:
 $N_{CS}=\Delta L^e=\Delta L^\mu=\Delta L^\tau=\Delta B/3$;
 at the same time, the change in the baryon charge $B$ and the change
 in the lepton charge  $L=L^e+L^\mu+L^\tau$ are related to each other in
such a way that $B-L$ is conserved, while
 $B+L$ is not invariant. Upon taking the sum of the equalities in
 (\ref{rub2}) over all doublets, one can  obtain $\Delta (B+ L)=12
 N_{CS}$ ~\cite{ufn}.

 We can evaluate the expectation value of the Chern-Simons
 functional (\ref{rub2})  (in the lowest order of perturbation
 theory in the coupling constant) in the Bogolyubov vacuum
 $b|0>_{\rm sq}=0$. Specifically, we have
 \be
 N_{CS}=N_{\rm W}+N_{\rm Z}\equiv
 -\sum_{{\rm v}=W,Z}\int\limits_0^{\eta_{L_{\rm v}}} d\eta \int \frac{d^3 x}{32\pi^2} \;
 {}_{\rm sq}\langle 0|{\rm Tr}\hat F^{\rm v}_{\mu\nu}
 {}^*\!{\hat F^{\rm v}_{\mu\nu}}|0\rangle{}_{\rm sq} ,
 \ee
 where $\eta_{L_W}$ and  $\eta_{L_Z}$ are the W- and the Z-boson
 lifetime, and $N_{\rm W}$ and $N_{\rm Z}$
 are the contributions
 of primordial W and Z bosons, respectively.
 The integral over the conformal spacetime bounded
 by three-dimensional hypersurfaces $\eta=0$ and $\eta =\eta_L$
  is given by
 $$N_{\rm v} =\beta_{\rm v}\frac{V_0}{2}
 \int_{0}^{{\eta_{L_{\rm v}}}} d\eta \int\limits_{0 }^{\infty }dk
 |k|^3 R_{\rm v}(k,\eta)
 $$
 where ${\rm v}=W,Z$;
 \be\beta_W=\frac{4{\alpha}_{\rm
 QED}}{\sin^{2}\theta_{(W)}}, ~~\beta_Z=\frac{{\alpha}_{\rm
 QED}}{\sin^{2}\theta_{(W)}\cos^{2}\theta_{(W)}},\ee
 and the rotation parameter
 $$R_{\rm v}=-\sinh(2r)\sin(2\theta)
 $$
  is
 specified by relevant solutions to the
 Bogolyubov equations (\ref{1g}). Upon a numerical calculation
 of this integral, we can estimate the expectation value of the
 Chern-Simons functional in the state of primordial bosons.

 At the vector-boson-lifetime values of
 $\tau_{L_W}= 15$, $\tau_{L_Z}= 30$,
 this yields the following result at $n_\gamma\simeq n_{\rm v}$
\be
\frac{N_{CS}}{V_{(r)}}=\frac{( N_W+ N_Z)}{V_{(r)}}\\
 =\frac{{\alpha}_{\rm QED}}{\sin^{2}\theta_{(W)}}
  T^3
  \left(4\times 1.44+\frac{2.41}{\cos^{2}\theta_{(W)}}\right)
  =1.2~  n_{\gamma}.
\ee On this basis, the violation of the fermion-number density in
the cosmological model being considered can be estimated as
\cite{114:a,039}
\begin{eqnarray}
\frac{\Delta F^{(i)}}{V_{(r)}}&=&\frac{N_{CS}}{V_{(r)}}
  =1.2  n_{\gamma},
\end{eqnarray}
 where $n_{\gamma}={ 2,402  \times T^3 }/{\pi^2}$ is the number
 density of photons forming cosmic microwave background radiation.

 According to Sakharov \cite{sufn} this violation of the
 fermion number is frozen by ${\rm CP}$ nonconservation, this
 leading to the baryon-number density
 \be\label{X} n_{\rm b}=
  X_{\rm CP}\frac{\Delta
  F^{(i)}}{V_{(r)}}\simeq X_{\rm CP}n_{\gamma}~.
  \ee
  where the factor $X_{\rm CP}$ is determined by the superweak
 interaction of $d$ and $s$ quarks,
 which
 is responsible for CP violation experimentally observed in
 $K$-meson decays \cite {o}.

 From the ratio of the number of baryons to the number of photons,
 one can deduce an estimate of the superweak-interaction coupling
 constant: $X_{\rm CP}\sim 10^{-9}$.
  Thus, the evolution of the Universe, primary
 vector bosons, and the aforementioned superweak interaction \cite{o}
 (it is responsible for CP violation and is characterized by a
 coupling-constant value of $X_{CP}\sim
10^{-9}$) lead to baryon-antibaryon
 asymmetry of the Universe, the respective baryon density being
 \be\label{data6} \rho_{\rm b}(\eta=\eta_{L})
 \simeq 10^{-9} \times 10^{-34}\rho_{\rm cr}(\eta=\eta_{L}).
 \ee
 In order to assess the
further evolution of the baryon density, one can take here the
W-boson lifetime for $\eta_{L}$.

 Upon the decay of the vector bosons in question,
their temperature is inherited by cosmic microwave background
radiation. The subsequent evolution of matter in a stationary cold
universe is an exact replica of the well-known scenario of a hot
universe \cite{three}, since this evolution is governed by
conformally invariant mass-to-temperature ratios $m/T$.

Formulas (\ref{life}), (\ref{F}), and (\ref{data6}) make it possible
to assess the ratio of the present-day values of the baryon density
and the density of the scalar field, which plays the role of
primordial conformal quintessence in the model being considered. We
have
  \be \Omega_{\rm
 b}(\eta_0)=\frac{\rho_{\rm b}(\eta_{0})}{\rho_{\rm cr}(\eta_{0})}=
 \left[\frac{\vh_0}{\vh_L}\right]^3=\left[\frac{\vh_0}{\vh_I}\right]^3
 \left[\frac{\vh_I}{\vh_L}\right]^3,
 \ee
 where we have considered that the baryon density increases in
 proportion to the mass and that the density of the primordial
 quintessence  decreases in inverse proportion to the mass
 squared. We recall that the ratio $[{\vh_0}/{\vh_I}]^3$
  is approximately equal
 to $10^{43}$ and that the ratio $[{\vh_I}/{\vh_L}]^3$ is
 determined by the boson
 lifetime in (\ref{lifes}) and by the equation of state
 $\vh(\eta)\sim \sqrt{\eta}$. On this basis, we can estimate
 $\Omega_{\rm b}(\eta_0)$ at
 \be\label{data7} \Omega_{\rm b}(\eta_0)
 =\left[\frac{\vh_0}{\vh_L}\right]^3 10^{-43}
 \sim 10^{43} \left[\frac{\eta_I}{\eta_L}\right]^{3/2}10^{-43}
 \sim \left[\frac{\alpha_{QED}}{\sin^2 \theta_{(W)}}\right] \sim
 0.03 ,
 \ee
 which is compatible with observational data \cite{fuk}.

 Thus, the general theory of relativity and the Standard Model,
 which are supplemented with a free scalar field   in a specific
 reference frame with the initial data $\vh_I=10^{4}$
 $H_I=2.7~{\rm K}$,
do not contradict the following scenario of the evolution of the
Universe within conformal cosmology \cite{114:a,039}:\\[1.5mm]
 $\eta \sim 10^{-12}s,$ {creation of vector bosons from a
 ``vacuum''};\\ [1.5mm] $10^{-12}s < \eta <
10^{-11}\div 10^{-10} s,$ {formation of baryon-antibaryon asymmetry;}\\
[1.5mm] $\eta \sim 10^{-10}s,$ {decay of vector bosons;}\\
[1.5mm] $10^{-10}c <\eta < 10^{11}s,$ { primordial chemical
evolution of matter;}\\ [1.5mm] $\eta \sim 10^{11}s,$ {recombination
or separation of cosmic
microwave background radiation;}\\
[1.5mm] $\eta \sim  10^{15}s,$ {formation of galaxies;}\\
[1.5mm]  $\eta > 10^{17}s,$ { terrestrial experiments and evolution
of supernovae.}

\end{document}